\documentclass[journal]{IEEEtran}

\usepackage{amsmath}
\usepackage{amssymb}
\usepackage{amsthm}
\usepackage{graphicx}
\usepackage{color}
\usepackage{mathrsfs}
 
\usepackage{epsfig}
\usepackage[numbers]{natbib}
\usepackage[linesnumbered,ruled]{algorithm2e}
\usepackage{url}

\begin{document}


 


\newcommand{\be}{\begin{equation}}
\newcommand{\ee}{\end{equation}}
\newcommand{\bes}{\begin{equation*}}
\newcommand{\ees}{\end{equation*}}
\newcommand{\beqn}{\begin{eqnarray}}
\newcommand{\eeqn}{\end{eqnarray}}
\newcommand{\beqns}{\begin{eqnarray*}}
\newcommand{\eeqns}{\end{eqnarray*}}
\newcommand{\begal}{\begin{align}}
\newcommand{\egal}{\end{align}}
\newcommand{\beals}{\begin{align*}}
\newcommand{\eeals}{\end{align*}}

\newcommand{\lkr}{\left(}
\newcommand{\lkv}{\left[}
\newcommand{\rkv}{\right]}
\newcommand{\rkr}{\right)}
\newcommand{\lfi}{\left\{}
\newcommand{\rfi}{\right\}}
\newcommand{\lnor}{\left\|}
\newcommand{\rnor}{\right\|}

\newcommand{\fr}[1]{(\ref{#1})}

\newcommand{\vart}{\vartheta}
\newcommand{\ro}{\varrho}
\newcommand{\ph}{\varphi}
\newcommand{\del}{\delta}
\newcommand{\Del}{\Delta}
\newcommand{\dn}{\delta_n}
\newcommand{\al}{\alpha}
\newcommand{\af}{\alpha}
\newcommand{\eps}{\epsilon}
\newcommand{\Ga}{\Gamma}
\newcommand{\ga}{\gamma}
\newcommand{\te}{\theta}
\newcommand{\om}{\omega}
\newcommand{\lam}{\lambda}
\newcommand{\Ups}{\Upsilon}
\newcommand{\up}{\upsilon}
\newcommand{\dz}{\zeta}
\newcommand{\sig}{\sigma}
\newcommand{\sgmd}{\sigma^2}
\newcommand{\Lam}{\Lambda}
\newcommand{\Om}{\Omega}
\newcommand{\Sig}{\Sigma}
\newcommand{\Te}{\Theta}

\newcommand{\EE}{\ensuremath{{\mathbb E}}}
\newcommand{\JJ}{\ensuremath{{\mathbb J}}}
\newcommand{\II}{\ensuremath{{\mathbb I}}}
\newcommand{\ZZ}{\ensuremath{{\mathbb Z}}}
\newcommand{\PP}{\ensuremath{{\mathbb P}}}
\newcommand{\QQ}{\ensuremath{{\mathbb Q}}}
\newcommand{\KK}{\ensuremath{{\mathbb K}}}
\newcommand{\RR}{{\mathbb R}}
\newcommand{\real}{{\mathbb R}}

\newcommand{\iter}{\mathrm{iter}}
\newcommand{\low}{\mbox{low}}
\newcommand{\vect}{\mbox{vec}}
\newcommand{\Pen}{\mbox{Pen}}
\newcommand{\Span}{\mbox{Span}}
\newcommand{\intg}{\mbox{int}}
\newcommand{\card}{\mbox{card}}
\newcommand{\Range}{\mbox{Range}}
\newcommand{\Var}{\mbox{Var}}
\newcommand{\Cov}{\mbox{Cov}}
\newcommand{\diag}{{\rm diag}}
\newcommand{\supp}{\mbox{supp}}
\newcommand{\etal}{{\it et  al. }}
\newcommand{\std}{\mbox{std}}
\newcommand{\SNR}{\mbox{SNR}}
\newcommand{\Tr}{\mbox{Tr}}
\newcommand{\proj}{\mbox{proj}}
\newcommand{\Expect}{\mathrm{Expect}}
\newcommand{\Skew}{\mathrm{Skew}}
\newcommand{\Kmax}{K_{\max}}
\newcommand{\Kmin}{K_{\min}}
\newcommand{\argmin}{\mbox{argmin}}
\newcommand{\argmax}{\mbox{argmax}}
\newcommand{\mat}{\mbox{mat}}
\newcommand{\tr}{\mbox{tr}}
\newcommand{\rank}{\mbox{rank}}
\newcommand{\Reg}{\mbox{Reg}}
\newcommand{\SVD}{\mbox{SVD}}
\newcommand{\sgn}{\mbox{sign}}
\newcommand{\iid}{\mbox{i.i.d.}}

\newtheorem{thm}{Theorem}
\newtheorem{lem}{Lemma}
\newtheorem{cor}{Corollary}
\newtheorem{rem}{Remark}
\newtheorem{ex}{Example}
\newtheorem{prop}{Proposition}

\newcommand{\bbA}{\boldsymbol{A}}
\newcommand{\bbB}{\boldsymbol{B}}
\newcommand{\bB}{\boldsymbol{B}}
\newcommand{\bbC}{\boldsymbol{C}}
\newcommand{\bL}{\mathbf{L}}
\newcommand{\bfe}{\mathbf{e}}
\newcommand{\bbP}{\boldsymbol{P}}
\newcommand{\bbQ}{\boldsymbol{Q}}
\newcommand{\bbX}{\boldsymbol{X}}
\newcommand{\bbY}{\boldsymbol{Y}}
\newcommand{\bbZ}{\boldsymbol{Z}}
\newcommand{\bbDelta}{\boldsymbol{\Delta}}

\newcommand{\bDelta}{\mathbf{\Delta}}

\newcommand{\ba}{\mathbf{a}}
\newcommand{\bb}{\mathbf{b}}
\newcommand{\bc}{\mathbf{c}}
\newcommand{\bd}{\mathbf{d}}
\newcommand{\boe}{\mathbf{e}}
\newcommand{\bof}{\mathbf{f}}
\newcommand{\bg}{\mathbf{g}}
\newcommand{\bi}{\mathbf{i}}
\newcommand{\bj}{\mathbf{j}}
\newcommand{\bk}{\mathbf{k}}
\newcommand{\bh}{\mathbf{h}}
\newcommand{\bp}{\mathbf{p}}
\newcommand{\bq}{\mathbf{q}}
\newcommand{\bt}{\mathbf{t}}
\newcommand{\bu}{\mathbf{u}}
\newcommand{\bv}{\mathbf{v}}
\newcommand{\bw}{\mathbf{w}}
\newcommand{\bx}{\mathbf{x}}
\newcommand{\by}{\mathbf{y}}
\newcommand{\bz}{\mathbf{z}}

\newcommand{\bA}{\mathbf{A}}
\newcommand{\boB}{\mathbf{B}}
\newcommand{\bC}{\mathbf{C}}
\newcommand{\bD}{\mathbf{D}}
\newcommand{\bE}{\mathbf{E}}
\newcommand{\bF}{\mathbf{F}}
\newcommand{\bG}{\mathbf{G}}
\newcommand{\bH}{\mathbf{H}}
\newcommand{\bI}{\mathbf{I}}
\newcommand{\bJ}{\mathbf{J}}
\newcommand{\bK}{\mathbf{K}}
\newcommand{\bM}{\mathbf{M}}
\newcommand{\bO}{\mathbf{O}}
\newcommand{\bP}{\mathbf{P}}
\newcommand{\bQ}{\mathbf{Q}}
\newcommand{\bR}{\mathbf{R}}
\newcommand{\bS}{\mathbf{S}}
\newcommand{\bU}{\mathbf{U}}
\newcommand{\bV}{\mathbf{V}}
\newcommand{\bW}{\mathbf{W}}
\newcommand{\bX}{\mathbf{X}}
\newcommand{\bY}{\mathbf{Y}}
\newcommand{\bZ}{\mathbf{Z}}

\newcommand{\bzero}{\mathbf{0}}
\newcommand{\bone}{\mathbf{1}}

 \newcommand{\blam}{\mbox{\mathversion{bold}$\lam$}}
 \newcommand{\bte}{\mbox{\mathversion{bold}$\te$}}
\newcommand{\bxi}{\mbox{\mathversion{bold}$\xi$}}
\newcommand{\boeta}{\mbox{\mathversion{bold}$\eta$}}
\newcommand{\bbe}{\mbox{\mathversion{bold}$\beta$}}
\newcommand{\bzeta}{\mbox{\mathversion{bold}$\zeta$}}
\newcommand{\bph}{\mbox{\mathversion{bold}$\ph$}}
\newcommand{\bpsi}{\mbox{\mathversion{bold}$\psi$}}
\newcommand{\beps}{\mbox{\mathversion{bold}$\eps$}}
\newcommand{\bobeta}{\mbox{\mathversion{bold}$\beta$}}
\newcommand{\bgamma}{\mbox{\mathversion{bold}$\gamma$}}
\newcommand{\bom}{\mbox{\mathversion{bold}$\om$}}

\newcommand{\bPsi}{\mbox{\mathversion{bold}$\Psi$}}
\newcommand{\bPhi}{\mbox{\mathversion{bold}$\Phi$}}
\newcommand{\bUp}{\mbox{\mathversion{bold}$\Up$}}
\newcommand{\bLam}{\mbox{\mathversion{bold}$\Lambda$}}
\newcommand{\bSig}{\mbox{\mathversion{bold}$\Sigma$}}
\newcommand{\bTe}{\mbox{\mathversion{bold}$\Theta$}}
\newcommand{\bXi}{\mbox{\mathversion{bold}$\Xi$}}
\newcommand{\bcalL}{\mbox{\mathversion{bold}$\calL$}}

\newcommand{\calA}{{\mathcal{A}}}
\newcommand{\calB}{{\mathcal{B}}}
\newcommand{\calC}{{\mathcal{C}}}
\newcommand{\calD}{{\mathcal{D}}}
\newcommand{\calE}{{\mathcal{E}}}
\newcommand{\calF}{{\mathcal{F}}}
\newcommand{\calG}{{\mathcal G}}
\newcommand{\calH}{{\mathcal H}}
\newcommand{\calJ}{{\mathcal{J}}}
\newcommand{\calK}{{\mathcal{K}}}
\newcommand{\calL}{{\mathcal{L}}}
\newcommand{\calM}{{\mathcal M}}
\newcommand{\calN}{{\mathcal N}}
\newcommand{\calO}{{\mathcal O}}
\newcommand{\calP}{{\mathcal{P}}}
\newcommand{\calR}{{\mathcal{R}}}
\newcommand{\calS}{{\mathcal{S}}}
\newcommand{\calT}{{\mathcal T}}
\newcommand{\calW}{{\mathcal W}}
\newcommand{\calX}{{\mathcal{X}}}
\newcommand{\calY}{{\mathcal{Y}}}
\newcommand{\calZ}{{\mathcal{Z}}}
\newcommand{\calV}{{\mathcal{V}}} 
\newcommand{\calU}{{\mathcal{U}}}
 
\newcommand{\lan}{\langle}
\newcommand{\ran}{\rangle}


\newcommand{\sumlL}{\sum_{l=1}^L}
\newcommand{\summM}{\sum_{m=1}^M}

\newcommand{\hbP}{\widehat{\bP}} 
\newcommand{\hbB}{\widehat{\bB}}
\newcommand{\hbC}{\widehat{\bC}}
\newcommand{\hbG}{\widehat{\bG}}
\newcommand{\hbH}{\widehat{\bH}}
\newcommand{\hbTe}{\widehat{\bTe}}
\newcommand{\hbD}{\widehat{\bD}}
\newcommand{\hbU}{\widehat{\bU}}
\newcommand{\hbV}{\widehat{\bV}}
\newcommand{\hbW}{\widehat{\bW}}
\newcommand{\hbQ}{\widehat{\bQ}}
\newcommand{\hbX}{\widehat{\bX}}
\newcommand{\hbZ}{\widehat{\bZ}}
\newcommand{\hUps}{\widehat{\Ups}}
\newcommand{\hbLam}{\widehat{\bLam}}
\newcommand{\hhbZ}{\widehat{\widehat{\bZ}}}
\newcommand{\hbd}{\hat{\bd}}

\newcommand{\hA}{\widehat{A}}
\newcommand{\hD}{\widehat{D}}
\newcommand{\hG}{\widehat{G}}
\newcommand{\hH}{\widehat{H}}
\newcommand{\hX}{\widehat{X}}
\newcommand{\hY}{\widehat{Y}}
\newcommand{\hPsi}{\widehat{\Psi}}
\newcommand{\hU}{\widehat{U}}
\newcommand{\hV}{\widehat{V}}
\newcommand{\hW}{\widehat{W}}
\newcommand{\hSig}{\widehat{\Sig}}
\newcommand{\hs}{\hat{s}}
\newcommand{\hte}{\hat{\te}}
\newcommand{\hTe}{\widehat{\Te}}
\newcommand{\hS}{\widehat{S}}
\newcommand{\hhU}{\widehat{\widehat{U}}}

\newcommand{\barU}{\overline{U}}
\newcommand{\barX}{\overline{X}}
 
\newcommand{\hL}{\widehat{L}} 
\newcommand{\hr}{\hat{r}} 
\newcommand{\hlam}{\widehat{\lam}}
\newcommand{\hK}{\widehat{K}} 
\newcommand{\hatc}{\hat{c}}

\newcommand{\barK}{\overline{K}}

\newcommand{\tilr}{\tilde{r}}
\newcommand{\tilA}{\widetilde{A}}
\newcommand{\tilC}{\widetilde{C}}
\newcommand{\tilL}{\widetilde{L}}
\newcommand{\tilR}{\widetilde{R}} 
\newcommand{\tilH}{\widetilde{H}}
\newcommand{\tilbG}{\widetilde{\bG}}
\newcommand{\tilbP}{\widetilde{\bP}}
\newcommand{\tilbA}{\widetilde{\bA}}
\newcommand{\tilbU}{\widetilde{\bU}}
\newcommand{\tilbV}{\widetilde{\bV}}
\newcommand{\tilbW}{\widetilde{\bW}}
\newcommand{\tilbTe}{\widetilde{\bTe}}
\newcommand{\tilbLam}{\widetilde{\bLam}}
\newcommand{\tbLam}{\widetilde{\bLam}}
\newcommand{\tOm}{\widetilde{\Om}}
\newcommand{\tH}{\widetilde{H}}
\newcommand{\tilTe}{\widetilde{\Te}}

\newcommand{\tilX}{\widetilde{X}}
\newcommand{\tilU}{\widetilde{U}}
\newcommand{\tilD}{\widetilde{D}}
\newcommand{\tilO}{\widetilde{O}}
\newcommand{\tilP}{\widetilde{P}}
\newcommand{\tilQ}{\widetilde{Q}}
\newcommand{\tilS}{\widetilde{S}}
\newcommand{\tilG}{\widetilde{G}}
\newcommand{\tilW}{\widetilde{W}}
\newcommand{\tilB}{\widetilde{B}}
\newcommand{\tilb}{\widetilde{b}}
\newcommand{\tilh}{\widetilde{h}}

\newcommand{\brX}{\breve{X}}
\newcommand{\brSig}{\breve{\Sig}}
\newcommand{\brB}{\breve{B}}
\newcommand{\brb}{\breve{b}}
\newcommand{\brP}{\breve{P}}

\newcommand{\bcalW}{\mbox{\mathversion{bold}$\calW$}}
\newcommand{\hbcalW}{\widehat{\bcalW}}
\newcommand{\tbcalW}{\widetilde{\bcalW}}

\newcommand{\bcalP}{\mbox{\mathversion{bold}$\calP$}}

\newcommand{\bcalV}{\mbox{\mathversion{bold}$\calV$}}
\newcommand{\hbcalV}{\widehat{\bcalV}}
\newcommand{\tbcalV}{\widetilde{\bcalV}}

\newcommand{\hcalP}{\widehat{\calP}}
\newcommand{\hcalG}{\widehat{\calG}}
\newcommand{\hcalH}{\widehat{\calH}}
\newcommand{\hcalW}{\widehat{\calW}}

\newcommand{\frS}{\mathfrak{S}}
\newcommand{\hfrS}{\widehat{\mathfrak{S}}}

\newcommand{\barbZ}{\overline{\bZ}}
\newcommand{\barbU}{\overline{\bU}}
\newcommand{\barbV}{\overline{\bV}}
\newcommand{\barbD}{\overline{\bD}}
\newcommand{\barbH}{\overline{\bH}}
\newcommand{\barbG}{\overline{\bG}}
\newcommand{\barbQ}{\overline{\bQ}}
\newcommand{\hbarbG}{\overline{\widehat{\bG}}}
\newcommand{\barbB}{\overline{\boB}}
\newcommand{\barcalH}{\overline{\calH}}
\newcommand{\barbR}{\overline{\bR}}
\newcommand{\barcalR}{\overline{\calR}}


\newcommand{\scrP}{\mathscr{P}}
\newcommand{\scrH}{\mathscr{H}}  

\newcommand{\minmM}{\displaystyle \min_{m=1, ....M}\ }
\newcommand{\maxmM}{\displaystyle \max_{m=1, ....M}\ }

 \newcommand{\minL}{\displaystyle \min_{l=1, ....L}\ }
 \newcommand{\maxL}{\displaystyle \max_{l=1, ....L}\ }
 \newcommand{\maxin}{\displaystyle \max_{i \in [n]} }

\newcommand{\lowc}{\underline{c}}
\newcommand{\highc}{\bar{c}}
\newcommand{\lowC}{\underline{C}}
\newcommand{\highC}{\bar{C}}
\newcommand{\lowd}{\underline{d}}
\newcommand{\highd}{\bar{d}}

\newcommand{\obrC}{\mbox{\"{C}}}
\newcommand{\uC}{\mbox{\u{C}}}
\newcommand{\lowcrho}{\underline{c}_{\rho}}
\newcommand{\highcrho}{\bar{c}_{\rho}}

\newcommand{\rhonl}{\rho_{n,l}}
\newcommand{\rhon}{\rho_{n}}
 
\newcommand{\sinTe}{\sin\Te}


\newcommand {\colred}[1] {\textcolor{red}{#1}}
\newcommand {\colblue}[1] {\textcolor{blue}{#1}}
\newcommand {\colyellow}[1] {\textcolor{yellow}{#1}}
\newcommand {\colblack}[1] {\textcolor{black}{#1}}
\newcommand {\colgreen}[1] {\textcolor{green}{#1}}
\newcommand {\colsred}[1] {\textcolor{OrangeRed}{#1}}
 

\long\def\ignore#1{}

\newcommand{\reals}{\mathbb{R}}

\newcommand{\upl}{^{(l)}}
\newcommand{\upm}{^{(m)}}
\newcommand{\upmi}{^{(m_i)}}
\newcommand{\upi}{^{(i)}}
\newcommand{\upj}{^{(j)}}
\newcommand{\tinf}{_{2,\infty}}

\newcommand{\linL}{l \in [L]}
\newcommand{\minM}{m \in [M]}
\newcommand{\kinK}{k \in [K]}
\newcommand{\kinKm}{k \in [K_m]}

\newcommand{\tio}{_{\times 1}}
\newcommand{\tit}{_{\times 2}}
\newcommand{\tir}{_{\times 3}}

\newcommand{\delo}{\del_1}
\newcommand{\delr}{\del_3}
\newcommand{\twin}{{2 \to \infty}}

\newcommand{\limn}{\lim_{n \to \infty}}
\newcommand{\Ctau}{C_{\tau}}
\newcommand{\Ctauo}{C_{\tau,\tau_0}}

\newcommand{\Omtaun}{\Om_{\tau,n}}
\newcommand{\epsrn}{\eps_{\rho, n}}

\newcommand{\Xiii}{\Xi_{(-i)}}
\newcommand{\xiil}{\xi^{(i,l)}}
\newcommand{\etaijl}{\eta\upl_{i,j}}

\newcommand{\scrE}{\mathscr{E}}


\newcommand{\dcF}{d_{\calF}}
\newcommand{\dUp}{d_{\Ups}}



\title{Signed Diverse Multiplex Networks: Clustering and Inference}

\author{Marianna Pensky,  University of Central Florida  
\thanks{Manuscript received June 18, 2024; revised  June 15, 2025. 
Corresponding author: M. Pensky (email: marianna.pensky@ucf.edu).}}

\markboth{IEEE TRANSACTIONS ON INFORMATION THEORY, VOL. ??, NO. ??,  }%
{  }

\IEEEtitleabstractindextext{ 
\begin{abstract}
The paper introduces a Signed Generalized Random Dot Product Graph (SGRDPG)  model,
which is a variant of the Generalized Random Dot Product Graph (GRDPG), where, in addition,  
edges can be positive or negative. The setting is extended to a multiplex version, where all 
layers   have  the same collection of nodes and follow the SGRDPG.  The only common feature 
of the layers of the network is that they can be partitioned into groups with common subspace structures,   
while  otherwise   matrices of connection probabilities can be all different.
The setting above is extremely flexible and includes a variety of existing multiplex network models, including GRDPG,
as its particular cases.
By employing novel methodologies, our  paper  ensures strongly consistent 
clustering of layers and highly accurate subspace  estimation,
which are significant improvements over the results of Pensky and Wang (2024).
All algorithms and theoretical results in the paper remain true 
for both signed and binary networks. In addition,  the paper shows that 
keeping signs of the edges in the process of network construction
leads to a better precision of estimation and clustering and, hence, is beneficial 
for tackling real world problems such as, for example, analysis of brain networks. 
\end{abstract}

}

\maketitle

\IEEEdisplaynontitleabstractindextext

\IEEEpeerreviewmaketitle



 
\section{Introduction}
\label{sec:introduction}


\subsection{Signed Generalized Random Dot Product Graph  (SGRDPG)  model   }
\label{sec:signed_model}

Stochastic network models appear in a variety of applications, including 
genetics, proteomics, medical imaging, international relationships, brain science and many more. 
In the last few decades, researchers developed a variety of ways to model those networks,
imposing some restriction on the matrix of probabilities that generates random edges. 
The  block models and the latent position graph settings are perhaps the most popular ways of 
modeling random graphs. In particular, the Generalized Random Dot Product Graph ({\bf GRDPG}) model 
is a latent position graph model which has great flexibility. 
However, the GRDPG requires assumptions that are difficult to satisfy. 
In order to remove this shortcoming, below  we introduce the 
Signed Generalized Random Dot Product Graph ({\bf SGRDPG}) model and extend it to a multilayer setting.

Consider a undirected network with $n$ nodes and without self-loops. 
The   GRDPG model refers to a network 
where the matrix of connection probabilities is of the form
\be \label{eq:GDPG}
 P  =  U   Q   U^T, \quad  Q = Q^T \in \RR^{K \times K}.
\ee
Here  $U \in \RR^{n \times K}$ is a  matrix  with orthonormal columns, and $U$ and $Q$ are such that 
all entries of $P$ are in $[0,1]$. The  GRDPG network model is very flexible and is 
known to have a variety  of the network models  as its particular cases:   
the Stochastic Block Model ({\bf SBM}), 
the Degree Corrected Block Model (\cite{gao2018}), 
the Popularity Adjusted Block Model  
(\cite{Minh_Tang_PABM}, \cite{Noroozi2021PABM}), 
the Mixed Membership Model ({\bf MMM}) 
(\cite{Airoldi:2008:MMS:1390681.1442798}, \cite{doi:10.1080/10618600.2016.1237362}),
or the Degree Corrected  Mixed Membership Model 
(\cite{ke2022optimal}).
If all eigenvalues of matrix $Q$ are positive, then model \fr{eq:GDPG} reduces to the Random Dot Product Graph  
({\bf  RDPG})  model (see, e.g., the survey in \cite{JMLR:v18:17-448}). The   RDPG   and the  
GRDPG  models  (\cite{JMLR:v18:17-448}, \cite{GDPG}) have recently  gained an 
increasing popularity. 

Nevertheless, the condition that elements of matrix $P$ in \fr{eq:GDPG} 
are indeed probabilities is not   easy  to satisfy.
Indeed, unless a network is generated using  one of the  above mentioned block model scenarios
and matrices $U$ and $Q$ are found later by using the Singular Value Decomposition ({\bf SVD}),  
it is very difficult to guarantee  that the elements of matrix $P$ are nonnegative.
The latter is due to the fact that,  in general, elements of matrix $U$ can be positive or negative. 
To the best of our knowledge, the only papers that  study  conditions for the elements of matrix $P$ to be non-negative, 
are \cite{NIPS2022_Jin} and its sequel \cite{Jin_Huang2024}. 
While these papers provide some sufficient conditions for non-negativity, these conditions  are far from trivial, and 
it is not clear how to satisfy them in a generic setting.
In addition, \cite{Jin_Huang2024} and the second half of  \cite{NIPS2022_Jin} focus on the case 
where the first (positive) eigenvalue is significantly larger than the rest of the singular values combined. 
This is the opposite scenario  from the majority of GRDPG studies 
where, similarly to our paper, it is assumed that all singular values are of the same order of magnitude.
In summary,   imposing conditions that ensure non-negativity is far from trivial and, 
for this reason, all simulation studies for the GRDPG always consider one of the existing block models, e.g., SBM or MMM.

To eliminate this difficulty and to add further flexibility to the model, here
we propose the Signed Generalized Dot Product Graph ({\bf SGRDPG}) model which generates an edge
between nodes $i$ and $j$ with probability $|P_{i,j}|$ and the corresponding element 
of the adjacency matrix $A(i,j)$ can be zero, 1~or~-1:
\begin{align} 
& \PP(|A(i,j)| = 1) = |P(i,j)|,\  \sgn(A(i,j)) = \sgn(P(i,j)), \nonumber \\
& A(i,j) = A(i,j) \quad 1 \leq i < j \leq n \label{eq:sign_adj} 
\end{align}
It is easy to ensure that $|P(i,j)|\leq 1$  since the latter is guaranteed by the 
row norms of matrix $U$ being bounded by one and  the eigenvalues of matrix $Q$ being in the interval $[-1,1]$. 
Observe that, if elements of matrix $P$ are nonnegative, the SGRDPG immediately reduces to the existing GRDPG model.
The setting \fr{eq:sign_adj}  can also be viewed as a particular case of a  Weighted Graph introduced in \cite{Priebe_WeightedGraph}.
However, the Signed GRDPG model  does not require special tools for its analysis, 
unlike the Weighted Graph model in 
\cite{Priebe_WeightedGraph}, and also can be easily extended to a multilayer context.

Note that the model above is motivated by situations where signs of edges are present but are not associated with 
with any intrinsic patterns. For example,  biological networks are often constructed on the basis of 
correlation (or precision) matrices, elements of which can be positive or negative. It often happens that, 
when data of this sort are modeled by a network, the signs of the edges are removed which leads to 
a significant loss of accuracy. We should mention that SGRDPG may not be a suitable tool to model social networks 
where relation between two nodes can be positive ("friendship"), or negative ("animosity"), or zero ("no relationship"),
and the distribution of signs is strongly associated with the structure of probability matrix. 
For example, modeling of social networks is often restricted by the so called 
balance theory which puts restrictions on the signs of the edges. In particular, the maxims 
 "an enemy of my enemy is my friend" and "two  of my friends cannot be enemies" lead to the fact that any 
closed  triangle in a   signed network can have either two or zero negative edges (strong balance). 
The allure  of the strong balance formulation lies in the fact that strongly balanced networks 
are clusterable (\cite{harary}), i.e., the nodes can be divided into some number of
disjoint sets such that all edges within sets are positive and all edges between sets are negative.
This is, however, a sufficient condition  while the necessary condition  for clusterability is  the 
so called weak balance (\cite{davis1967clustering}), which requires that there are no closed loops in the network with exactly one negative edge.  However, many of the existing real world networks are not balanced.
This  particularly applies to biological networks, where there 
is no intrinsic reason for the balance.

Therefore, in this paper we do not associate the signs distribution with any intrinsic pattern, 
as it is done in, e.g., \cite{SPONGE_AISTATS_2019}, and  instead  just preserve the sign of each edge, 
while modeling the probabilities of connections using the most flexible GRDPG model.


\subsection{DIMPLE-SGRDPG: the   SGRDPG-equipped Diverse MultiPLEx   network   }
\label{sec:DIMPLE-SGRDPG_formulation}

While in the early years of the field of stochastic networks, research  mainly focused  on studying a
single network, in recent years the frontier moved to investigation of collection of networks, the
so called {\it multilayer network}, which allows to study relationships between nodes
with respect to various modalities (e.g., relationships between species based on food or space),
or consists of  network data collected from different individuals (e.g., brain networks).
Although there are many different ways of modeling a multilayer network (see, e.g.,
an excellent review article of  \cite{10.1093/comnet/cnu016}), in this paper, we consider a 
{\it Multiplex Network Model} where
all layers have the same set of  nodes, and all the edges between nodes are drawn within  layers, i.e.,
there are no edges connecting the nodes in different layers. 
 Multiplex network models have a variety of applications 
such as brain networks  (\cite{Sporns_2018}) where nodes  are   brain regions and layers are individuals, or
trade networks (\cite{doi:10.1038/ncomms7864})  where nodes 
and layers represent, respectively, various countries and commodities in which they are trading.

Many authors researched  multiplex network models, imposing an array of restrictions of varying degree of 
severity. 
The simplest of the multiplex network models,  the ``checker board model''  of \cite{JMLR:v21:18-155},
assumes that all layers are equipped with the Stochastic Block Models, 
all layers have the same communities  and there are only few distinct
block connectivity matrices.  
On the other hand,  the DIverse MultiPLEx    Generalized Random Dot 
Product Graph ({\bf DIMPLE-GRDPG}) in \cite{pensky2021clustering} considers  layers that follow the  
GRDPG model (\cite{GDPG}) and allows utmost variability in layers' structures 
and connectivity patterns throughout the network.

The present paper follows  this most general   multiplex network model of \cite{pensky2021clustering},
and, in addition, allows  edges to be equipped with signs. 
The objective of the paper is twofold. First, it demonstrates that, 
even for this extremely complex 
multilayer network, one can achieve {\bf similar or better inference precision}, 
than has been obtained  for  simpler multiplex network  models. 
Second, it shows that keeping signs of the edges in the process of network construction
leads to a more accurate estimation and clustering.

Specifically, this paper  studies  a multiplex network  where each layer follows the SGRDPG model introduced in 
\fr{eq:sign_adj}. Consider    an $L$-layer network on the same set of 
$n$ vertices $[n] :=  \{1,\cdots,n\}$.
The tensor of connection probabilities  $\bP \in [0,1]^{n \times n \times L}$
is formed by layers $P^{(l)}$, $\linL$, which can be partitioned into   $M$  groups,  
so that there exists a label function $s: [L] \to [M]$.
It is  assumed that the probability matrices $P\upl$ of the layers of the network follow the SGRDPG  model defined in  
\fr{eq:GDPG}   and \fr{eq:sign_adj}, and that  each group of layers is embedded in its own ambient subspace, but otherwise all matrices 
of connection probabilities can be different. In particular,  $P\upl$, $l \in [L]$, are given by
\be \label{eq:DIMPLE_GDPG}
 P\upl =  U\upm  Q\upl (U\upm)^T, \quad m = s(l),    
\ee
where $Q\upl = (Q\upl)^T $, and  $U\upm$, $\minM$,  are matrices  with orthonormal columns.  
Here, however,  matrices  $U\upm$ and $Q\upl$ are not required to ensure 
that elements of matrices $P\upl$ 
are probabilities, but just that they  lie in the interval $[-1,1]$, 
which is a much easier condition to satisfy. 
We shall call this setting the  DIverse MultiPLEx  Signed Generalized Random Dot 
Product Graph ({\bf DIMPLE-SGRDPG}).

Here,   one observes the adjacency tensor $\bA \in \{0,1,-1\}^{n \times n \times L}$ with layers $A\upl$ such that 
$A{\upl} (i,j)$  are independent from each other for $1 \leq i <  j \leq n$ and $\linL$, 
$A{\upl} (i,j) = A\upl (j,i)$. 
and  follow  \fr{eq:sign_adj}, i.e.,  
\begin{align} \label{eq:SGDPG_adjacen}
& \PP(|A\upl(i,j)| = 1) = |P\upl(i,j)|, \\
& \sgn(A\upl(i,j)) = \sgn(P\upl(i,j)) \nonumber
\end{align}
The objective is to recover the between-layer clustering function $s: [L]\to[M]$ with  the corresponding clustering 
matrix $S$, and  the bases  $U\upm$ of the  ambient subspaces in \fr{eq:DIMPLE_GDPG}.
Note that, however,   the main task in the DIMPLE-SGRDPG model is the between-layer clustering.   
If the groups of layers are correctly identified, then the layers with common
subspace structures can be pooled 
together in a variety of ways, as it is done in  \cite{JMLR:v22:19-558} or   \cite{pensky2021clustering}.

Introduction of a model, where groups of layers can be embedded into different subspaces, is undeniably very crucial.
Indeed, after embedding layers of a network into an appropriate subspace, further inference usually relies on asymptotic 
normality of estimators of matrices $Q\upl$, which, in turn, requires that the ratios between 
the highest and the lowest eigenvalues of matrices $Q\upl$ are bounded above.  
However, this is not true  if a network follows model \fr{eq:DIMPLE_GDPG}, where  matrices 
$U\upm$   represent different subspaces, but are modeled using only one common subspace $U$. 
As it is easy to see, matrix $U$ will have higher rank than 
matrices $U\upm$, so that the lowest eigenvalues of all matrices $Q\upl$ will be equal to zero.

Nevertheless, in  this paper,  we are not interested in estimating the matrices $Q\upl$ in \fr{eq:DIMPLE_GDPG} 
since they act as nuisance parameters, as far as the inference on the structure of the network is concerned. 
If desired, those matrices can be recovered  by standard methods (see, e.g., \cite{JMLR:v18:17-448}).


\subsection{ Existing results and technical challenges  }
\label{sec:existing}

The DIMPLE-SGRDPG model is a relatively straightforward extension  of the DIMPLE-GRDPG model of 
\cite{pensky2021clustering},  which, in turn,  generalizes a multitude of  multiplex network models.

In particular, if all layers share the same ambient subspace structure, i.e., $M=1$  in \fr{eq:DIMPLE_GDPG}, the
DIMPLE-GRDPG reduces to the COSIE model in  \cite{JMLR:v22:19-558} and  \cite{MinhTang_arxiv2022}. 
Furthermore, there is a multitude of papers that study SBM-equipped multiplex networks,
which constitute particular cases of the  DIMPLE-GRDPG (see an extensive discussion in \cite{noroozi2022_SSC} and references therein). 
Moreover, the multiplex network  equipped with the Degree-Corrected Stochastic Block Models, featured in 
 \cite{agterberg_Multilayer_DCBM}, also reduces to the COSIE model with $M=1$, due to the fact that the positions of
the diagonal weight matrices and the membership matrices can be interchanged  in the presentation of the layer probability matrices.

Furthermore, we should comment on the relationship of the DIMPLE-SGRDPG model to the low-rank tensor models.
In those models, one can  represent  a low rank  tensor using the Tucker decomposition with a low-dimensional core tensor
and subsequently recover the subspaces and the core tensor using the Higher Order Orthogonal Iterations  ({\bf HOOI})
(see, e.g., \cite{agterberg2022_HOOI}, \cite{Han_Luo_JRSSB_2022_TBM}, \cite{JMLR:AZhang21},  \cite{ZhangXiaIEEE2018}). 
The papers cited above, however,  assume the noise components to be Gaussian 
(or sub-Gaussian with particular parameters)
which makes their results not transferable to  sparse binary networks.
Recently, \cite{fan2021alma}  and  \cite{TWIST-AOS2079}  considered 
the Mixed MultiLayer Stochastic Block Model ({\bf MMLSBM}), where the  layers can be partitioned into $M$ groups,
with each group of layers  equipped with its unique  SBM. The MMLSBM is a particular case of the DIMPLE-GRDPG in \fr{eq:DIMPLE_GDPG},
where $Q\upl \equiv Q\upm$, $\linL$, depend  on $m = s(l)$ only.

Note that in all the papers cited above, the probability tensor 
$\bP$ is of low rank, with the subspace structures associated with the problems of interest, and the inference relies on application of  HOOI 
(except in \cite{fan2021alma}  which uses an Alternating Minimization Algorithm). 
Nonetheless, our framework is not generated by a low rank tensor. 
Instead, \fr{eq:DIMPLE_GDPG} and \fr{eq:SGDPG_adjacen}  
can be viewed as a signed binary tensor which, in expectation, 
due to matrices $Q\upl$ being arbitrary, reduces to a partial multilinear structure 
 as in,  e.g., Section~5 of \cite{JMLR:AZhang21}. 
In this case, one can recover the subspaces associated with the low-rank modes with  much better precision than the 
high-rank mode  which, in our case, is associated with the network layer clustering.

However, so far it appeared that the  flexibility of the DIMPLE-SGRDPG network comes at a price. 
Indeed, if $\rhon$ is the sparsity parameter of network,
so that $\displaystyle \max_{i,j,l} |P\upl(i,j)| \geq C \rhon \geq C n^{-1} \log n$,
then \cite{pensky2021clustering} show that the proportion of mis-clustered layers is bounded above by 
$C (n \rhon)^{-1}$ and, unfortunately, the between-layer clustering error dominates the 
error of estimating of matrices $U\upm$ in \fr{eq:DIMPLE_GDPG}, or community detection in groups of layers when 
the layers of the network are equipped with the SBMs.

The goal of the present paper is to eliminate this deficiency, and to develop a strongly consistent 
 between-layer clustering algorithm that leads to the same  errors of subspaces recoveries that are 
obtained in simpler models like COSIE. The latter is achieved by removing the common components from subspaces and 
employing the $\ell_{2,\infty}$-norm based evaluations. The new technique delivers strongly consistent between-layer 
clustering under assumptions that are very similar to the ones in \cite{pensky2021clustering}. The latter guarantees 
highly accurate estimators of the subspace basis matrices $U\upm$, or community assignments in groups of layers,
for a SBM-equipped multiples network. The specific error rates are 
thoroughly discussed in Remarks~\ref{rem:between_comparison} and \ref{rem:within_comparison}
in Section~\ref{sec:theory}.

We should mention that all algorithms and theoretical assessments in the paper remain correct 
for un-signed  multiplex  networks with binary adjacency matrices. 
Nevertheless, as simulations and real data examples show, an opportunity of 
retaining signs of edges when they are present  allows to achieve better clustering precision.

\begin{rem} \label{rem:time-varying}
{\bf  Relation to time-varying networks.\ }
{\rm
The DIMPLE-GRDPG  can also be viewed as a generalization of a time-varying network with the GRDPG-equipped layers
(e.g., \cite{arxiv.2205.06877}).  However, the inference  in a multiplex network   is much more difficult than in the respective dynamic 
network since, in a dynamic network, the layers are ordered according to time instances, 
while in a multiplex network the enumeration of layers is completely arbitrary.  
The latter makes  techniques designed for dynamic networks unsuitable for the inference in the 
DIMPLE-GRDPG setting.
}
\end{rem}

\ignore{
Our paper can   also be  viewed as  related to the papers that study shared structures in the layers, specifically, 
the common and the uncommon parts of the bases for the ambient subspaces, associated with groups of layers. 
However, the publications known to us, such as 
\cite{levina_shared_structures_JMLR2022} and \cite{multiplex_shared_structures},
consider a very different setting where the probabilities of connections are the same in all layers while 
noise distribution varies from one layer to another. 
}


\subsection{Our contributions  }
\label{sec:contributions}

\noindent
%
Our paper makes several key contributions. 

\begin{enumerate} 

\item
To the best of our knowledge, our paper is the first one to introduce the very flexible signed SGRDPG network model 
and to extend it to the multilayer case. The difference between our SGRDPG model and other signed network settings 
is that it immediately reduces to GRDPG when no signs are present, and hence provides a true generalization of 
the popular GRDPG model. Hence, we are considering the most broad case of the multiplex networks 
with latent position/block model equipped layers.

\item 
Our paper offers a novel {\bf  strongly consistent} between-layer clustering algorithm that works for 
the GRDPG and SGRDPG equipped multiplex networks, when the sparsity $\rhon$ satisfies $n \rhon \geq C \log n$. 
This result has been previously achieved only in much  simpler multiplex network models, where layers are equipped with the SBMs
(see \cite{noroozi2022_SSC}).

\item
Our paper develops novel theoretical results under very simple and intuitive assumptions. 
First, it establishes theoretical guarantees for the  strongly consistent  between-layer clustering
for the new very general model.
In addition, it provides upper bounds on the subspace estimation errors in both 
operator and $(2,\infty)$ norms.  While the former are derived by techniques similar to those, employed in 
\cite{pensky2021clustering}, the latter employ novel ideas and are {\bf significantly better} than the rates obtained  in 
\cite{MinhTang_arxiv2022} for the simpler COSIE model.
All results in the paper remain valid  for ``un-signed'' multiplex networks with binary adjacency matrices.

\item
Simulations in the paper establish that the between-layer and the within-layer clustering 
algorithms deliver high precision in a finite parameter settings.
In addition,  simulations and real data example confirm that, in the case 
when edges of the network can naturally be positive or negative, 
 keeping signs of the edges in the process of network construction
leads to better accuracy of estimation and clustering, and, hence, is beneficial 
for tackling real world problems such as analysis of brain networks.

\ignore{
our subspace recovery error compares favorably to 
the ones,  derived for the simpler COSIE model in \cite{JMLR:v22:19-558}  and \cite{MinhTang_arxiv2022}, 
due to employment of a different algorithm.
}

\item
Since our setting includes majority of the popular network models as its particular cases,
our approach removes the need for testing, which of the particular network models should be used 
for the layers. After the groups of layers with similar ambient subspace structures are 
identified, one can recover those common structures and subsequently estimate community assignments,
if one suspects that a block model generated networks in the layers.

\end{enumerate}


\subsection{Notations  and organization of the paper}
\label{sec:notation}

We denote $a_n = O(b_n)$ if $a_n \leq C b_n$,  $a_n = \om(b_n)$ if $a_n \geq c b_n$,
$a_n \propto b_n$ if $c b_n \leq a_n \leq C b_n$,  where $0<c\leq C <\infty$ are absolute constants independent of $n$.
Also, $a_n = o(b_n)$  and $a_n = \Om(b_n)$ if, respectively, $a_n/b_n \to 0$ and  $a_n/b_n \to \infty$
as $n \to \infty$.  We use $C$ as a generic absolute constant independent of $n$, $L$, $M$ and $K$.

For any vector $v \in \RR^p$, denote  its $\ell_2$, $\ell_1$, $\ell_0$ and $\ell_\infty$ norms 
by $\|   v\|$, $\|   v\|_1$,  $\|   v\|_0$ and $\|  v\|_\infty$, respectively. 
Denote by $1_m$  the $m$-dimensional column vector with all components equal to one. 
 
For any matrix $A$,  denote its spectral, Frobenius and $(2,\infty)$ norms by, respectively,  $\|  A \|$, $\|  A \|_F$,
$\|  A \|_{2,\infty} = \max\|A(i,:)\|$  and $\|  A \|_{1,\infty} = \max\|A(i,:)\|_1$. 
Denote the eigenvalues and singular values of $A$ by $\lam(A)$ and $\sig(A)$, respectively.
Let $\SVD_r(A)$ be $r$ left leading eigenvectors of $A$.
The column $j$ and the row $i$ of a matrix $A$ are denoted by $A(:, j)$ and $A(i, :)$, respectively.
Let $\vect(A)$ be the vector obtained from matrix $A$ by sequentially stacking its columns. 
Denote by $A \otimes B$ the Kronecker product of matrices $A$ and $B$. 
Denote the diagonal of a matrix $A$ by $\diag(A)$. Also, with some abuse of notations, denote the $K$-dimensional 
diagonal matrix with $a_1,\ldots,a_K$ on the diagonal by $\diag(a_1,\ldots,a_K)$, and the diagonal matrix consisting of only the diagonal of a square matrix $A$ by $\diag(A)$.
Denote 
$\calO_{n,K} = \left \{A \in \RR^{n \times K} : A^T A = I_K \right \}$,   $\calO_n=\calO_{n,n}$.
Denote by $\calC_{n,K}$ the set of clustering matrices  $X \in \{0,1\}^{n \times K}$ that  partition $n$ objects into $K$ clusters, 
where $X$ is a clustering matrix  if it is binary and has exactly one 1 per row.  
Also, we denote an absolute constants independent of $n, K, L$ and $M$, which can take different values at different instances, by $C$.
 

The rest of the paper is organized as follows. Section~\ref{sec:DIMPLE-SGDPG_clust_est} 
introduces a natural generative mechanism for a network that follows the DIMPLE-SGRDPG model and 
presents algorithms for its analysis. Specifically, Section~\ref{sec:between_layer}
suggests a novel between-layer clustering algorithm, while Section~\ref{sec:Fitting_subspaces}
examines techniques for fitting subspaces in groups of layers. Section~\ref{sec:theory}  
provides theoretical guarantees for the methodologies in Section~\ref{sec:DIMPLE-SGDPG_clust_est}:
the upper bounds on the between-layer clustering errors (Section~\ref{sec:error_between}),
the subspace fitting errors (Section~\ref{sec:error_fitting}), and the errors in the special case 
of the  Stochastic Block Model and the Mixed Membership Model (Section~\ref{sec:SBM_MMM}).
Section~\ref{sec:simul_study} considers finite sample simulation  studies of the proposed model and techniques. 
Section~\ref{sec:real data} illustrates the material in the paper with the real data example which
demonstrates the advantage of keeping the edges' signs in the process of network construction.  
Proofs for all  statements of the paper can be found in  Appendix~\ref{sec:proofs}.




\section{Clustering and estimation techniques }  
\label{sec:DIMPLE-SGDPG_clust_est}


\subsection{{The DIMPLE-SGRDPG network: generation and assumptions  } }
\label{sec:net_form}

We generate a DIMPLE-SGRDPG network by first sampling   group labels and subsequently 
obtaining   SGRDPG networks in each group of  layers using  the standard protocols of  \cite{JMLR:v22:19-558}.

Specifically, for each layer $l \in [L]$, we generate its group membership as 
\begin{align} \label{eq:group_memb}
&  s(l) \sim {\rm Multinomial}(\vec{\pi}), \\
& \vec{\pi}  = (\pi_1, ..., \pi_M) \in [0,1]^M, \ 
\pi_1 + ... + \pi_M = 1. \nonumber
\end{align}
In order to create the   probability tensor $\bP$, we generate $M$ matrices $X\upm \in [-1,1]^{n \times K_m}$ with i.i.d. rows:
\begin{align} \label{eq:Xm_gen}
& X\upm(i,:) \sim \iid \, f_m, \ \  \EE(X\upm(i,:)) = \mu\upm, \\  
& \Cov(X\upm(i,:)) = \Sig\upm, \ \ \minM, \ i \in [n], \nonumber
\end{align}
where $f_m$ is a probability distribution on  a subset  of $[-1,1]^{K_m}$.
Subsequently, we consider  a collection of symmetric matrices $B\upl \in \RR^{K_m \times K_m}$, 
 $m = s(l)$, $\linL$, and   form the layer probability matrices $P\upl$ as 
\be \label{eq:DIMPnew_GDPG}
 P\upl =  X\upm  B\upl (X\upm)^T.  
\ee
It is easy to see that \fr{eq:DIMPnew_GDPG} is equivalent to the model \fr{eq:DIMPLE_GDPG}.
Indeed, if $X\upm = U_X\upm D_X\upm  O_X\upm$ is the SVD of $X\upm$, then \fr{eq:DIMPnew_GDPG}
can be rewritten as  \fr{eq:DIMPLE_GDPG} with 
\begin{align} \label{eq:UupmQupl}
& Q\upl = D_X\upm  O_X\upm B\upl (O_X\upm)^T D_X\upm, \\
& U\upm = U_X\upm, \ \  m = s(l), \ \linL. \nonumber
\end{align} 
Note that we do not assume the elements of matrices $P\upl$ to be non-negative: we merely 
require that entries of $P\upl$ lie between -1 and 1. It is relatively easy to satisfy this 
assumption. In particular, if $f_m$ is a distribution on the $K_m$-dimensional unit ball $\calB (K_m)$
and all singular values of $B\upl$ lie in $[-1,1]$, then $P\upl(i,j) \in [-1,1]$, $i,j \in [n]$. 
Since $X\upm$ and  $B\upl$ are  defined up to a scaling constant, 
in what follows, we naturally assume that    $X\upm(i,:) \in  \calB (K_m)$, and model sparsity via matrices $B\upl$, $\linL$. 
We impose the following assumptions: 
\\ 

\noindent
{\bf A1.}  Vector $\vec{\pi}  = (\pi_1, ..., \pi_M) $ is  such that  
$\lowc_{\pi}\, M^{-1}  \leq \pi_m \leq \highc_{\pi}\, M^{-1}$, $\minM$.   
\\

\noindent
{\bf A2.} Distributions $f_m$, $\minM$, are supported on the $K_m$-dimensional unit balls $\calB (K_m)$. 
Here, $\displaystyle {K =  \max_m K_m}$ and $\displaystyle {K \leq C_K  \min_m K_m}$ for some  constant  $C_K$.
\\

\noindent
{\bf A3.}  Covariance matrices $\Sig\upm$, $\minM$,  satisfy 
\be \label{eq:lam_Sig} 
0 < \lowc  \leq \lam_{\min} (\Sig\upm)  \leq \lam_{\max} (\Sig\upm) \leq  \highc < \infty.
\ee

\noindent
{\bf A4.}  The number of layers $L$ grows at most polynomially with respect to $n$. Specifically, for some constant $\tau_0$
\be \label{eq:nLtau}
L \leq n^{\tau_0}, \quad 0 < \tau_0 < \infty .
\ee

\noindent
{\bf A5.} For some positive constants $\lowC_B$ and $\highC_B$, $0 <   \lowC_B \le \highC_B < \infty$ , one has
\begin{align} 
& B^{(l)} = \rho_n B^{(l)}_0 \quad \mbox{with} \nonumber \\
&  0 < \lowC_B \leq \sig_{\min}(B^{(l)}_0) \leq \sig_{\max}(B^{(l)}_0) \leq \highC_B < \infty  \label{eq:Bol}
\end{align} 

\noindent 
Note that Assumption  {\bf A1} as well as conditions on $K_m$  are always satisfied if $M$ is a constant independent of $L$ and $n$.  
The constants in Assumptions~{\bf A3}  and {\bf A5} 
can be modified to depend on $K_m$. However, this will make the notations too cumbersome. For this reason, we 
impose uniform bounds in Assumptions~{\bf A3}  and {\bf A5}. Again, if  $M$  and $K_m$  are constants 
independent of $L$ and $n$, those constraints are easily justifiable.
Assumption~{\bf A4} is hardly restrictive and is relatively common. Indeed,    \cite{TWIST-AOS2079} assume  that $L \leq n$, so, in their paper, 
\eqref{eq:nLtau}  holds with $\tau_0=1$. We allow any polynomial growth of $L$ with respect to~$n$.
\\


There are many ways of generating rows $X\upm(i,:)  \in \calB(K_m)$ of matrices $X\upm$, 
$i \in [n]$, $\minM$, satisfying conditions \fr{eq:Xm_gen}
and \fr{eq:lam_Sig}. 
Below are some examples, where we suppress  the index~$i$. 

\begin{enumerate}

\item
 $f_m$ is a truncated normal distribution: $\xi \sim f_m$ is obtained as 
$\xi = \eta/\|\eta\|$  where $\eta \sim \calN(\mu, \Sig)$, the normal distribution 
with the mean $\mu$ and covariance $\Sig$.

\item 
$f_m$ is a truncated $T$-distribution  with $\nu$ degrees of freedom: $\xi \sim f_m$ is obtained as 
$\xi = \eta/\|\eta\|$  where $\eta \sim T(\nu)$, the  $T$-distribution  with $\nu$ degrees of freedom.

\item 
Generate  a vector $\eta \in [0,1]^{K+1}$ as $\eta \sim {\rm Dirichlet} (\vec{\alpha})$,  where
$\vec{\alpha} = (\alpha_1, ..., \alpha_{K+1})$, $\alpha_k >0$, $k \in [K+1]$.
Set $\xi$ to be equal to the first $K$ components of $\eta$

\item 
 Generate  a vector $\eta \in [0,1]^{K+1}$ as $\eta \sim {\rm Multinomial} (\vec{\varpi})$, where
$\vec{\varpi}= (\varpi_1, ..., \varpi_{K+1})$, $\varpi_k > 0$, $\varpi_1 + ...+ \varpi_{K+1} = 1$.
Set $\xi$ to be equal to the first $K$ components of $\eta$.

\end{enumerate}

Note that Examples~3 and 4 bring to mind, respectively, the Mixed Membership Model (MMM) and the Stochastic Block Model (SBM),
where the last column of the community assignment matrix $X\upm$ is missing. Indeed, in the case of the SBM or the MMM,
rows of matrix $X\upm$ sum to one, which means that covariance matrix of the rows of $X\upm$ is singular and does not satisfy Assumption {\bf A3}.
However, as we see later in Section~\ref{sec:SBM_MMM}, our methodology can handle the SBM or the MMM equipped networks, 
although Assumption {\bf A3} does not hold in those cases.


\subsection{Disentangling the DIMPLE-SGRDPG network: the between-layer clustering    }
\label{sec:untangle}

Although matrices $X\upm$ and, hence, $U_X\upm$ are different for different values of $m$, they are too 
similar for ensuring a reliable separation of layers of the network, due to the  common mean $\mu\upm$
of rows of $X\upm$, which implies that $\EE(X\upm) = 1_n  \mu\upm$, $\minM$.
For this reason, we ``de-bias'' the layers by removing this common mean  and replacing $X\upm$ with 
$\tilX\upm = X\upm - 1_n  \bar{X}\upm$, where $\bar{X}\upm$ is the sample mean of the rows of $X\upm$.
By introducing a projection operator $\Pi= n^{-1} 1_n 1_n^T$, we can rewrite $\tilX\upm$ as
\be    \label{eq:tX}
\tilX\upm =   X\upm - \Pi   X\upm = \Pi^{\bot} X\upm, \  \Pi := n^{-1} 1_n 1_n^T.  
\ee 
It is easy to see that for $\minM$,\   $\linL$ 
\be \label{eq:tilbp1}
\tilP\upl = \Pi^{\bot} P\upl  \Pi^{\bot} = \tilX\upm   B\upl (\tilX\upm)^T, \ \  s(l)=m.
\ee 
Consider the SVDs of $\tilX\upm$ and $\tilP\upl$:
\be \label{eq:X_P_svds}
\tilX\upm = \tilU\upm \tilD\upm  (\tilO\upm)^T, \ \   \tilP\upl = \tilU_P\upl D_P\upl  (\tilU_P\upl)^T, 
\ee
where $\tilO\upm \in \calO_{K_m}$. Recall that $\tilP\upl = \tilU\upm \tilB\upl \tilU\upm$, 
where $\tilB\upl = \tilD\upm (\tilO\upm)^T   B\upl  \tilO\upm \tilD\upm$.  Hence,
one can rewrite the SVD of  $\tilP\upl$ using the SVD $\tilB\upl = \tilU_B\upl \tilD_B\upl (\tilU_B\upl)^T$ 
of $\tilB\upl$:
\be    \label{eq:tilbp2} 
\tilP\upl = \tilU\upm  \tilU_B\upl \tilD_B\upl  (\tilU\upm  \tilU_B\upl)^T, 
\ee  
where $\tilU_B\upl \in \calO_{K_m}$.
Hence, it follows from  \fr{eq:X_P_svds}  and   \fr{eq:tilbp2}  that  $D_P\upl = \tilD_B\upl$ and
\be \label{eq:tilU_rel}
\tilU\upm (\tilU\upm)^T = \tilU_P\upl (\tilU_P\upl)^T,\quad s(l)=m, 
\ee
so that, the values of $\tilU_P\upl (\tilU_P\upl)^T$ are unique for each group of layers $\minM$.

This  seems to be a trivial consequence of the observation,  made in \cite{pensky2021clustering}, 
that the values of $U_P\upl (U_P\upl)^T$, where $U_P\upl$ are the matrices of $K\upm$ singular vectors of $P\upl$, $l = s(m)$, 
are unique for each group of layers $\minM$.  However, we are going to use this fact in a completely different way 
from \cite{pensky2021clustering}. Specifically, we show below that the scalar products of vectors
$\vect(\tilU_P\upl (\tilU_P\upl)^T)$  are large if they belong to the same group of layers and are small
otherwise. This actually means that vectors $\vect(\tilU_P\upl (\tilU_P\upl)^T)$ for different groups of layers 
form almost orthogonal subspaces. The latter however is not true for subspaces formed by vectors 
$\vect(U_P\upl (U_P\upl)^T)$  since all matrices $U_P\upl$ include the vector $n^{-1/2}\, 1_{n}$  as a column (up to a rotation).

Therefore, in order to carry out between-layer clustering, for $l, l_1, l_2 \in [L]$, we denote 
\be  \label{eq:te_Te}
\te\upl = \vect(\tilU_P\upl (\tilU_P\upl)^T), \quad \Te(l_1, l_2) =  \lan \te^{(l_1)},  \te^{(l_2)} \ran.
\ee
Furthermore, we calculate the sample-based versions of $\te\upl$ and exploit the fact the values of $\Te(l_1, l_2)$ 
are small when $s(l_1) \neq s(l_2)$ and the same is true for their sample-based versions
under the assumptions in Section~\ref{sec:net_form}.
The following statement confirms that $\te\upl$ are dissimilar enough to allow one to separate  the network layers
into groups.

\begin{lem} \label{lem:true_Theta}  
Let Assumptions~{\bf A1}--{\bf A4} hold. If $n$ is large enough, 
then, for any fixed constant $\tau$, with probability 
at least $1 - C M \, n^{-\tau}$,   one has 
  \begin{align} 
   &       \Te(l_1, l_2) = K_m,     \quad  \hspace{18mm}   \mbox{if}\   s(l_1) = s(l_2) = m,  \label{eq:Theta1} \\
  &      \max_{l_1, l_2}   |\Te(l_1, l_2)| \leq C\,   K  \, n^{-1}   \log n, \ \   \   \mbox{if}\   s(l_1) \neq s(l_2), \label{eq:Theta2}
  \end{align} 
where constants $C$   depend only on $\tau$ and the constants in Assumptions~{\bf A1}--{\bf A4}.
\end{lem}

\noindent
The formal proof of Lemma~\ref{lem:true_Theta} is given in the Appendix. Below, we give a brief explanation why it is true. 
If $s(l_1) = s(l_2) = m$, then $\Te(l_1, l_2) = \|\te\upm \|^2 = \|  \vect(\tilU\upm (\tilU\upm)^T)\|^2$,
and it is easy to show that the first relation in \fr{eq:Theta1} holds. In order to justify  \fr{eq:Theta2},
consider  the sample covariance matrices $\hSig\upm$ with $\EE(\hSig\upm) = \Sig\upm$. If $n$ is large enough,
matrices $\hSig\upm$ are both non-singular, and it follows from \fr{eq:X_P_svds} that 
\be \label{eq:hSig_def}
\hSig\upm =  \frac{(\tilX\upm)^T \tilX\upm}{n-1} 
= \frac{\tilO\upm\, (\tilD\upm)^2  (\tilO\upm)^T}{n-1}.
\ee 
Denoting $( \hSig\upm )^{1/2}  = (n-1)^{-1/2}\, \tilD\upm  (\tilO\upm)^T$ and 
$(\hSig\upm )^{-1/2} = (n-1)^{1/2}\, \tilO\upm (\tilD\upm)^{-1}$, derive that 
\be \label{eq:tilUX}
\tilU\upm = (n-1)^{-1/2}\, \tilX\upm\, ( \hSig\upm )^{-1/2} .
\ee 
It turns out that matrices $ \tilX\upm$, $\minM$, are dissimilar enough to allow separation of layers.
In particular, the proof of \fr{eq:Theta2} is based on the fact that, if $n$ is large enough,
the singular values of the true and the sample covariance matrices are close to each other and that,
with high probability
\\
\beqns
 \max_{l_1 \neq l_2}\,  |\Te(l_1, l_2)| & = 
  O \lkr  n^{-2}\,  \| \lkr \tilX^{(m_1)}\rkr^T  \tilX^{(m_2)}   \|_F^2  \rkr \\
  & = 
  O \lkr   n^{-1}\, \log n  \rkr.
\eeqns 
Note that, since the values of $\Te(l_1, l_2)$ in \fr{eq:Theta1} and \fr{eq:Theta2}  do not depend on matrices $B\upl$, they   
are not affected by sparsity factor $\rhon$, so their separation relies  on the number $n$ of nodes being large enough. 
Nevertheless, the accuracy of their sample counterparts  depends on the sparsity factor $\rhon$.


\subsection{The  between-layer clustering  algorithm    }
\label{sec:between_layer}

It is easy to see that Lemma~\ref{lem:true_Theta} offers a path for the between-layer clustering.
Consider   proxies $\tilA\upl = \Pi^{\bot} A\upl  \Pi^{\bot}$ for $\tilP\upl$ in \fr{eq:tilbp1} 
and their full SVDs
$\tilA\upl = \tilU_A\upl \tilD_A\upl  (\tilU_A\upl)^T$, $\linL$.
Note that $\rank(P\upl) = K_m$ where $m = s(l)$, so columns of $\tilU_A\upl$, beyond the first $K_m$ ones, estimate pure noise. 
Since we do not know the grouping of  the layers in advance, we instead have to deal with $K\upl = \rank(P\upl)$,  where, of  course,
$K\upl = K_m$ if $s(l)=m$. Specifically, we form matrices 
\be   \label{eq:tilA_svd}
\hU_A\upl = SVD_{K\upl} (\tilA\upl), \quad  \tilA\upl = \Pi^{\bot} A\upl  \Pi^{\bot},
\ee
and, subsequently,  
construct vectors $\hte\upl$ and a matrix $\hTe$ with components $\hTe(l_1, l_2)$, $l, l_1, l_2 \in [L]$
\be \label{eq:hte_hTe}
\hte\upl = \vect \lkr \hU_A\upl (\hU_A\upl)^T \rkr, \quad
\hTe(l_1, l_2) =  \lan \hte^{(l_1)},  \hte^{(l_2)} \ran.
\ee

Note that, since the values of $K_m$ are not 
truly known before the between-layer clustering has been carried out, one can just replace $K_m$ in the definition of $T$ by 
$K\upl$. 
The procedure  above is summarized by  Algorithm~\ref{alg:between}.

%
\begin{algorithm} [t] 
\caption{\ The between-layer clustering}
\label{alg:between}
\begin{flushleft} 
{\bf Input:} Adjacency tensor $\bA \in \{0,1,-1\}^{n \times n \times L}$; number of groups of layers $M$;  
ambient dimension $K^{(l)}$ of each layer $\linL$;  parameter $\eps$ \\
{\bf Output:} Estimated clustering function $\hs: [L] \to [M]$ for groups of layers\\
{\bf Steps:}\\
{\bf 1:} Find matrices $\hU_A\upl = \SVD_{K\upl} (\tilA\upl)$ 
of $K\upl$ left leading singular vectors  of matrices $\tilA\upl = \Pi^{\bot} A\upl  \Pi^{\bot}$, $\linL$ \\
{\bf 2:} Form  vectors $\hte\upl =  \vect(\lkr \hU_A\upl (\hU_A\upl)^T \rkr$, $\linL$ \\
{\bf 3:} Form matrix $\hTe$ with elements $\hTe(l_1, l_2) =  \lan \hte^{(l_1)},  \hte^{(l_2)} \ran$, $l_1, l_2 \in [L]$\\
%
%
{\bf 4:} Obtain matrix $\hV = \SVD_M (\hTe) \in \calO_{L,M}$  of $M$  leading singular vectors of matrix 
$\hTe$ \\
{\bf 5:}  Cluster  $L$ rows of  $\hV$ into $M$ clusters using   $(1+\eps)$-approximate $K$-means clustering. Obtain 
estimated clustering function $\hs$   
\end{flushleft} 
\end{algorithm}
%



\begin{rem} \label{rem:common_U}
{\bf Recovering  common left singular vectors. }
{\rm
Note that in the model above, one can recover the common singular subspace $U$ of all matrices $P\upl$, $\linL$,
with high precision. This can be achieved, for example, by concatenating matrices $A\upl$  into $A = [A^{1)}|...|A^{L)}]$ 
and estimating $U$ by the set of the left singular vectors $\hU$ of $A$. 
Alternatively, $U$ can be estimated by the set of the left singular vectors  of the unbiased sum of squares $(A\upl)^2$.
It is known (see, e.g., \cite{Lei_2023_JASA}) that matrix $U$ can be recovered with  high precision  
if $\rhon\,  n\, L > C\, (\log n)^{\nu}$ for $C$ and $\nu$ large enough.
The latter however does not translate into easy clustering of layers.  

To illustrate this issue, consider the case when $M=2$, $K_1 = K_2 = K$ and $U^{(1)}$ and $U^{(2)}$ are the matrices 
of singular vectors in the two groups of layers. If $\tilU =  [U^{(1)}|U^{(2)}] \in R^{n \times 2K}$  has the SVD $\tilU = U \Lam_U O_U^T$ 
with $U \calO_{n, 2K}$ and $O_U \in \calO_{2K}$, then $P = U\, \tilP$, where
$\tilP =  \Lam_U \, O_U^T\, [\tilQ^{(1)}\,  \tilU^T |\, ...\,| \tilQ^{(L)}\,  \tilU^T]$
and  $\tilQ\upl$ are $(2K \times 2K)$ block-diagonal matrices with $\tilQ\upl_{1,1} \in \RR^{K \times K}$ 
and $\tilQ\upl_{2,2} \in \RR^{K \times K}$ on the diagonal.
Here, $\tilQ\upl_{1,1} = Q\upl$,   $\tilQ\upl_{2,2} = 0$ if $s(l)=1$,  
and $\tilQ\upl_{1,1} = 0$, $\tilQ\upl_{2,2} = Q^{1)}$  if $s(l)=2$.
Now, suppose that matrix $U$ is known (but, of course, we do not know $U^{(1)}$ and $U^{(2)}$).
Then, our original goal would be replaced by the task of partitioning  $(2K \times 2K)$ matrices 
$\Lam_U O_U^T \, \tilQ\upl\, O_U\,  \Lam_U$ into two clusters. Since matrices $Q\upl$ are arbitrary, 
this is not an easy undertaking. Indeed, matrix $U$ acts in this setting as a nuisance parameter. 
It can easily be estimated but does not help much in 
clustering of layers into groups.
}
\end{rem}


\begin{rem} \label{rem:unknownM_Km}
{\bf Unknown number of layers and ambient dimensions. }
{\rm
While Algorithm~\ref{alg:between} requires knowledge of  $M$ and ambient dimensions 
$K^{(l)}$ of each layer network, they may not be known in the case of real data. 
The assumption that these parameters are known is overwhelmingly common 
in methodological works, since  the problem of finding the number of clusters 
is a very distinct problem from the process of actually solving the clustering problem
with a known number of clusters.
Since in the setting of this paper, the number of groups of layers and the ambient dimensions of those groups of layers 
coincide  with the ranks of the respective matrices
(specifically,   $\Te$  and $U\upm$), there are many possible ways of accomplishing this task.
In this paper, we estimate those quantities using the ScreeNOT technique of \cite{donoho_screenot}.

We further comment on using $K\upl$ instead of $K_m$ in  Algorithm~\ref{alg:between}.
While one can possibly  assume that the values of $K_m$, $\minM$, are known, this knowledge 
is hard to exploit  since group labels are interchangeable and, therefore, 
in the case of non-identical subspace dimensions, 
it is hard to choose, which of the values corresponds to which of the  groups.
This is actually the reason why   \cite{fan2021alma} and \cite{TWIST-AOS2079}, who imposed this assumption,
used it only in theory, while their simulations and real data examples are all restricted to the case 
of equal number of communities in all layers $K_m = K$, $\minM$.  On the contrary, knowledge of $K\upl$ allows one to deal 
with different ambient dimensions  in the groups of layers in simulations and real data examples.
After the layers are partitioned into groups,  due to  $K^{(l)}= K_m$  when $m = s(l)$, the values of $K_m$ should be re-fitted. 
}
\end{rem}

 

\subsection{Fitting invariant subspaces in groups of layers } 
\label{sec:Fitting_subspaces}

If we knew the true clustering matrix $S \in \calC_{L,M}$ and the true probability tensor  
$\bP \in \RR^{n \times n \times L}$ with layers $P\upl$ given by \eqref{eq:DIMPLE_GDPG},
then  we could average  layers with identical subspace structures.  
Precision of estimating matrices $U\upm$, however, depends on whether the eigenvalues 
of $Q\upl$ with $s(l)=m$ in \fr{eq:DIMPLE_GDPG} add up.  
Since the latter is not guaranteed, one can alternatively add the squares   $G\upl= (P\upl)^2$, 
obtaining matrices  $H\upm$, $\minM$:
\be  \label{eq:Hm_def}
H\upm \! = \! 
\sum_{s(l) = m}\! (P\upl)^2 \! =  \!  U\upm \!  \lkv \sum_{s(l) = m}\!  (Q\upl)^2 \rkv \!  (U\upm)^T.
\ee 
In this case, the eigenvalues of $(Q\upl)^2$ are all positive which ensures successful recovery of matrices $U\upm$. 
Note that, however, $(A\upl)^2$ is not an unbiased estimator of $(P\upl)^2$.
Indeed, while $\EE [(A\upl)^2  (i,j)] = [(P\upl)^2  (i,j)]$  for $i \neq j$, for the diagonal elements, one has 
\bes 
\EE [(A\upl)^2  (i,i)] = (P\upl)^2  (i,i) + \sum_j [|P\upl (i,j)| - (P\upl)^2 (i,j)].
\ees 
Therefore, 
we   de-bias the squares $(A\upl)^2$ by 
subtracting the diagonal matrices $\hD\upl$ with elements
\be \label{eq:hDl}
\hD\upl (i,i) = \sum_{j=1}^n |A\upl (i,j)|,\quad \hD\upl (i,j)=0 \ \ \mbox{if}\ \ i \neq j,
\ee 
and estimate $G\upl = (P\upl)^2$, $\linL$, and  $\hH\upm$, $\minM$,  by, respectively, 
\be \label{eq:hGhH}
\hG\upl =   \lkr A\upl \rkr^2 - \hD\upl, \quad 
\hH\upm = \sum_{s(l) = m} \hG\upl. 
\ee
After that, estimate $U\upm$   by the matrix of $K_m$ leading singular vectors of $\hH\upm$:
\be \label{eq:hUupm}
\hU\upm = \SVD_{K_m} (\hH\upm), \quad \minM.
\ee
 The procedure is summarized in Algorithm~\ref{alg:subspace_est}.

%
%
\begin{algorithm} [t] 
\caption{\ Estimating invariant subspaces}
\label{alg:subspace_est}
\begin{flushleft} 
{\bf Input:} Adjacency matrices $\bA\upl \in \{0,1,-1\}$, $\linL$; number of groups of layers $M$;
ambient  dimensions $K_m$ of a  group of layers  $\minM$;
estimated layer  clustering matrix $\hS \in \calC_{L,M}$  \\  
{\bf Output:} Estimated invariant subspaces $\hU\upm$, $\minM$ \\
{\bf Steps:}\\
{\bf 1:} Construct matrices   $\hG\upl$, $\linL$, and $\hH\upm$, $\minM$ using  \eqref{eq:hGhH}\\
{\bf 2:}  For each $\minM$, find $\hU\upm$ using formula \fr{eq:hUupm}
\end{flushleft} 
\end{algorithm}
%
%


\begin{rem} \label{rem:spectr_embedding}
{\bf  Alternative techniques for estimating ambient subspaces.\ }
{\rm
Note that one can estimate  bases $\hU\upm$ of the ambient subspaces using alternative techniques. 
For example, in the COSIE model,   which corresponds to $M=1$ in our setting,
\cite{JMLR:v22:19-558} used  spectral embedding.  The methodology is based on constructing matrices 
$\hU\upl= \SVD_{K} (A\upl)$, $\linL$, and subsequently placing them, side by side, forming a new matrix 
$\widetilde{U} = [U^{(1)}|...|U^{(L)}] \in \RR^{n \times  KL}$. The estimator $\hU$ of $U$ is then obtained as 
$\hU = \SVD_K(\widetilde{U})$.   Another    way of estimating $\hU\upm$ is to follow the technique of, e.g., 
\cite{arxiv.2007.10455} where, in the case of $M=1$, adjacency matrices $A\upl$ are placed side by side to form one 
matrix $\tilA$, and the common matrix $U$ of  the  left singular vectors of $P\upl$, $\linL$, is then  estimated by
the matrix of the left singular vectors $\hU$ of $\tilA$. We shall compare the latter technique with our Algorithm~2 
in Section~\ref{sec:simul_study}. 
} 
\end{rem}



\section{Theoretical analysis}
\label{sec:theory}
 
\subsection{Error measures }
\label{sec:errors}

In this section, we study the  between-layer clustering error rates of  Algorithm~\ref{alg:between} and 
 the error of estimation of invariant subspaces  of Algorithm~\ref{alg:subspace_est}.
Since clustering is unique only up to a permutation of cluster labels, 
denote the set of $K$-dimensional permutation functions of $[K]$ by $\aleph(K)$, and the set of $(K \times K)$ permutation matrices by 
$\mathfrak{F} (K)$. 
The misclassification error rate of the between-layer clustering is then given by
\be \label{eq:err_betw_def}
R_{BL} = (2\,L)^{-1}\ \min_{\scrP \in \mathfrak{F}  (M)}\  \|\hS  - S \, \scrP\|^2_F.
\ee
In order to assess the accuracy of estimating matrices $U\upm$ by $\hU\upm$, $\minM$, 
we use the $\sinTe$ distances that allow to take into account that  matrices $\hU\upm$ 
are unique only up to an arbitrary  rotation. 
The $\sinTe$ distances   between two subspaces with orthonormal 
bases  $U \in \calO_{n,K}$ and $\hU  \in \calO_{n,K}$, respectively, are defined as follows  (\cite{10.1214/17-AOS1541}). 
Suppose the singular values of $U^T \hU$ are 
$\sig_1 \geq \sig_2 \geq ... \geq \sig_K>0$. 
The quantitative measures of the distance between the
column spaces of $U$ and $\hU$ are then 
\begin{align*}    
&  \| \sin \Te(U, \hU) \| =  \lkr 1 - \sig_{K}^2 (U^T \hU) \rkr^{1/2},\\ 
& \| \sin \Te(U, \hU)  \|_F = \lkr K - \|U^T \hU)\|^2_F\rkr^{1/2}. 
\end{align*}  
We also define
\be \label{eq:sinT2inf}
D(U,\hU; 2,\infty) = \min_{W \in \calO_K} \|\hU \, W - U \|_{2,\infty}
\ee
and observe  that
$D(U,\hU; 2, \infty) \leq \sqrt{2}\, \| \sin \Te(U, \hU) \|$
 (see, e.g., \cite{10.1214/17-AOS1541}).
Since the numbering of groups of layers is defined   up to a permutation, we need to 
minimize those errors over the permutation of labels. In particular, we measure the precision
of subspace estimation by 
\be \label{eq:err_subspace_2inf}
R_{U,2,\infty}    =   \min_{\aleph(M)}\ \max_{\minM} \, 
  D  \lkr U\upm, \hU^{(\aleph(m))}; 2,\infty \rkr, 
\ee 
\be \label{eq:err_subspace_ave}
R^2_{U,ave}    = \frac{1}{M} \ \min_{\aleph(M)}\ \sum_{m=1}^M \, 
\left\| \sin \Te \lkr U\upm, \hU^{(\aleph(m))} \rkr \right\|_F^2,
\ee
\be \label{eq:err_subspace_max}
R_{U,\max}    =   \min_{\aleph(M)}\ \max_{\minM} \, 
 \left\| \sin \Te \lkr U\upm, \hU^{(\aleph(m))} \rkr \right\|.
\ee


\subsection{The accuracy of the between-layer clustering}
\label{sec:error_between}

Successful between-layer clustering relies on the  fact that the elements of the true  Gramm matrix $\Te(l_1, l_2)$ in \fr{eq:Theta1}
and  \fr{eq:Theta2}  are large, when $ s(l_1) = s(l_2)$, and are small otherwise. 
In practice, those values are unavailable, and one has to use their sample substitutes in Algorithm~\ref{alg:between}. 
It is obvious that the between-layer clustering errors depend on
whether matrix $\hTe$ in \fr{eq:hte_hTe} inherits the properties of $\Te$. It turns out that the latter is true if $n$ is large enough
and the layers are not too sparse. Specifically, the following statement holds.

\begin{lem} \label{lem:est_Theta}  
Let Assumptions {\bf A1}--{\bf A5} hold and  matrix $\hTe$ be defined in \fr{eq:hte_hTe}.  
If the layers of the network are moderately sparse, so that 
\be \label{eq:rhon_cond}
n\, \rhon \geq C_{\rho}\, \log n,
\ee 
then, if $n$ is large enough, for any fixed constant $\tau$, uniformly, with probability at least $1 - C  \, n^{-(\tau- \tau_0)}$,
one has
\be \label{eq:hTheta1} 
     \hTe(l_1, l_2) \!  \geq \! K_m \! \lkv 1 - C (n  \rhon)^{-\frac12} \rkv \     \mbox{if}\ 
     s(l_1)=s(l_2)=m,  
\ee
\be \label{eq:hTheta2}
 |\hTe(l_1, l_2)|    \leq C \,  K \, (n \rhon)^{-1/2}  \quad  \quad     \mbox{if}\ s(l_1) \neq s(l_2), 
\ee
where constants $C$  depend only on $\tau$ and the constants in Assumptions~{\bf A1}--{\bf A5}.
\end{lem}

While we state that $n$ has to be large enough in order Lemma~\ref{lem:est_Theta} holds, we 
provide an exact inequality for the lower bound on $n$ in formula \fr{eq:n_lowbound} of Section~\ref{sec:proof_suppl},
which ensures that Lemma~\ref{lem:est_Theta} and subsequent statements are valid.   

Note that condition \fr{eq:rhon_cond} is very common in the random networks models. If a random network follows the SBM, 
then condition \fr{eq:rhon_cond} is necessary and sufficient for adequate community recovery (\cite{Gao:2017:AOM:3122009.3153016}).
Since in our multiplex network the layers are grouped on the basis of subspace similarities, it is only natural that assumption
\fr{eq:rhon_cond}  appears in our theory. Indeed, this assumption is present in a number of papers that consider simpler multilayer 
GRDPG models, such as COSIE in  \cite{JMLR:v22:19-558} and  \cite{MinhTang_arxiv2022},  or the Multilayer Degree-Corrected Stochastic Blockmodels
in \cite{agterberg_Multilayer_DCBM}.

\begin{thm} \label{thm:clust_between}  
Let layer assignment function $\hs$ be obtained by Algorithm~\ref{alg:between} with $K\upl = K_m$, $m=s(l)$.
Let Assumptions {\bf A1}--{\bf A5}  and condition \fr{eq:rhon_cond}   hold, and   
\be \label{eq:extra_cond}
\limn  K \, M^2\, \log n\, (n \, \rhon)^{-1}   = 0.
\ee
If $n$ is large enough, so that condition  \fr{eq:n_lowbound} is valid,  
then, for any fixed constant $\tau$ and $\Ctau$ that depends only on  $\tau$ and constants in Assumptions  {\bf A1}--{\bf A5}, one has
\be \label{eq:error_between}
\PP (R_{BL} = 0) = \PP (\hs = s)   \geq 1  - C\, n^{- \tau}.
\ee
Hence, Algorithm~\ref{alg:between} delivers strongly consistent clustering with high probability.
\end{thm}

\begin{rem} \label{rem:between_comparison}
{\bf  Comparison with results of Pensky and Wang (2024).\ }
{\rm
Observe that the between-layer  clustering error rates  are much smaller than in 
\cite{pensky2021clustering} where it is shown, that under similar assumptions, with high probability, 
$R_{BL} = O \lkr  \sqrt{\log n}\, (n \rhon)^{-1/2}  \rkr$  as $n \to \infty$.  
The rates in Theorem~\ref{thm:clust_between} are similar to the ones, obtained  in \cite{noroozi2022_SSC}, 
for a much simpler  network with the SBM-equipped sign-free layers. 
In addition, \cite{noroozi2022_SSC} used a methodology that does not work for a network with the   
GRDPG-equipped layers.  
}
\end{rem}


\subsection{The subspace  fitting errors   in the groups of layers}
\label{sec:error_fitting}

In this section, we provide upper bounds for the divergence between matrices $U\upm$ and their estimators
$\hU\upm$, $\minM$, delivered by Algorithm~\ref{alg:subspace_est}.
We measure their discrepancies by $R_{U,ave}$ and $R_{U,\max}$ defined in \eqref{eq:err_subspace_ave}
and \eqref{eq:err_subspace_max},
as well as by  $R_{U,2,\infty}$   defined in \eqref{eq:err_subspace_2inf}.  
The upper bounds for $R_{U,ave}$ and $R_{U,\max}$ are presented below in 
Theorem~\ref{th:error_Um_est}.

\begin{thm} \label{th:error_Um_est}
Let Assumptions of Theorem~\ref{thm:clust_between} hold. Let   matrices  $\hU\upm$, $\minM$, 
be obtained using   Algorithm~\ref{alg:subspace_est} and
\be \label{eq:L_lowbound}
L \geq 2\, (\lowc_\pi)^{-2}\, M^2 \log(n^{\tau+2 \tau_0}).
\ee
Then,
for any $\tau >0$, and absolute constants $\Ctau$ that depend on $\tau$ and constants 
in Assumptions~{\bf A1}--{\bf A5} only,  with probability at least 
$1 - C \, n^{-(\tau - 2 (\tau_0 + 1))}$, one has
\begin{align}
& R_{U,\max}  \leq   
\Ctau\,   \lkr \frac{\sqrt{M\, \log n} }{\sqrt{\rhon\, n \, L}}   + \frac{1}{n}  \rkr, \nonumber\\
&  \label{eq:error_Save_max}\\
& R_{U,ave} \leq  
\Ctau\,   \lkr \frac{\sqrt{K\, M\, \log n} }{\sqrt{\rhon\, n \, L}}   + \frac{\sqrt{K} }{n}  \rkr. 
\nonumber
\end{align}
\end{thm}

While the proof of Theorem~\ref{th:error_Um_est} follows the lines of the proofs in  \cite{pensky2021clustering} 
and   \cite{noroozi2022_SSC}, construction of an upper bound for  $R_{U,2,\infty}$ requires novel techniques. 
\ignore{
While we consider   symmetric layer networks in this paper, we believe that, with   minimal adjustments, 
the same methodology can be applied for directed networks with $P\upl = U\upm B\upl V\upm$, $\minM$, $\linL$.
Setting $M=1$ yields  Corollary~\ref{cor:error_2inf} that provides the $(2,\infty)$ error bounds in undirected COSIE model. 
}

\begin{thm} \label{th:error_2inf}
Let Assumptions of Theorem~\ref{th:error_Um_est} hold. Then,
for any $\tau >0$, and absolute constant  $\Ctau$ that depends on $\tau$ and constants 
in Assumptions~{\bf A1}--{\bf A5} only,  with probability at least 
$1 - C \, n^{-(\tau - \tau_0 - 1)}$, one has
\be \label{eq:error_2inf}
R_{U,2,\infty}  \leq   
\Ctau  \frac{\sqrt{K}}{\sqrt{n}}\! \lkv \frac{\sqrt{M  \log n} }{\sqrt{n  \rhon  L}}  \!     +  \! \frac{1}{n} \rkv \! 
\lkr 1 \! + \! \frac{\sqrt{M  \log n}}{\rhon  \sqrt{n  L   K}} \rkr.
\ee
\end{thm}

\medskip

\begin{cor} \label{cor:error_2inf}
Let Assumptions of Theorem~\ref{th:error_Um_est} hold with $M=1$. Then, 
with probability at least $1 - C \, n^{-(\tau - \tau_0 - 1)}$, one has
\be \label{eq:error_2inf_COSIE}
R_{U,2,\infty}  \leq   
\Ctau  \frac{\sqrt{K}}{\sqrt{n}}  \lkv \frac{\sqrt{\log n} }{\sqrt{n  \rhon  L}}  \!     +  \! \frac{1}{n} \rkv   
\lkr 1 + \frac{\sqrt{\log n}}{\rhon\, \sqrt{n  L   K}} \rkr.
\ee
\end{cor}

\medskip
 
Observe that the errors in \fr{eq:error_2inf} and \fr{eq:error_2inf_COSIE} are products of three terms. 
The middle terms in the square parentheses are the upper bounds of the errors of subspace recovery in operational norm.
The first factor, $\sqrt{K/n}$ is  the main improvement effect of using the $(2,\infty)$ rather than operational norm.
The last multipliers are the correction terms, that are  due to the fact that, in the expansion for the $(2,\infty)$ errors, the 
operational norm is multiplied by a factor that can be potentially larger than $\sqrt{K/n}$. The value of this term
depends on sparsity $\rhon$, the number of layers $L$ and on $n$ and $K$. It is easy to see that the term is bounded above 
if the  network is not too sparse, or if $L$ is large enough.

Note that, although  we consider   symmetric layer networks in this paper, we believe that, with   minimal adjustments, 
the same methodology can be applied to directed networks with $P\upl = U\upm B\upl V\upm$, $\minM$, $\linL$.
Setting $M=1$ would then  provide the $(2,\infty)$ error bound  in the undirected COSIE model of \cite{MinhTang_arxiv2022}.


\begin{rem} \label{rem:est_Ql}
{\bf Estimation of the loading matrices.\ }
{\rm
While in our setting, matrices $Q\upl$ in \fr{eq:GDPG} are just nuisance parameters,
one can imagine a scenario where those matrices may exhibit some behavior of interest and,
hence, need to be estimated. Since under conditions of Theorem~\ref{thm:clust_between}, clustering is 
perfect with high probability when $n$ is large enough,  the model reduces to the COSIE model 
for each group of clusters. Therefore, one can essentially repeat all results that were obtained for 
the COSIE model. Specifically, since one can estimate $U\upm$, $\minM$, with high precision, the errors 
of estimating $Q\upl$ will be dominated by the portions originated from the differences between $A\upl$
and $P\upl$. 
}
\end{rem}


\begin{rem} \label{rem:within_comparison}
{\bf  Comparison with competitive results.\ }
{\rm
Observe also that the average subspace recovery errors in \fr{eq:error_Save_max}  are similar to the rates,
obtained in \cite{JMLR:v22:19-558}     for the COSIE model. They are considerably lower than the errors 
in \cite{pensky2021clustering}, where it is proved that,   with high probability,  
\begin{align*}  
R_{U,ave} & \leq \Ctau \lfi   
I (M>1)\, \frac{\sqrt{K^5  M \log n}}{\sqrt{n \, \rhon}}\,  \lkv 1 + \frac{K \, \log  n}{\sqrt{n \, \rhon}} \rkv \right.\\
& + \left.\frac{\sqrt{K^5  M \log n}}{\sqrt{n L \rhon}} + \frac{K^{5/2}}{n}  
\rfi.
\end{align*}
Moreover, due to similarities between the algorithms, they are similar to the errors in  
\cite{Lei_2023_JASA}: the seeming difference comes from the fact that \cite{Lei_2023_JASA} study sparser networks with $n\, \rhon \leq C$
where $C$ is a constant.

As it is expected, the error rate  $R_{U,2,\infty}$  of subspace recovery in $(2,\infty)$-norm is consistently smaller that the rate $R_{U,\max}$
in operational norm. Specifically, it follows from \fr{eq:error_Save_max} and \fr{eq:error_2inf} that, under condition \fr{eq:rhon_cond},
one has, as $n \to \infty$
\bes 
R_{U,2,\infty}/R_{U,\max} \propto \lkv \sqrt{K/n} + \sqrt{\log n}/(\rhon\, n \, \sqrt{L})\rkv  \to 0 
\ees
for any $L$. Note that the latter is not true for \cite{MinhTang_arxiv2022}, who studied the   COSIE model, corresponding to $M=1$.
 Specifically, in our notations,  under  the assumption that  $\|U \|_{2,\infty} \leq C\,  K^{1/2}\, n^{-1/2}$,
 they derived that 
 \begin{align}
  & \min_{W \in \calO_K} \|\hU \, W - U \|_F \leq   C\, \sqrt{K}\, (L n \rhon)^{-1/2}, \label{eq:minh1} \\
 & \min_{W \in \calO_K} \|\hU \, W - U \|_{2,\infty} \leq C\, \lkv  (K\, \log n)^{1/2} (L\, n^2 \rhon)^{-1/2} \right. \nonumber \\
 & \left. \hspace{20mm}  + \ 
K^{1/2}\, n^{-1/2}\, \log n\, (n \rhon)^{-1} \rkv.
 \label{eq:minh2} 
 \end{align}
Therefore, \fr{eq:minh1} and \fr{eq:minh2} imply that, in their case, 
\be \label{eq:minh_frac}
\frac{\displaystyle \min_{W \in \calO_K} \|\hU \, W - U \|_{2,\infty}}{\displaystyle \min_{W \in \calO_K} \|\hU \, W - U \|_F}
\propto
\frac{\sqrt{K}}{\sqrt{n}} + \frac{\sqrt{L   \log n}}{\sqrt{n^2\, \rhon}}.
\ee 
Observe that  the second term  tends to infinity if $L$ grows polynomially faster than $n$ and $n \rhon \asymp (\log n)^{a+1}$
for some constant $a >0$. 
%
%
Note also that the $R_{U,2,\infty}$ error of Algorithm~\ref{alg:subspace_est} is {\bf significantly smaller} than that of 
\cite{MinhTang_arxiv2022}.  We do not know whether the latter is due to a shortcoming of their algorithm, or the deficiency of their proof. 
}
\end{rem}


\subsection{The  Stochastic Block Model and the Mixed Membership Model}
\label{sec:SBM_MMM}

Observe that in the case of the  multiplex network   where layers are equipped with the SBM or MMM,  
one has $X\upm = Z\upm$, where  $Z\upm$ are membership matrices,
so that each row of matrix $Z\upm$ sums to one. The latter leads to  all
covariance matrices $\Sig\upm$ being  singular. Nevertheless, 
Algorithms~\ref{alg:between} and \ref{alg:subspace_est} can still be 
successfully applied, albeit  with very minor modifications.

Indeed,  Algorithm~\ref{alg:between}  starts with replacing matrices $A\upl$ with $\tilA\upl = \Pi^{\bot} A\upl  \Pi^{\bot}$,
$\linL$. In the case when Assumption~{\bf A3} holds, matrices  $\tilA\upl$ are proxies for matrices $\tilP\upl$ of   rank 
$K_m$ where $m=s(l)$. If the layer graphs are generated by the SBM or the MMM, 
one has $\rank(\tilP\upl) = K_m-1$. This, however,
does not matter  as long as one treats $K\upl$ as the rank of  $\tilP\upl$, $\linL$, so that $K\upl= K_m -1$ for $s(l)=m$.

In order to see why this is true, consider the situation where, similarly to  \fr{eq:Xm_gen}, one has $X\upm(i,:) \sim \iid \, f_m$, $i \in [n]$,
and $f_m$,  $\minM$, are such that $X\upm(i,:) 1_{K_m} =1$  but covariance matrices of the first $(K_m -1)$ components of row vectors 
$X\upm(i,:)$ are non-singular.  In order to formalize the discussion above, we introduce matrices 
$\brX\upm = X\upm(:,1:K_m-1)$ which consist of the first $K_m -1$ columns of matrices $X\upm$
and denote their covariance matrices by $\brSig\upm = \Cov(\brX(i,:)) \in \RR^{(K_m-1)\times (K_m-1)}$.
We assume that $f_m$ are such that 
\begin{align} 
& X\upm 1_{K_m} = 1_n, \quad, \ \  \minM, \label{eq:new_Xm_assump}\\
\nonumber
& 0 < \lowc  \leq \lam_{\min} (\brSig\upm)  \leq \lam_{\max} (\brSig\upm) \leq  \highc < \infty.
\end{align}
Note that the first condition in \fr{eq:new_Xm_assump} holds for both the SBM and the MMM. 
The second condition in \fr{eq:new_Xm_assump} can be guaranteed by a variety of assumptions.
In this paper, as one of the options, we assume that, in the cases of the SBM
and the MMM, $f_m$  are, respectively, the Multinomial($\al\upm$) or
the Dirichlet($\al\upm$) distributions, where vectors $\al\upm$ are such that
\be \label{eq:alpha_cond} 
\lowc_{\al}/K \leq \al\upm_k \leq \highc_{\al}/K, \ \minM.
\ee
Partition each matrix $B\upl$, $s(l)=m$,  into the  three portions: matrix $\tilB\upl = B\upl(1:K_m-1,1:K_m-1)$, 
vector $\tilb\upl = B\upl(1:K_m-1,K_m)$ and a scalar $B\upl(K_m,K_m)$. Introduce 
$\brb\upl = \tilb\upl - B\upl(K_m,K_m) \, 1_{K_m-1}$ and
\begin{align} \label{eq:brb} 
\brB\upl & = \tilB\upl -  \tilb\upl \, (1_{K_m-1})^T - 1_{K_m-1}\, (\tilb\upl)^T \\
& +  B\upl(K_m,K_m)\, 1_{K_m-1}\, (1_{K_m-1})^T.  \nonumber
\end{align}
Observe  that  $\brB\upl = J\upm B\upl (J\upm)^T$ where matrix $J\upm$ is a horizontal 
concatenation of $I_{K_m-1}$, the 
$(K_m -1)$-dimensional identity matrix, and the vector $-1_{K_m-1}$. 
Rewrite $P\upl$ as 
$P\upl = \brX\upm \brB\upl (\brX\upm)^T + \brX\upm  \brb\upl 1_n^T + 1_n (\brX\upm  \brb\upl)^T + B\upl(K_m,K_m)$. 
While $P\upl$ cannot be expressed via $\brX\upm$ and $\brB\upl$, matrices $\tilP\upl$ in \fr{eq:tilbp1} can:
\begin{align} \label{eq:tilPupl}
\tilP\upl & =  \brX\upm \,  \brB\upl\, (\brX\upm)^T, \\
  \brB\upl & = J\upm B\upl (J\upm)^T, 
\    J\upm = [I_{K_m-1}| -1_{K_m-1}].  \nonumber
\end{align}
Here, matrices $\brX\upm$ are such that  condition \fr{eq:new_Xm_assump} holds and can  replace  Assumption~{\bf A3}. 
Moreover, since all singular values of $J\upm$ lie between 1 and $\sqrt{K_m}$, matrices $\brB\upl$ satisfy the assumption 
which is equivalent to the Assumption~{\bf A5} 
\be \label{eq:brBol}
\brB^{(l)} = \rho_n \brB^{(l)}_0 \quad \mbox{with} \quad \sig_{\min}(\brB^{(l)}_0)  \geq \lowC_{B} > 0.
\ee 
Therefore, one obtains an immediate corollary of Theorem~\ref{thm:clust_between}.

\begin{cor} \label{cor:clust_between}  
Let distributions $f_m$ and matrices $B\upl$ satisfy Assumptions {\bf A1}, {\bf A2}, {\bf A4}, {\bf A5} and \fr{eq:new_Xm_assump}.
Let the layer assignment function $\hs$  be obtained by Algorithm~\ref{alg:between}  with $K^{(l)} = K_m-1$, $m = s(l)$, where 
$\displaystyle \min_m K_m \geq 2$. 
Let   \fr{eq:rhon_cond}   and   \fr{eq:extra_cond} hold.
Then, if $n$ is large enough, for any fixed constant $\tau$, \fr{eq:error_between} is valid, 
so that Algorithm~\ref{alg:between} delivers strongly consistent layer clustering with high probability. 
\end{cor}




\section{Simulation study }  
\label{sec:simul_study}

%
\begin{algorithm}   [t]  
\caption{\ The between-layer clustering of \cite{pensky2021clustering} }
\label{alg:Wang}
\begin{flushleft} 
{\bf Input:} Adjacency tensor $\bA \in \{0,1\}^{n \times n \times L}$; number of groups of layers $M$;  
ambient dimension $K^{(l)}$ of each layer $\linL$; parameter $\eps$ \\
{\bf Output:} Estimated clustering function $\hs: [L] \to [M]$ for groups of layers\\
{\bf Steps:}\\
{\bf 1:} Find matrices $\hhU_A\upl = SVD_{K\upl} (A\upl)$ of $K\upl$ 
left leading singular vectors  of matrices $A\upl$ \\
{\bf 2:} Form matrix $\tilTe \in \RR^{n^2 \times L}$ with columns $\tilTe(:,l) = \vect(\hhU_{A,l} (\hhU_{A,l})^T)$,  $\linL$\\
{\bf 3:} Find matrix $\hcalW = \SVD_M (\tilTe^T) \in \calO_{L,M}$  of $M$  leading singular vectors of matrix 
$\tilTe^T$ \\
{\bf 4:}  Cluster  $L$ rows of  $\hcalW$ into $M$ clusters using   $(1+\eps)$-approximate $K$-means clustering. Obtain 
estimated clustering function $\hs$   
\end{flushleft} 
\end{algorithm}
%

\subsection{Simulations settings}
\label{sec:simul_setting}

A limited simulation study, described in this section,   has a dual purpose.
The first goal is to show that our techniques deliver accurate cluster assignments and 
ambient subspace estimators in a finite sample case. Since to the best of our knowledge, 
the only other paper that considers the multilayer GRDPG with groups of layers  is 
\cite{pensky2021clustering},  we carry out   finite sample comparisons with the between-layer 
clustering in  \cite{pensky2021clustering} (Algorithm~\ref{alg:Wang}).
We do not exhibit the comparison of the ambient subspaces recovery  for those cases 
since the construction of the unbiased tensor of squared averages in \cite{pensky2021clustering}  
needs to be modified  due to the presence of signs, and Algorithm~\ref{alg:subspace_est} 
carries out this modification.

Instead, we compare our within-layer clustering technique with the one 
used in \cite{arxiv.2007.10455} where adjacency matrices $A\upl$ are placed side by side to form one 
matrix $\tilA$, and the common matrix of  the  left singular vectors of $P\upl$, $\linL$, is then  estimated by
the matrix of the left singular vectors of $\tilA$ (see Remark~\ref{rem:spectr_embedding}).
In order the comparison is fair, we use Algorithm~\ref{alg:between} for the between-layer clustering for 
both their method and Algorithm~\ref{alg:subspace_est}.

The second goal is to demonstrate advantages of the model itself. Indeed,
in many applications, when a network is constructed and edges are drawn, the associated 
edges' signs are removed, and  the adjacency matrix has 0/1 entries only. In our simulations examples, we show that
by just keeping the sign and allowing the entries of the adjacency matrix take $0/1/-1$ values,
one can significantly improve the accuracy of the analysis of the network.
%


\begin{figure*} [!t] 
\centering
\[\includegraphics[width=6.0cm, height=3.8cm]{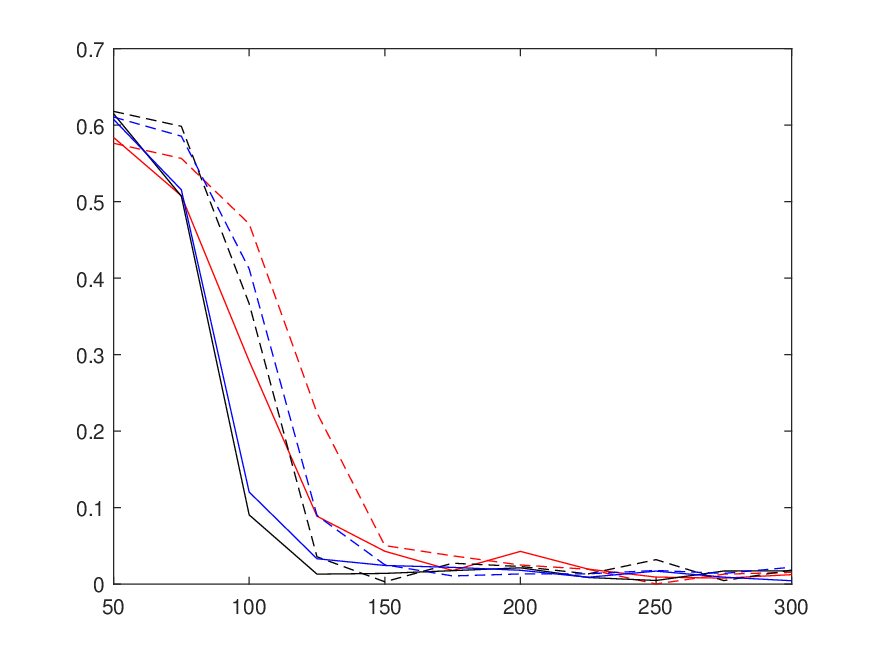} 
\includegraphics[width=6.0cm, height=3.8cm]{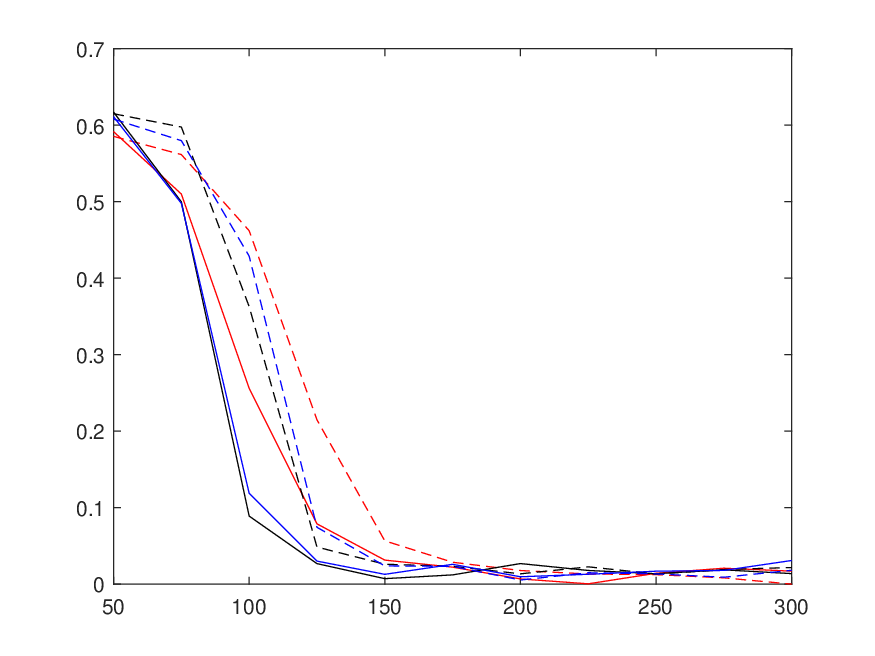} 
\includegraphics[width=6.0cm, height=3.8cm]{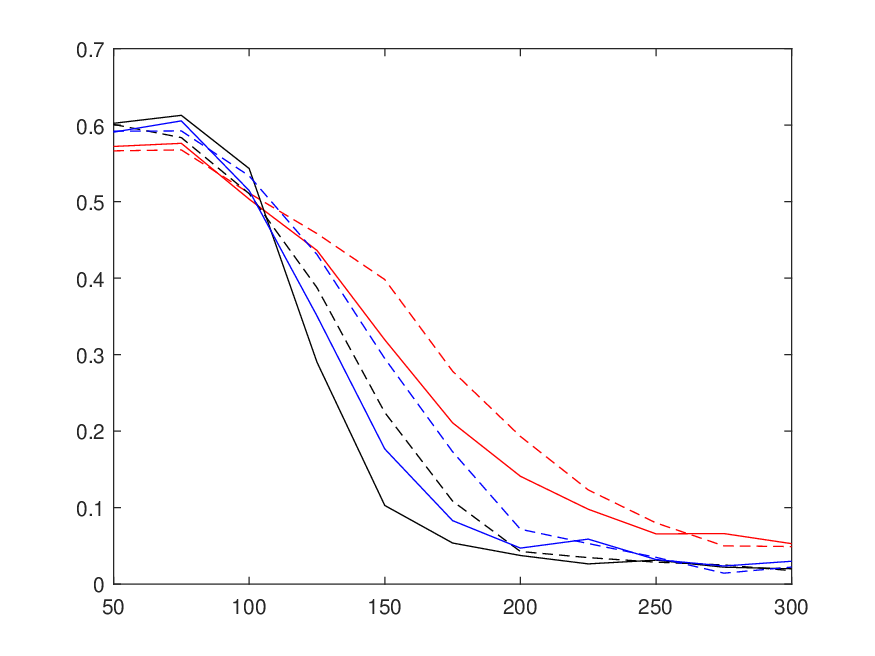} \] \\
\[\includegraphics[width=6.0cm, height=3.8cm]{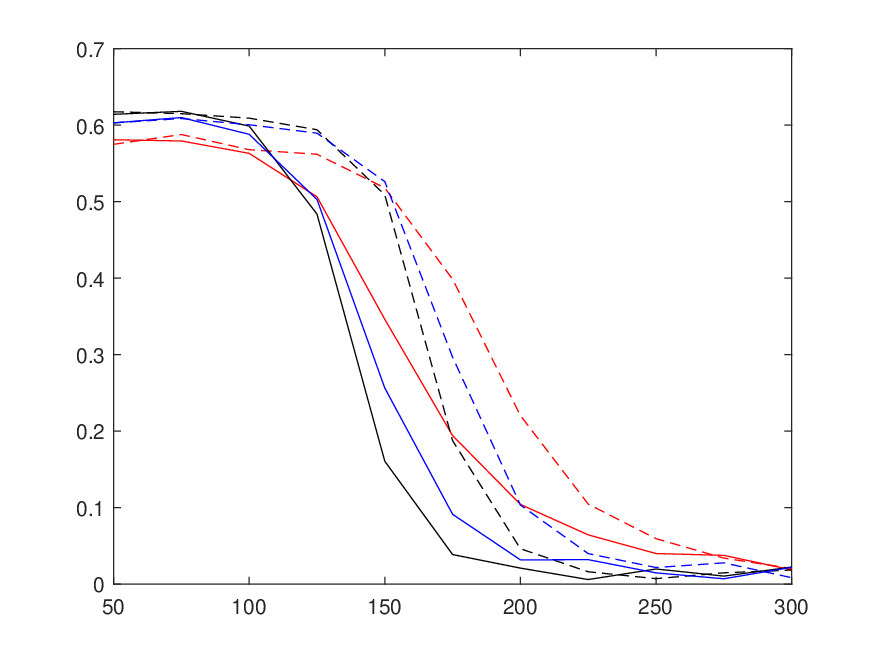} 
\includegraphics[width=6.0cm, height=3.8cm]{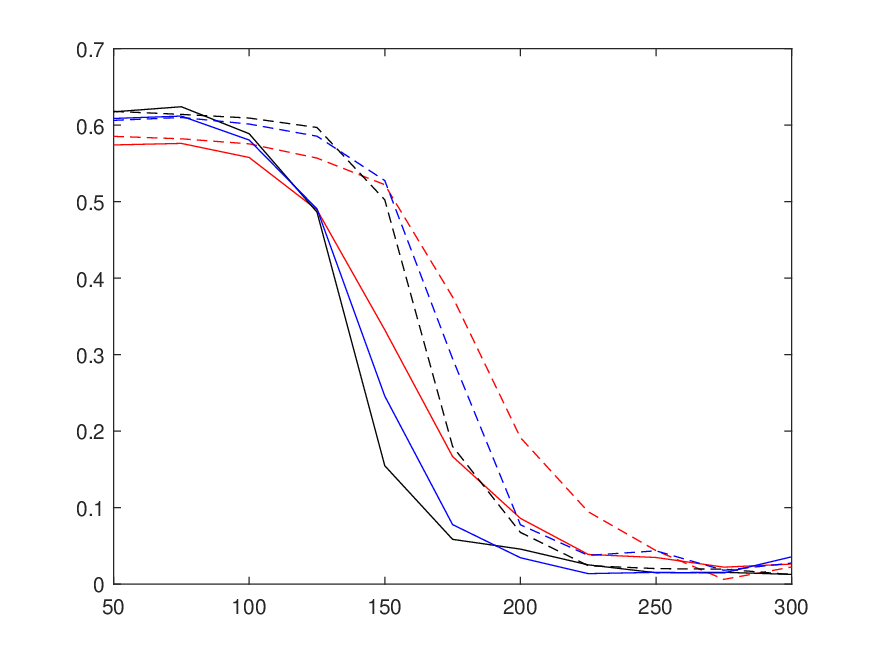} 
\includegraphics[width=6.0cm, height=3.8cm]{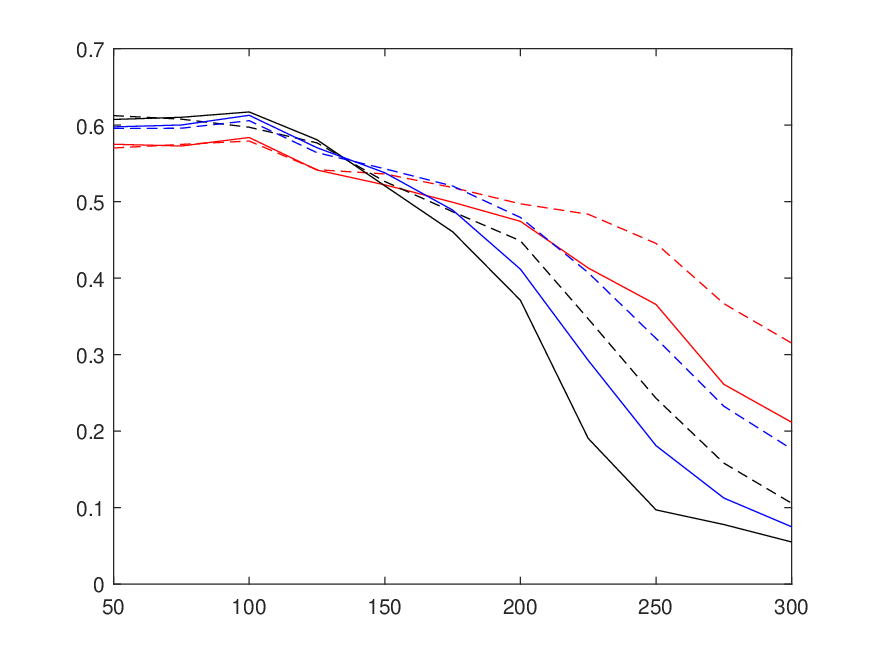} \] \\
\caption{\footnotesize{The between-layer clustering errors $R_{BL}$  in \fr{eq:err_betw_def} of Algorithms~\ref{alg:between}
 (solid lines) and Algorithms~\ref{alg:Wang} (dash lines). 
Matrices $X\upm$  are generated by the truncated normal distribution  with $\sig =1$, $\mu =0$  (left panels),
the  truncated  $T$-distribution with $\nu =2$  (middle panels) and 
the truncated Dirichlet distribution with  $\alpha_k = \alpha =0.1$, $k \in [K_m+1]$ (right panels). 
In all cases, $M=3$ and  $K_m=K=3$. 
The entries of $B\upl$ are generated  as uniform random numbers  between   $c=-0.05$ and
$d = 0.05$ (top row) and  $c=-0.03$ and $d = 0.03$  (bottom row). 
Errors are averaged over 100 simulation runs,  $L=50$ (red),  $L=100$ (blue),   $L=150$ (black).
}}
\label{fig:Wang_between_compare}
\end{figure*}


In order to attain those objectives, we examine the  SGRDPG multiplex networks 
with matrices $X\upm$, $\minM$,  generated by  the following four distributions described in Section~\ref{sec:net_form}:

{\it Case 1.}  $f_m$ is a truncated normal distribution, 
so that $\xi \sim f_m$ is obtained as $\xi = \eta/\|\eta\|$,  where  $\eta$ is a $K$-dimensional 
 vector  with components being i.i.d. normal with mean zero and variance $\sig^2$.

{\it Case 2.} $f_m$ is a truncated $T$-distribution  with $\nu$ degrees of freedom: $\xi \sim f_m$ is obtained as 
$\xi = \eta/\|\eta\|$  where $\eta \sim T(\nu)$, the  $T$-distribution  with $\nu$ degrees of freedom.

{\it Case 3.} $f_m$ is a  reduced Dirichlet  distribution,   so that $\xi \sim f_m$ is obtained as 
 the first $K$ components of  vector $\eta\in [0,1]^{K+1}$, where  $\eta \sim {\rm Dirichlet} (\vec{\alpha})$   with
$\vec{\alpha} = (\alpha_1, ..., \alpha_{K+1})$, $\alpha_k > 0$, $k \in [K+1]$.

{\it Case 4.} $f_m$ is a the multinomial distribution, so that $\xi \sim f_m$ is obtained as 
$\eta \sim {\rm Multinomial} (\vec{\varpi})$, where
$\vec{\varpi}= (\varpi_1, ..., \varpi_K)$, $0\leq \varpi_k \leq 1$, $\varpi_1 + ...+ \varpi_{K} = 1$.
Observe that in this case, distribution $f_m$ does not satisfy Assumption {\bf A3} and leads to the 
Signed SBM  studied in the Section~\ref{sec:SBM_MMM}.  
\\


In order to obtain the DIMPLE-SGRDPG network, we fixed the number of groups of layers $M$
and the ambient dimension of each group of layers $K_m$. 
We generate  group membership of  each layer $\linL$, using the 
Multinomial distribution as described in \fr{eq:group_memb} with $\pi_m = 1/M$. 
After that, we generate $M$ matrices $X\upm \in [-1,1]^{K_m}$ with i.i.d. rows using \fr{eq:Xm_gen}, 
where $f_m$ is one of the four distributions above.
Next, we generate the above diagonal entries of symmetric matrices $B\upl$, $\linL$, in \fr{eq:DIMPnew_GDPG} 
as uniform random numbers  between $c$ and $d$.
Then, connection probability matrices $P\upl$, $\linL$, are of the form \fr{eq:DIMPnew_GDPG}.
The common bases of the ambient subspaces $U\upm$  in groups of layers are formed by the $K$ left singular vectors 
of matrices  $X\upm$,  and matrices $Q\upl$ in \fr{eq:DIMPLE_GDPG} are of the forms \fr{eq:UupmQupl}. 
Subsequently, the layer adjacency matrices $A\upl$ are  generated according to \fr{eq:sign_adj}.
The code is available from the author's website at\\
{\scriptsize {\tt https://sciences.ucf.edu/math/mpensky/recent-publications/}}


\subsection{Between-layer clustering simulations   comparisons with the  technique of Pensky\& Wang (2024)}
\label{sec:results_comparison}

In this section, we compare the between-layer clustering assignments 
using Algorithm~\ref{alg:between} and Algorithm~\ref{alg:Wang} of \cite{pensky2021clustering} 
in the cases of the  truncated normal distribution, the  truncated $T$-distribution and the  truncated 
Dirichlet distribution with $M=3$, $K_m=K=3$ and two sparsity settings. Results are presented in 
Figure~\ref{fig:Wang_between_compare}.


\begin{figure*} [!t] 
\centering
\[\includegraphics[width=6.0cm, height=3.8cm]{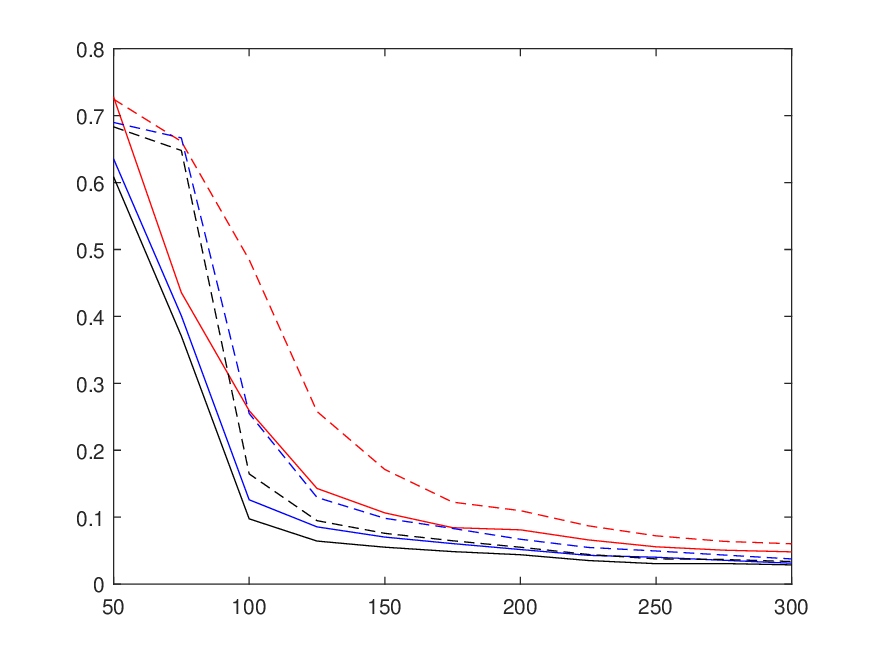} 
\includegraphics[width=6.0cm, height=3.8cm]{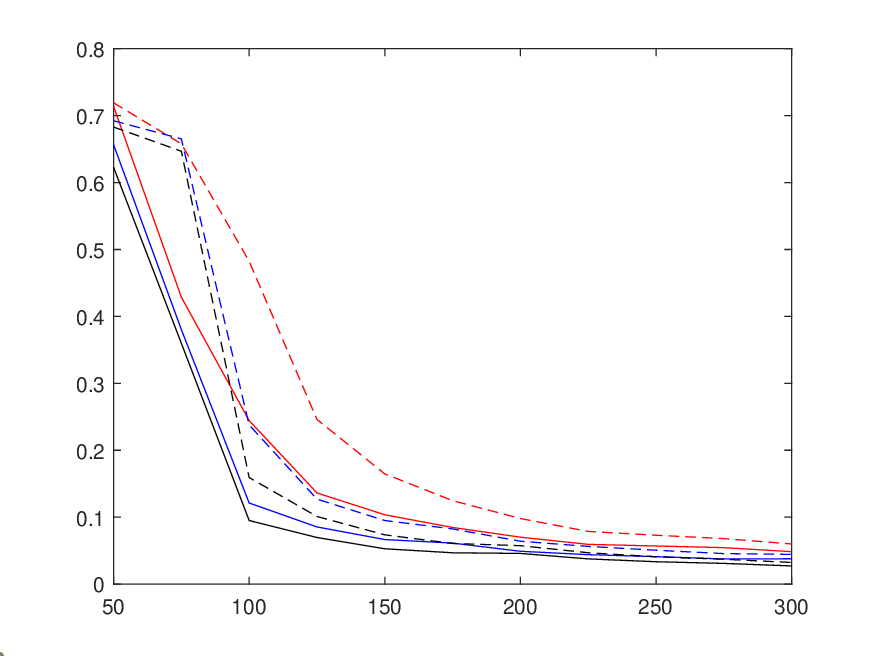} 
\includegraphics[width=6.0cm, height=3.8cm]{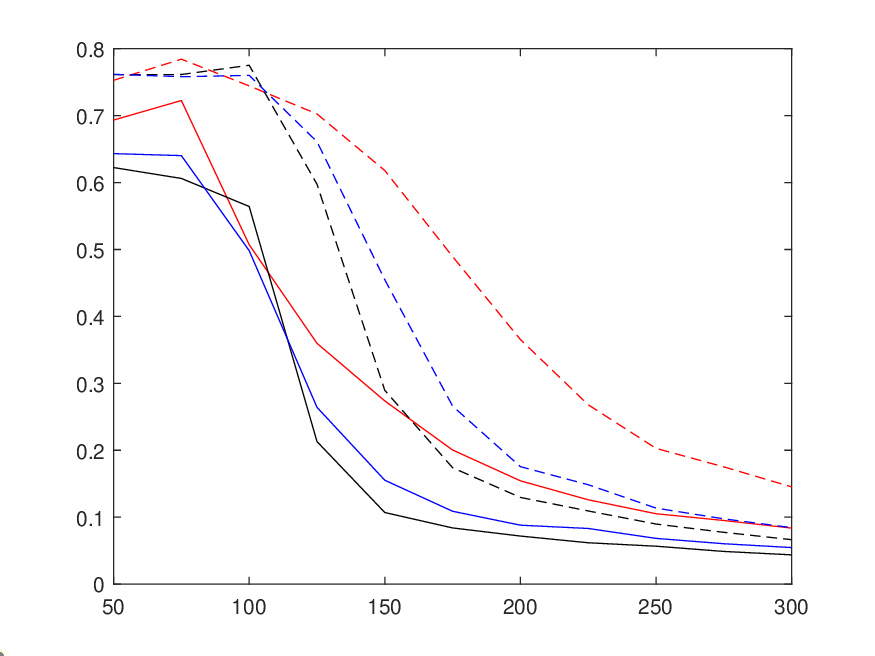} \] \\
\[\includegraphics[width=6.0cm, height=3.8cm]{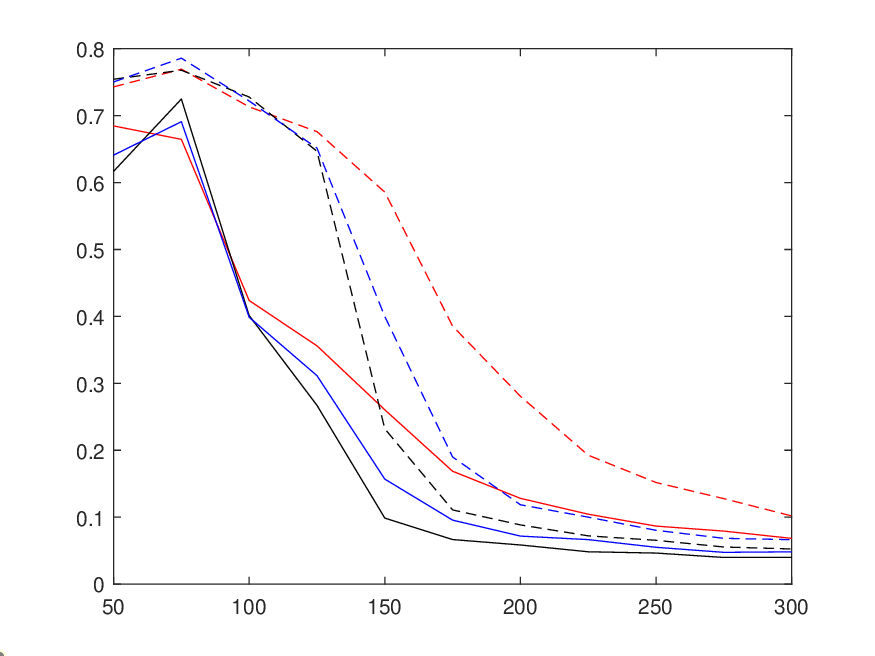} 
\includegraphics[width=6.0cm, height=3.8cm]{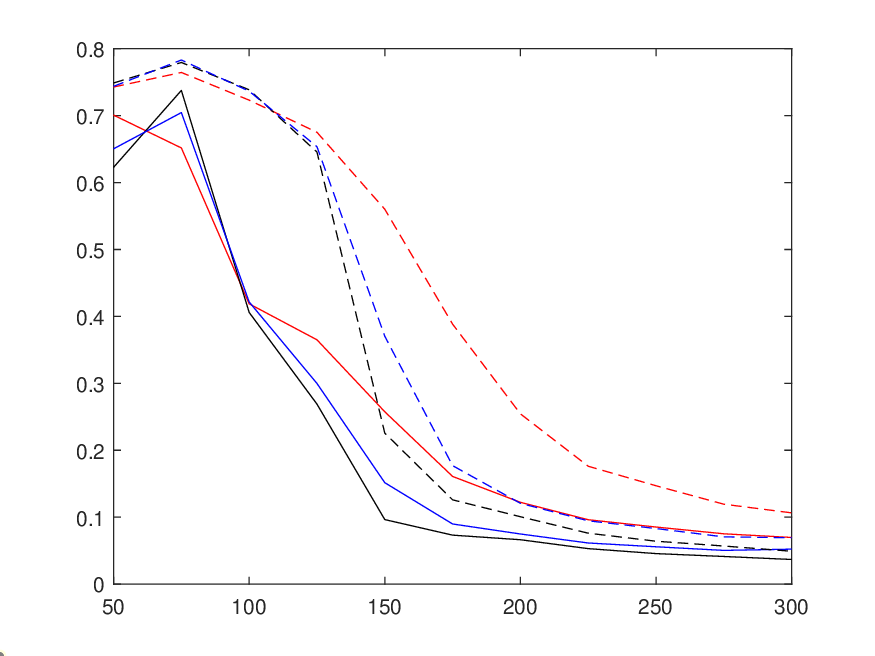} 
\includegraphics[width=6.0cm, height=3.8cm]{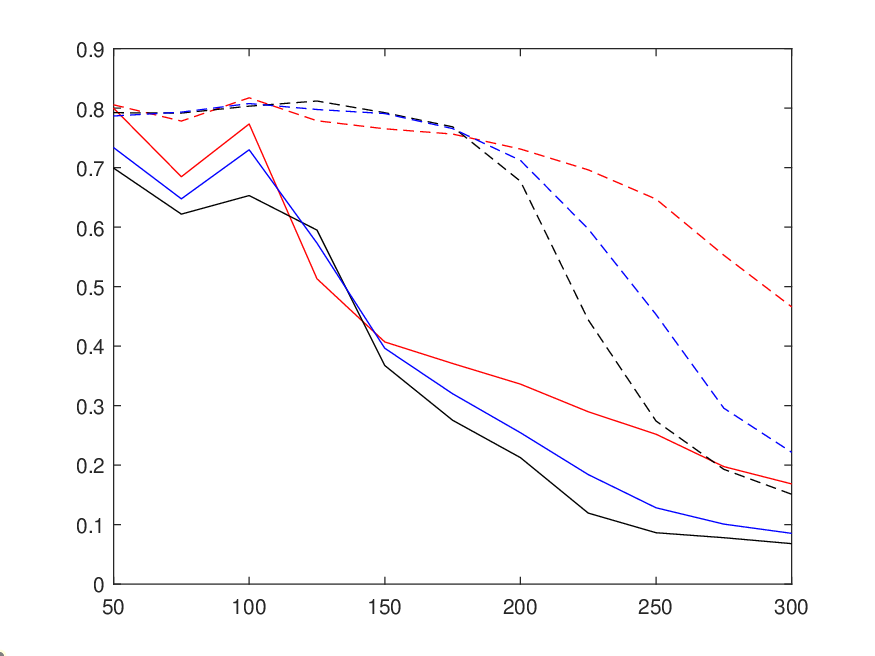} \] \\
\caption{\footnotesize{The subspace estimation errors $R_{2,\infty, ave}$  in \fr{eq:subspace_est_simul_er} 
of Algorithms~\ref{alg:subspace_est}  (solid lines) and of the Algorithm described in Remark~\ref{rem:spectr_embedding}  (dash lines). 
Matrices $X\upm$  are generated by the truncated normal distribution  with $\sig =1$, $\mu =0$  (left panels),
the  truncated  $T$-distribution with $\nu =2$  (middle panels) and 
the truncated Dirichlet distribution with  $\alpha_k = \alpha =0.1$, $k \in [K_m+1]$ (right panels). 
In all cases, $M=3$ and  $K_m=K=3$. 
The entries of $B\upl$ are generated  as uniform random numbers  between   $c=-0.05$ and
$d = 0.05$ (top row) and  $c=-0.03$ and $d = 0.03$  (bottom row). 
Errors are averaged over 100 simulation runs,  $L=50$ (red),  $L=100$ (blue),   $L=150$ (black).
}}
\label{fig:Rubin-Delanchy_within_compare}
\end{figure*}


As   Figure~\ref{fig:Wang_between_compare} demonstrates, even for such a sparse networks the between 
layer clustering errors are decreasing rapidly, as $n$ grows, and are declining slowly as $L$ increases. 
It is also easy to see that  the between layer clustering errors of 
 Algorithm~\ref{alg:Wang} in \cite{pensky2021clustering}   remain larger than the respective errors 
of Algorithm~\ref{alg:between}.


\subsection{Comparison of the ambient subspace estimation techniques}
\label{sec:simul_ambient_est}

In this section we compare the ambient subspace estimation precision of Algorithm~\ref{alg:subspace_est} 
and the algorithm of \cite{arxiv.2007.10455} described in Remark~\ref{rem:spectr_embedding}.
Here, we exhibit results for the examples considered in the previous section, that is,  
the cases of the normal distribution, the $T$-distribution and the Dirichlet distribution 
with $M=3$, $K_m=K=3$ and two sparsity settings. In all examples in this section,  the 
between-layer clustering is carried out using Algorithm~\ref{alg:between}. 

We measure the deviations between ambient subspaces in the two-to-infinity norm defined in \fr{eq:sinT2inf}. 
Specifically, instead of trying to find a matrix $W$ that delivers the minimum in \fr{eq:sinT2inf},
we are using its Frobenius proxy $W_U$ defined as follows. If $\hU^T U = O_1 D_U O_2^T$ is the SVD of  $\hU^T U$,
then $W_U = O_1 \, O_2^T$. Hence, for $\minM$, we replace $D(U\upm,\hU\upm; 2,\infty)$ with  
$\|\hU\upm \, W_U\upm - U\upm \|_{2,\infty}$, where $\hU\upm$ is the best match for $U\upm$ over all permutations of $\minM$. 
Subsequently, we evaluate  the errors as the averages of those quantities:
\be \label{eq:subspace_est_simul_er}
R_{2,\infty, ave} = M^{-1} \sum_{m=1}^M \|\hU\upm \, W_U\upm - U\upm \|_{2,\infty}, 
\ee
where $W_U\upm = O_1\upm \, (O_2\upm)^T$ and 
$(\hU\upm)^T U\upm = O_1\upm D_U\upm (O_2\upm)^T$ is the SVD of  $(\hU\upm)^T U\upm$.

Results are presented in Figure~\ref{fig:Rubin-Delanchy_within_compare}. 
As  it is evident from Figure~\ref{fig:Rubin-Delanchy_within_compare},
Algorithm~\ref{alg:subspace_est} always delivers higher accuracy than the algorithm suggested in 
\cite{arxiv.2007.10455}. The differences between the performances between the algorithms become
more acute when the networks are getting sparser.  The superior accuracy of Algorithm~\ref{alg:subspace_est}
follows from de-biasing, as it is shown in \cite{Lei_2023_JASA}.


\subsection{Comparison with the   networks where signs are removed}
\label{sec:simul_no_sign}

In order to study the effect of removing the signs of the entries of matrices  $A\upl$,
for each simulation run, we carry out a parallel inference with 
matrices $P\upl$ and $A\upl$ replaced by  $|P\upl|$ and $|A\upl|$,
where $|Y|$ denotes the matrix which contains absolute values of the entries of matrix $Y$.
We  denote the  tensors with absolute values, corresponding to $\bP$ and $\bA$,   by $|\bP|$ 
and  $|\bA|$,  respectively.

 
 \begin{figure*}[t!] 
\[\includegraphics[width=6.0cm, height=3.8cm]{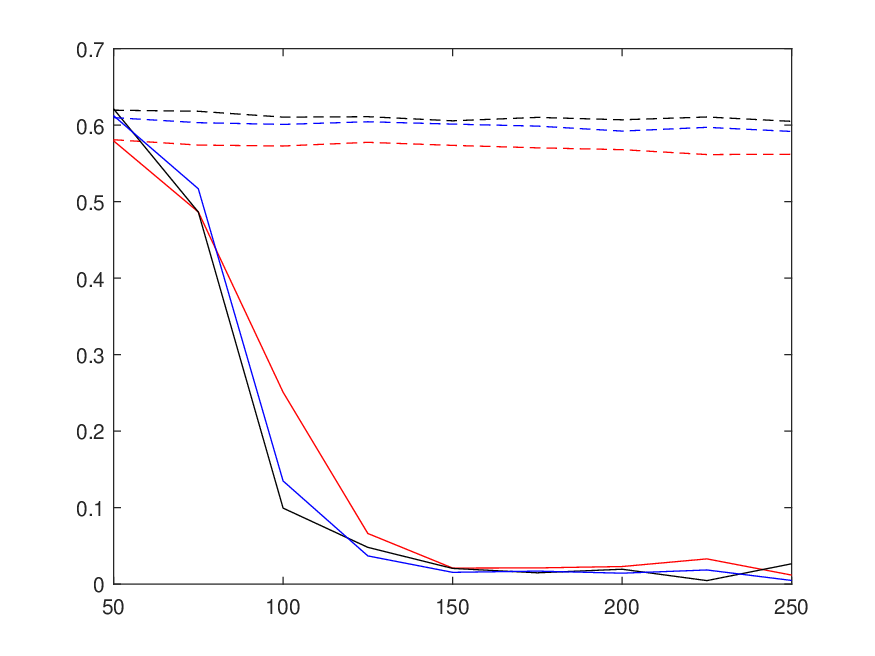}  
\includegraphics[width=6.0cm, height=3.8cm]{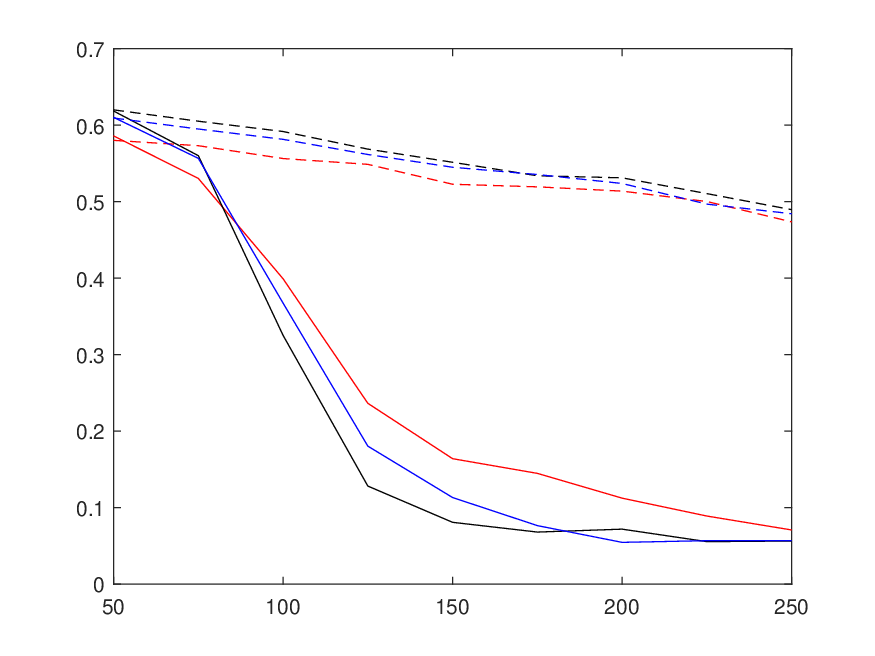} 
\includegraphics[width=6.0cm, height=3.8cm]{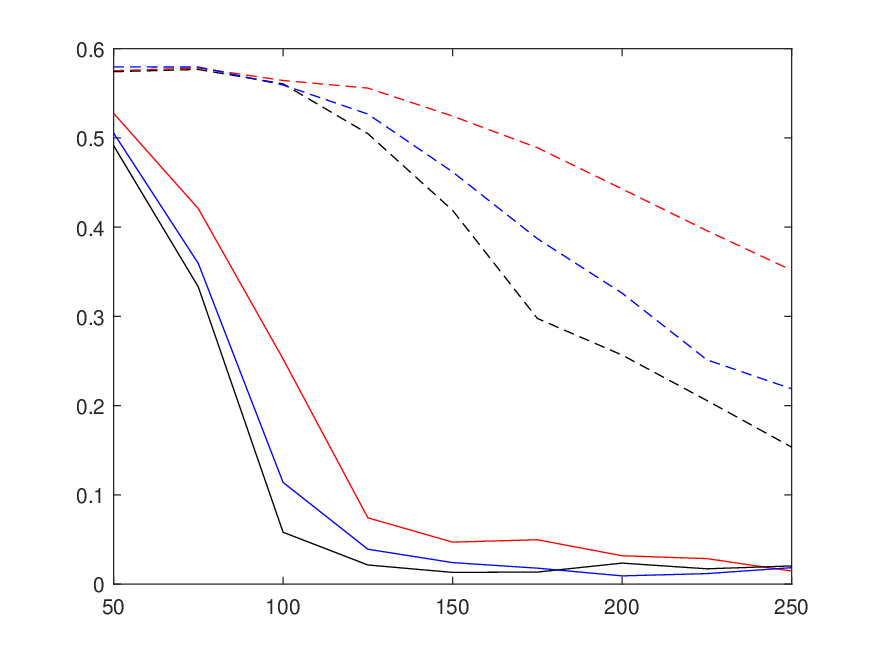}  \] \\
\[\includegraphics[width=6.0cm, height=3.8cm]{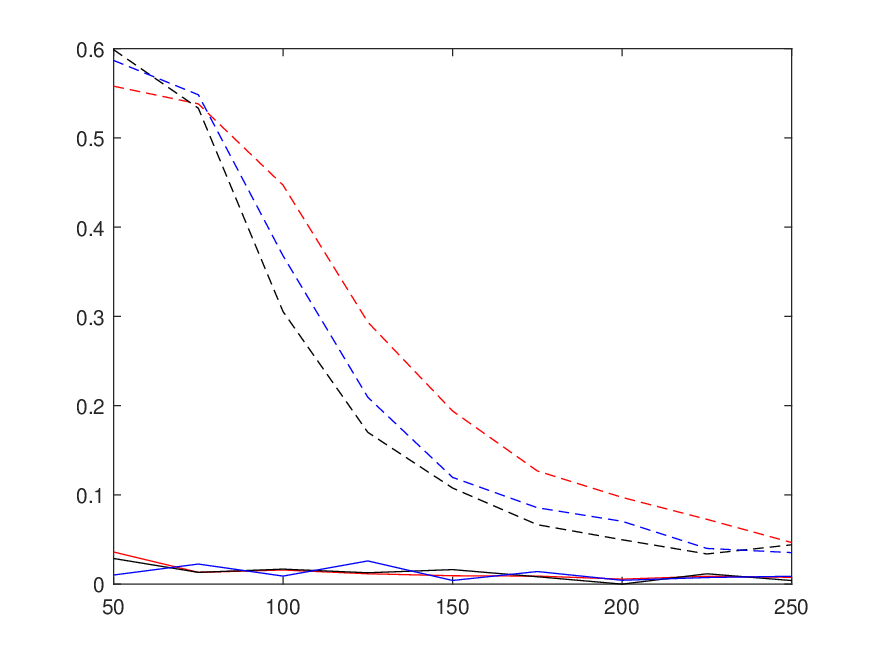}
\includegraphics[width=6.0cm, height=3.8cm]{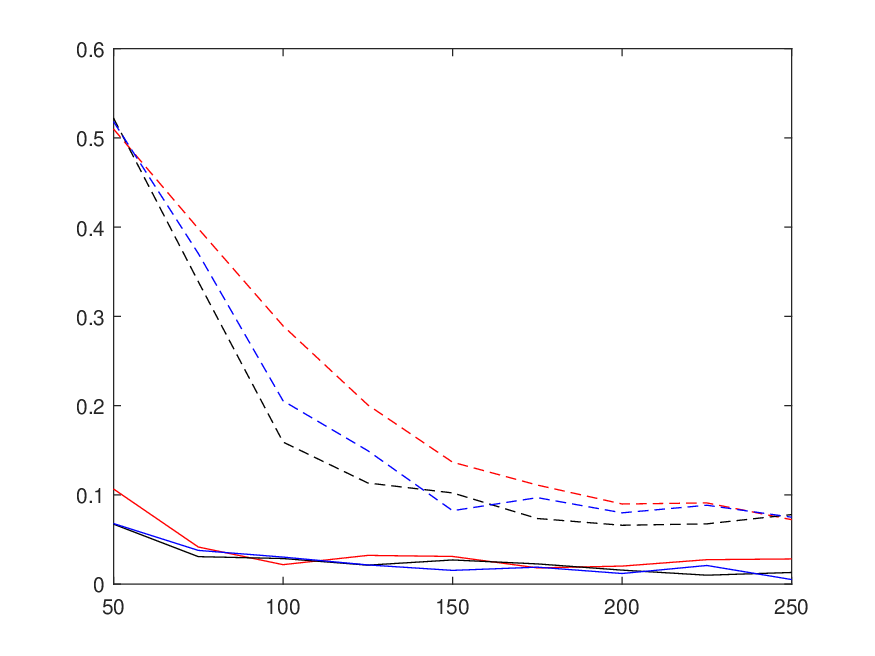}
\includegraphics[width=6.0cm, height=3.8cm]{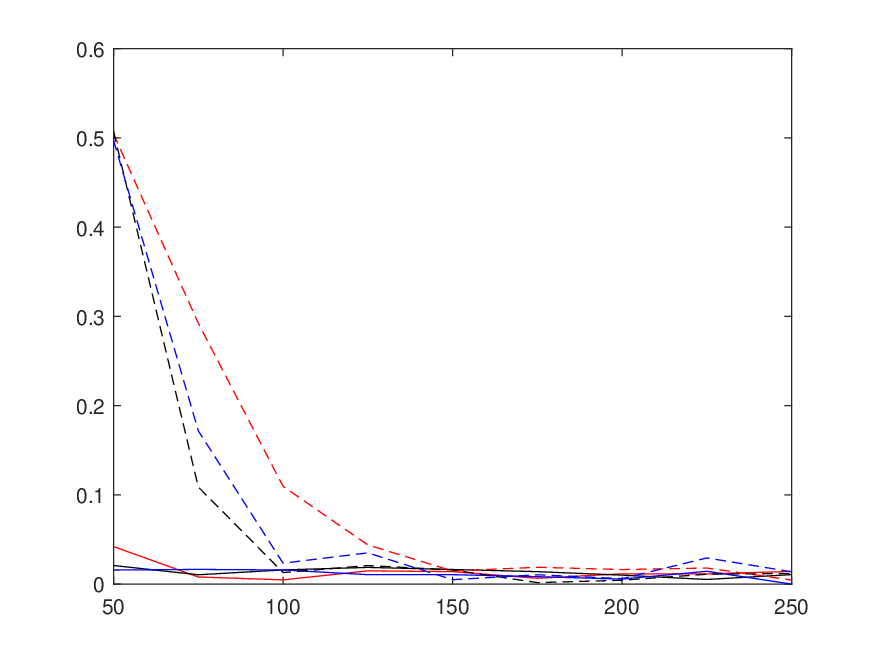} \] \\
\caption{\footnotesize{Comparison of clustering errors when signs are present (solid lines) or removed    (dash lines).  
The between-layer clustering errors $R_{BL}$  in \fr{eq:err_betw_def} of Algorithm~\ref{alg:between}
when matrices $X\upm$  are generated by the truncated Dirichlet distribution with  $M=3$, $K_m =K= 3$, $\varpi_k = \alpha=0.1$, $k \in [K+1]$
(left panels).  The between-layer clustering errors $R_{BL}$  in \fr{eq:err_betw_def} of Algorithm~\ref{alg:between}
(middle panels) and the within-layer clustering errors $R_{WL}$  in \fr{eq:err_within_ave} (right panel)
when layers are equipped with the SBMs where matrices $X\upm$. $\minM$,  are generated by the multinomial 
distribution  with probabilities $(1/K_m, ..., 1/K_m)$,    $M=3$ and $K_m= K = 3$.
The entries of $B\upl$ are generated  as uniform random numbers  between $c$ and $d$ with  $c=-0.05$, $d = 0.05$  
(top panels)  and  $c=-0.2$, $d = 0.2$ (bottom panels).
Errors are averaged over 100 simulation runs,  $L=50$ (red),  $L=100$ (blue),   $L=150$ (black).
}}
\label{fig:signs_compare}
\end{figure*}


We consider two examples of the SGRDPG  networks, one where the  ambient subspaces 
of the layers of the network are generated using
by Dirichlet distribution (Case 2),
and another, where layers are  equipped with the true SBM (Case 3). 
The reason for those choices is that, in both  cases, it is easy 
to trace the impact of removing signs on the probability tensor. 
Specifically, the latter will correspond to removing the 
signs in the matrices of block probabilities  $B\upl$, $\linL$, in \fr{eq:DIMPnew_GDPG}.

Same as before, in our simulations,  we use  Algorithm~\ref{alg:between} for clustering layers $A\upl$   and $|A\upl|$, $\linL$,
into $M$ groups, and determine the precision of the partition using the  between-layer 
clustering error rate  \fr{eq:err_betw_def}. 
In the case of the SBM, subsequently, we estimate the bases of the ambient 
subspaces in each of $M$ groups of layers, using Algorithm~\ref{alg:subspace_est},
and cluster the rows by $(1+\eps)$-approximate $K$-means clustering, obtaining 
community assignments for each group of layers. 
We measure the accuracy of the community assignment in groups of layers by the average  within-layer clustering error
 given by
\begin{align} \label{eq:err_within_ave}
& R_{WL}  =   \frac{1}{M} \ \min_{\aleph(M)}\ \sum_{m=1}^M R_{WL} (m) 
\\ 
& = \frac{1}{2  M n} \ \min_{\aleph(M)} 
\sum_{m=1}^M  \lkv \min_{\scrP_m \in \mathfrak{F}  (K_m)}   \|\hbZ\upm  - \bZ\upm  \scrP_m \|^2_F \rkv.  \nonumber
\end{align}
Simulations results are presented in Figure~\ref{fig:signs_compare}.

It is clear from all three of the figures that the removal of the signs of the edges 
is extremely detrimental to network analysis. 
The effects are much more dramatic when a network is extremely sparse (left panels) and are 
more moderate for denser networks (right panels). 
Indeed, while the between-layer clustering of the layers is very accurate  
for $c=-0.05$, $d = 0.05$ if edges' signs are kept, it fails completely when these signs
are removed. 
Moreover, even in the denser case, when $c=-0.2$, $d = 0.2$, 
the differences in clustering precision  are very dramatic.



\section{Real data example}  
\label{sec:real data}

In this section we present an application of our model to the brain fMRI data set obtained 
from the Autism Brain Imaging Data Exchange (ABIDE I) located at  \\
{\scriptsize {\tt  http://fcon\underline{ }1000.projects.nitrc.org/indi/abide/abide\underline{ }I.html }} \\
Specifically, we use 37  fMRI brain images obtained by California Institute of Technology
and featured in \cite{tyszka_kennedy_paul_adolphs_2014}. The dataset contains resting-state fMRI data 
at the whole-brain level of  18 high-functioning adults with an Autism Spectrum Disorders (ASDs) and 19 neurotypical
controls with no family history of autism, who were recruited under a protocol approved by the Human Subjects Protection
Committee of the California Institute of Technology. The acquisition and initial processing of the data is described in 
\cite{tyszka_kennedy_paul_adolphs_2014}, leading to a data matrix of the fMRI recordings over 116 brain regions 
(see, e.g., \cite{Sun2009MiningBR}) at 145 time instances.

The goal of our study was to create a signed multilayer network of brain connectivity, 
with layers being  individuals, participating in the  study, 
and to partition the layers of the network into groups using our methodology. 
We would like to point out that the objective of this  example is illustration of our technique and of the advantage
of keeping correlations' signs rather than in-depth investigation of brain networks associated with the ASDs.
Since classification of noisy data without model selection is known to be as bad as pure guessing (see \cite{Fan_Fan_AOS2008}),   
we further pre-processed  the fMRI signals, using wavelet denoising with 
the Daubechies-4 wavelets and soft thresholding to reduce noise and, subsequently,    
identified 64 brain regions by removing the regions that are not relates to cognitive functions.
The necessity of removing  irrelevant brain regions was motivated by the fact that vast majority 
of brain regions are not related to cognitive functions and therefore would make partitioning of layers 
of the network impossible (which was, in essence, the conclusion that \cite{tyszka_kennedy_paul_adolphs_2014} drew).
Afterwards, for each individual, we created a signed network on the basis of pairwise correlations. 
Specifically, we  used 40\% of the highest correlations by absolute value, 
and created a signed adjacency tensor with values 
0,1 and -1. For the  sake of comparison, we also created a 
corresponding adjacency tensor with  signs removed.

\begin{figure*}[t]   
\hspace*{-3cm}
\[\includegraphics[width=7.0cm, height=4.3cm]{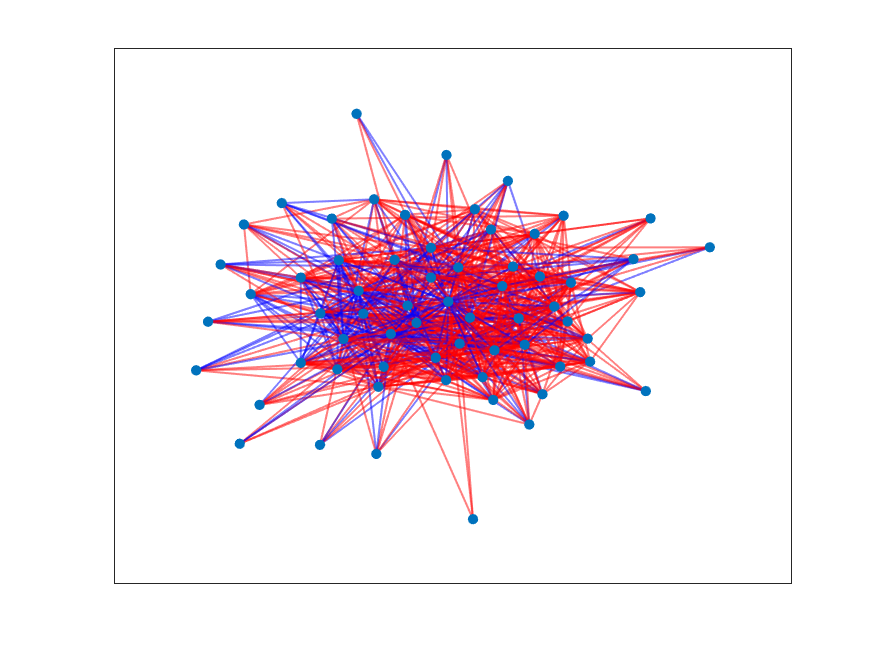}  \hspace{3mm} 
\includegraphics[width=7.0cm, height=4.3cm]{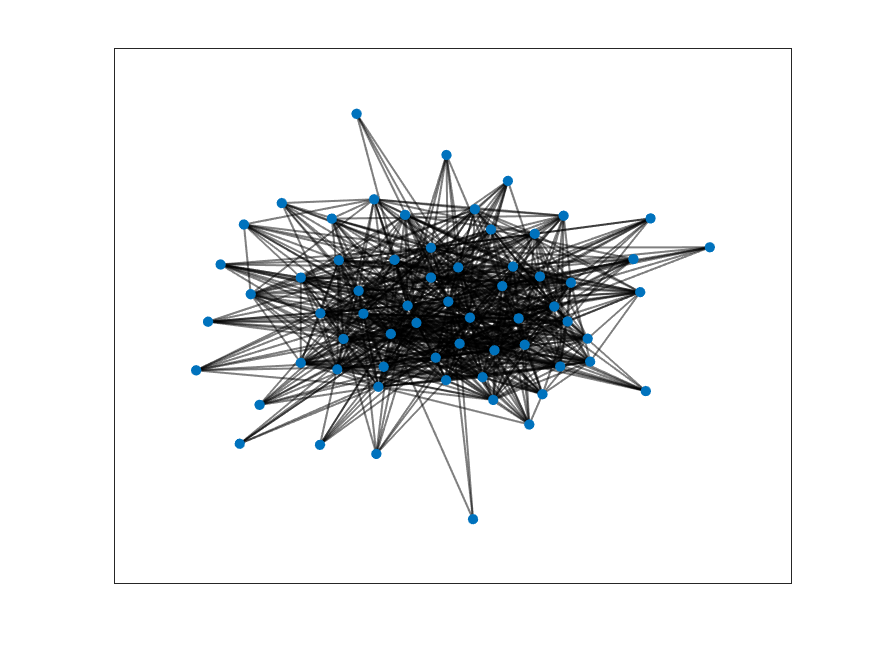}   \]
\caption{{The  graphs of a network for   brain regions of an individual with signs (left) and without signs (right).  
Positive edges are red, negative edges are blue. The networks are constructed using 
the brain fMRI data set obtained by the California Institute of Technology.
Available as a part of  the Autism Brain Imaging Data Exchange (ABIDE I) located at
{\tt  http://fcon\underline{ }1000.projects.nitrc.org/indi/abide/abide\underline{ }I.html }
}}
\label{fig:S_NS_graph}
\end{figure*}

Figure~\ref{fig:S_NS_graph} pictures a brain network of one of the individuals in the study 
with signs (left panel) and sign-free (right panel). 
It is easy to see that negative and positive connections patterns are different, 
which helps to divide layers of the network into two groups. 
Specifically, on the average, networks  have 1638.4   edges 
with 170 edges being negative. If we consider the two groups of layers (individuals with and without ASDs),
the average numbers of edges are almost identical for the two groups (1638.4  for group~1 and 1638.5 for group~2), 
while the average number of negative edges exhibits a minor variation (173.0 for group~1 and 166.7 for group~2). 
Therefore, one cannot use the differences in the network densities for partitioning of layers into two groups,
even if the signs of the edges are taken into account.

 
 \begin{figure*} [t] 
\hspace*{-3cm}
\[\includegraphics[width=7.5cm, height=4.3cm]{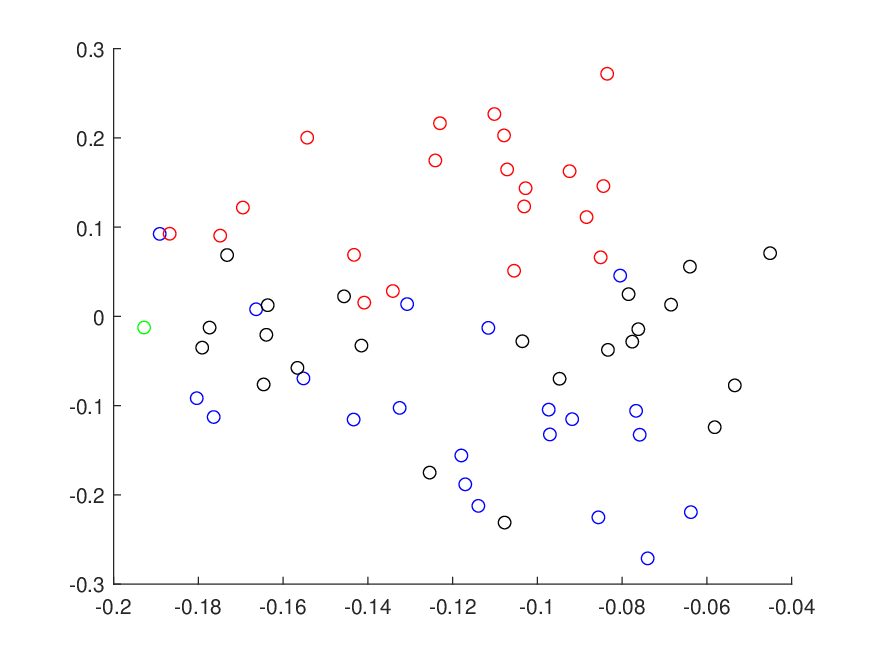}  \hspace{3mm} 
\includegraphics[width=7.5cm, height=4.53cm]{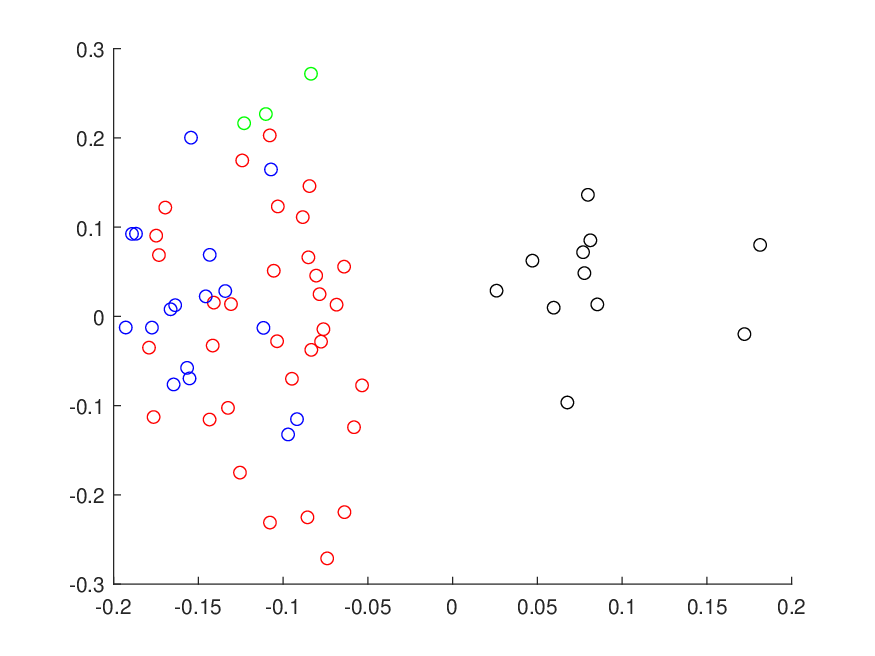}   \]
\[\includegraphics[width=8.0cm, height=4.3cm]{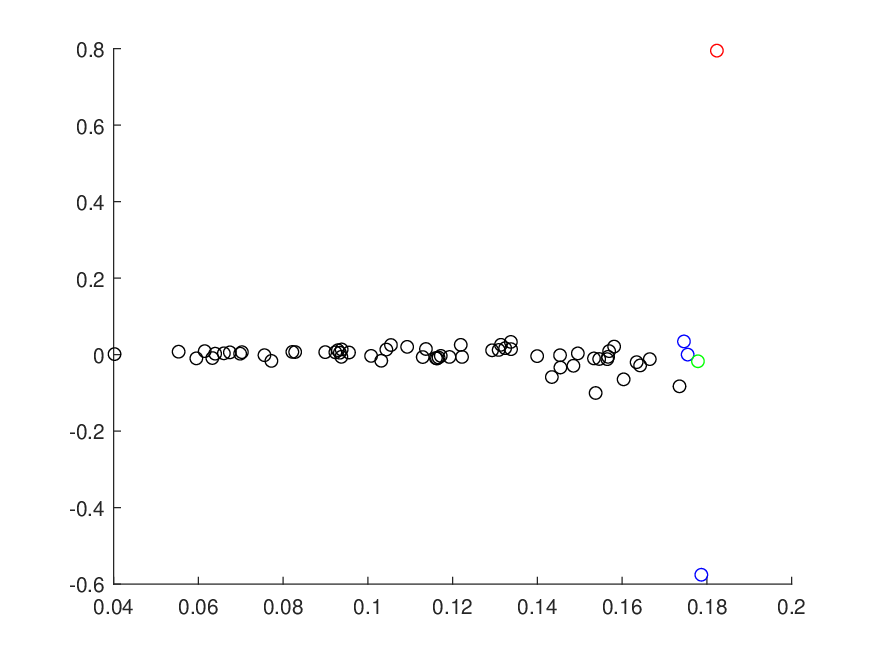}  \hspace{3mm} 
\includegraphics[width=8.0cm, height=4.3cm]{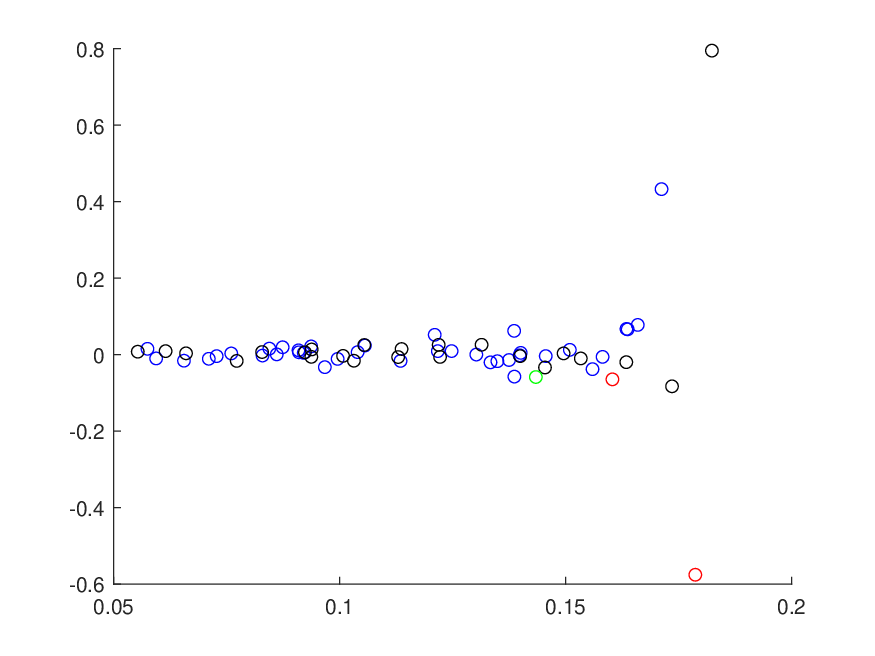}   \]
\caption{{The first two components of the eigenvectors, used for the community assignments in the two groups of
layers of the multilayer network. Communities in the layers are identified by different colors. 
Top row: community assignments in the signed network. Bottom row: 
community assignments in the  network with signs removed. Left column: group of layers 1, right column: 
group of layers 2. 
Networks are constructed using 
the brain fMRI data set obtained by the California Institute of Technology.
Available as a part of  the Autism Brain Imaging Data Exchange (ABIDE I) located at
{\tt  http://fcon\underline{ }1000.projects.nitrc.org/indi/abide/abide\underline{ }I.html }
}}
\label{fig:S_communities}
\end{figure*}


In our analysis, we naturally used $M=2$ (for individuals with and without ASDs),
and chose $K=4$.
We scrambled the layers of the adjacency tensor and applied Algorithm~\ref{alg:between}  
to its signed and signed-free versions. Since the results of the approximate $k$-means clustering  are random,  
we repeated the process over 100 simulations runs. Thus, 
we obtained 18.9\% average clustering error 
over signed multilayer network, and  35.2\%  if we ignore signs (with the standard deviations of
about 0.6\% in the signed network case and 2.7 \% for the case without signs).

Finally, with an additional assumption that layers of the network follow the SBM,  
we partitioned the nodes in each of the two groups of layers into $K=4$ communities.
For this purpose, using Algorithm~\ref{alg:subspace_est}, we constructed estimators
$\hV_{1,S}, \hV_{2,S}$, $\hV_{1,NS}$ and $\hV_{2,NS}$ for the  the ambient subspace bases matrices  
$V_{1,S}, V_{2,S}$, $V_{1,NS}$ and $V_{2,NS}$, for the two groups  of layers in the signed and sign-free cases, respectively.  
We carried out the $k$-means algorithm to partition rows of 
each of $\hV_{1,S}, \hV_{2,S}$, $\hV_{1,NS}$ and $\hV_{2,NS}$ into four communities and obtained 
four sets of community assignments, one for each of the  two groups of layers and for the signed and the sign-free networks. 
Figure~\ref{fig:S_communities} presents components of the first two columns  of $\hV_{1,S}, \hV_{2,S}$, and $\hV_{1,NS}$, $\hV_{2,NS}$,
respectively, sorted by the assigned communities. 
In particular, Figure~\ref{fig:S_communities} confirms that keeping the signs of the edges leads to both more balanced communities and better 
community separation. Communities in Figure~\ref{fig:S_communities} are identified by different colors.






\appendices

\section{Appendix.  Proofs}
\label{sec:proofs}

\subsection{Proofs of Lemmas~\ref{lem:true_Theta} and \ref{lem:est_Theta} }
\label{sec:proof_lem_1_2} 


\noindent
{\bf Proof of Lemma~\ref{lem:true_Theta}.\ }  
If $s(l_1) = m_1$ and $s(l_2)= m_2$, then, by \fr{eq:tilU_rel}, one has 
\begin{align*} 
\Te(l_1, l_2) & = \lkv \vect\! \lkr \tilU^{(m_1)}\! (\tilU^{(m_1)})^T \rkr \rkv^T \! \lkv \vect \! \lkr \tilU^{(m_2)} (\tilU^{(m_2)}\rkr^T) \rkv \\
& = \|\tilU^{(m_1)} (\tilU^{(m_2)})^T \|^2_F . 
\end{align*}
Therefore, if  $s(l_1) = s(l_2) = m$, then,  $\Te(l_1, l_2) = \|\tilU\upm (\tilU\upm)^T \|^2_F = \|I_{K_m}\|^2_F$,
where $I_k$ is the identity matrix of size $k$, and \fr{eq:Theta1} holds. 
If $m_1 \neq m_2$, then, by \fr{eq:tilUX}, derive
\bes 
\Te(l_1, l_2) =   
\frac{\left\|\tilX^{(m_1)}  \lkr  \hSig^{(m_1)} \rkr^{-\frac12} 
\lkv \tilX^{(m_2)} \lkr  \hSig^{(m_2)} \rkr^{-\frac12} \rkv^T \right\|^2_F}{(n-1)^{2}} .
\ees 
Then, by   Lemma~\ref{lem:SighSigrel}, obtain that, if condition \fr{eq:n_lowbound} holds, then, 
for $s(l_1) = m_1$ and $s(l_2) = m_2$, 
 with probability at  least $1 - 2   \, M\, n^{-\tau}$ one has
one has
\bes 
\max_{m_1 \neq m_2}\,  |\Te(l_1, l_2)| \leq 4\, \lowc^{-2}\,  (n-1)^{-2}\, \left\|\tilX^{(m_1)} \lkr \tilX^{(m_2)} \rkr^T \right\|^2_F. 
\ees 
Application of  Lemma~\ref{lem:true_Delta} and the fact that matrices $\tilX\upm$ have ranks at most $K$, 
yield  that, with probability at  least $1 -  C \, M\, n^{-\tau}$,  one has
\bes 
\max_{l_1, l_2}\,  |\Te(l_1, l_2)| \leq C\, K\,  n^{-1}\, \log n
\ees 
which completes the proof of \fr{eq:Theta2}.
\\



\noindent
{\bf Proof of Lemma~\ref{lem:est_Theta} \ }  
Note that, for $l_1 \neq l_2$, one has 
\begin{align*} 
 |(\hte^{(l_1)})^T \hte^{(l_2)} - (\te^{(l_1)})^T \te^{(l_2)}|   
 & \leq |(\hte^{(l_1)})^T [ \hte^{(l_2)} - \te^{(l_2)}] |\\
 & + |[ \hte^{(l_1)} - \te^{(l_1)}]^T \te^{(l_2)} |  
\end{align*}
Since   $\|\te\upl\|^2= K_m$ for $s(l)=m$, obtain
\begin{align*} 
& \max_{l_1, l_2}\, |(\hte^{(l_1)})^T \hte^{(l_2)} - (\te^{(l_1)})^T \te^{(l_2)}| \\
& \leq \max_{\linL} \|\hte\upl - \te\upl\| 
\lkv 2 \sqrt{K_m} + \max_{\linL} \|\hte\upl - \te\upl\| \rkv
\end{align*}
Hence, due to
\begin{align*} 
\|\hte\upl - \te\upl\|  & = \left\| \hU_A\upl (\hU_A\upl)^T -  \tilU_P\upl (\tilU_P\upl)^T  \right\|_F \\
& = \sqrt{2}\, \left\| \sinTe (\hU_A\upl, \tilU_P\upl) \right\|_F
\end{align*}
and Lemma~\ref{lem:U_hU}, derive $|(\hte^{(l_1)})^T \hte^{(l_2)} - (\te^{(l_1)})^T \te^{(l_2)}|$ \\ 
$\leq C\,  \sqrt{K}\, (n \rhon)^{-1/2}
\lkv  2 \sqrt{K} + C  \sqrt{K}\, (n \rhon)^{-1/2} \rkv$, so that
\begin{align}  \nonumber
  \PP  \lfi
 \max_{l_1, l_2}  |(\hte^{(l_1)})^T \hte^{(l_2)}\! -\! (\te^{(l_1)})^T \te^{(l_2)}|  \!  \leq  \! C  K  (n \rhon)^{-\frac{1}{2}}\!
 \rfi & \\
    \ge 1- 2 n^{-(\tau - \tau_0)} \hspace{1.0cm} & \label{eq:difference}  
\end{align} 
Now, observe that, if $s(l_1) = s(l_2) =m$, then, due to $K_m \geq K/C_K$, $\Te(l_1,l_2)= K_m$ and
$\hTe(l_1,l_2) \geq \Te(l_1,l_2) -  |(\hte^{(l_1)})^T \hte^{(l_2)} - (\te^{(l_1)})^T \te^{(l_2)}|$,
one obtains \fr{eq:hTheta1}  with probability at least $1- 2 n^{-(\tau - \tau_0)}$.
  
In order to derive \fr{eq:hTheta2}, note that 
$\hTe(l_1,l_2) \leq |\Te(l_1,l_2)| + |(\hte^{(l_1)})^T \hte^{(l_2)} - (\te^{(l_1)})^T \te^{(l_2)}|$.
Then \fr{eq:hTheta2} follows from the combination of  \fr{eq:Theta2} and \fr{eq:difference}
and the fact that $n^{-1}\, \log n = o( (n \rhon)^{-1/2})$ for large $n$.



\subsection{Proofs of Theorems 1, 2 and 3 }
\label{sec:proof_theorems}


\noindent
{\bf Proof of Theorem~\ref{thm:clust_between}.\ }  
Consider matrices $\calF \in \RR^{n^2 \times M}$ and $\Ups \in \RR^{n^2 \times L}$  with rows 
$\calF(m,:) = \vect \lkr U\upm (U\upm)^T \rkr$, $\minM$, and $\Ups(l,:) = \vect \lkr U\upl (U\upl)^T \rkr$, $\linL$,
respectively.   If $S$ is the true clustering matrix and $W = S D_S^{-1/2}$, where $D_S = S^T S$, then 
\bes
\Ups = S\, \calF = W   D_S^{1/2}, \quad   \Te = \Ups \, \Ups^T  =  W \, D_S^{1/2} \calF  \calF^T\, D_S^{1/2} \, W^T.
\ees
Since Lemma~\ref{lem:true_Theta} implies that 
\begin{align*}
\lam_{\min} (\calF\,  \calF^T) & \geq \min_m K_m - C K M \, (n \rhon)^{-1/2} \\
 & \geq C \, K (1 - M \, (n \rhon)^{-1/2}),
\end{align*}
obtain that, under assumption \fr{eq:extra_cond}, one has 
\be \label{eq:lammin_Te}
(dUp^2 = \sig_{\min}^2 (\Ups) = \lam_{\min} (\Te) \geq \ C\, K\, L \,M^{-1}.
\ee 
Consider matrix $\hUps \in \RR^{n^2 \times L}$  with rows  
\bes
\hUps(l,:) = \vect \lkr \hU\upl (\hU\upl)^T \rkr, \ \linL.
\ees
Denote  $\Xi = \hUps - \Ups$ and observe that, by the Davis-Kahan theorem, with probability at  least 
$1 - n^{-\tau}$, one has
\bes
\|\hUps - \Ups\|^2\tinf \asymp \max_{\linL}\, \|\sinTe (\hU\upl, U\upl)\|^2_F \leq \frac{\Ctau\, K^2 \log L}{n\, \rhon}.
\ees 
Hence, due to condition \fr{eq:nLtau}, with probability at  least $1 - n^{-\tau}$, obtain
\begin{align} 
\|\Xi\|\tinf & \leq   \Ctauo\, K \, \sqrt{\log n}\, (n\, \rhon)^{-1/2},\nonumber \\
&  \label{eq:cl_bet_Xi_bounds}\\
\|\Xi\|_F & \leq  \Ctauo\, K \, \sqrt{L\, \log n}\, (n\, \rhon)^{-1/2} \nonumber
\end{align}

Apply the second part of Proposition~1 of \cite{pensky2024daviskahan} with $U$ being replaced by $W$, $\hU$ by $\hW$, $n$ by $L$,
$r$ by $M$, $X$ by $\Ups$, $\hX$ by $\hUps$ and $d_r$ by $\dUp$. Consider the case of no hollowing, i.e., $\tilh=0$, and define,
similarly to \cite{pensky2024daviskahan},  
\bes
\scrE_1 = \Xi\, \Xi^T, \ \ 
\scrE_2 = \Xi\, \Ups^T, \ \ 
\scrE_3 = \Ups\, \Xi^T, \ \ 
\scrE  = \scrE_1 + \scrE_2 + \scrE_3.
\ees
To apply Proposition~1 of \cite{pensky2024daviskahan}, we need to check that, with probability at  least $1 - n^{-\tau}$, one has 
$\sqrt{M}\, \dUp^{-2}\, \|\scrE\| = o(1)$ as $n \to \infty$ and that conditions in formula (5.13) of Proposition~1 in 
\cite{pensky2024daviskahan} are valid. 
By observing that $\|\Ups\| \asymp \dUp$, derive that 
\bes 
\dUp^{-2} \! \|\scrE\| \leq \dUp^{-2}  \lkr \|\Xi\|^2  \! +  \! 2\, \|\Xi\| \|\Ups \| \rkr 
\leq C  \dUp^{-1}   \! \|\Xi\|  \! (\lkr 1  \! +  \! \dUp^{-1}\,  \|\Xi\| \rkr 
\ees
Therefore, since $\|\Xi\|  \leq \| \Xi\|_F$, obtain 
\be \label{eq:tepsEo}
\PP \lfi 
\frac{\|\Xi\| }{\dUp} \leq \frac{\Ctau\, \sqrt{K\, M\, \log n}}{\sqrt{n\, \rhon}}
\rfi \geq 1 - n^{-\tau},
\ee
so that \fr{eq:extra_cond} guarantees that $\sqrt{M}\, \dUp^{-2}\, \|\scrE\| = o(1)$ as $n \to \infty$.

Now, we need to check that conditions in formula (5.13) of Proposition~1 in \cite{pensky2024daviskahan}
hold. In our notations, the latter is true if, with probability at  least $1 - n^{-\tau}$, as $n \to \infty$, one has
\begin{align}  
& L^{\frac12} M^{-\frac12} \, \dUp^{-2}\, \|\Xi\, \Xi^T\, W\|\tinf \to 0, \nonumber\\ 
& L^{\frac12} M^{-\frac12} \, \dUp^{-2}\,\|\Xi\, \Ups^T\, W\|\tinf  \to 0, \label{eq:cond_in_513}\\
& L^{\frac12} M^{-\frac12} \, \dUp^{-2}\, \|\scrE_1 + \scrE_2\|\tinf   \to 0. \nonumber
\end{align}
We shall show that assumption \fr{eq:extra_cond} guarantees all of the statements in \fr{eq:cond_in_513}.
It is easy to see that, by \fr{eq:lammin_Te} and \fr{eq:cl_bet_Xi_bounds}, one has
\bes
\frac{\|\Xi\, \Xi^T\, W\|\tinf}{\dUp^2}  \leq \frac{\|\Xi\|\tinf \, \|\Xi\|}{\dUp^2} 
\leq \Ctau\, \frac{\sqrt{M}}{\sqrt{L}}\, \frac{K\, \sqrt{M} \, \log n}{n\, \rhon},
\ees
so that the first condition in \fr{eq:cond_in_513} is ensured by \fr{eq:extra_cond}.
Similarly, 
\bes
\frac{\|\Xi\, \Ups^T\, W\|\tinf}{\dUp^2}  \leq \frac{C\, \|\Xi\|\tinf}{\dUp}
\leq \Ctau\, \frac{\sqrt{M}}{\sqrt{L}}\, \frac{\sqrt{K\, \log n}}{\sqrt{n \, \rhon}},
\ees
so that the second condition in \fr{eq:cond_in_513} is also ensured by \fr{eq:extra_cond}.
Finally, note that 
\begin{align*}
\dUp^{-2}\,\|\scrE_1 + \scrE_2\|\tinf & \leq \dUp^{-1}\, \|\Xi\|\tinf 
\lkr 1 + \dUp^{-1}\, \|\Xi\| \rkr \\
& \leq 
\Ctau\, \frac{\sqrt{M}}{\sqrt{L}}\, \frac{K\, \log n\, \sqrt{M}}{n\, \rhon}.
\end{align*}
Hence, the last  condition in \fr{eq:cond_in_513} is true, due to \fr{eq:extra_cond},
so that the validity of Theorem~\ref{thm:clust_between} follows from 
Proposition~1 in \cite{pensky2024daviskahan}.
\\



\noindent
{\bf Proof of Theorem~\ref{th:error_Um_est} \ }  
Fix $\minM$ and consider $H\upm$ and $\hH\upm$ defined in \fr{eq:Hm_def} and \fr{eq:hGhH}, respectively. 
%
Application of the  Davis-Kahan theorem yields
\be \label{eq:davis-kahan}
\left\| \sin\Te \lkr \hU\upm,  U\upm \rkr \right\|  \leq 
\frac{C\,  \|\hH\upm - H\upm\|}{\sig_{\min} (H\upm)}
\ee   
Therefore, in order to assess $R_{U,max}$ and $R_{U,ave}$, one needs to examine the spectral structure of matrices  $H\upm$ 
and their deviation from the sample-based versions $\hH \upm$.
We start with the first task.

It follows from \fr{eq:DIMPnew_GDPG} and Assumption~{\bf A5} that 
\be \label{eq:P2upl}
(P\upl)^2 = \rhon^2\, X\upm\, B_0\upl \, (X\upm)^T \,  X\upm\, B_0\upl \, (X\upm)^T  
\ee  
Recall that, by \fr{eq:tX},  $X\upm = \tilX\upm + 1_n \barX\upm$, where $\tilX\upm = \Pi^{\bot} X\upm$.
Since $(\tilX\upm)^T 1_n \barX\upm= 0$, by Lemma~2 of \cite{10.1214/17-AOS1541}, obtain that
$\sig^2_{K_m} (X\upm) \geq \sig^2_{K_m} (\tilX\upm)$. Furthermore, using \fr{eq:hSig_def} and Lemma~\ref{lem:SighSigrel}, derive
\bes
\sig^2_{K_m} (\tilX\upm) = \lam_{\min} ((\tilX\upm)^T \tilX\upm) \geq \lowc n/4 
\ees 
The latter, together with \fr{eq:P2upl} and  Assumption~{\bf A5}, yields that
$$
\sig_{K_m} ([P\upl]^2) \geq \rhon^2\, \sig^4_{K_m} (\tilX\upm) \, \sig^2_{\min} (B_0\upl) \geq C\, \rhon^2\, n^2.
$$
Since matrices $[P\upl]^2$ are all positive definite, one has $\sig_{K_m} (H\upm) \geq C\,  L_m\, \rhon^2\, n^2$.
Applying Lemma~\ref{lem:balanced_groups} with  
\be  \label{eq:new_t_cond}
 1+ \log (n^{\tau})/ \log L  \leq t \leq (\lowc_{\pi})^2\, L/(2 M^2 \log L) -1,
\ee  
and the union bound, derive from \fr{eq:hLm} that, with probability at least $1 - 2  n^{-(\tau - \tau_0)}$ 
\be \label{eq:sigmin_Hm} 
  \min_m  \sig_{_{K_m}} (H\upm) \geq C  L  M^{-1}  \rhon^2  n^2.  
\ee

Now, we examine deviations $\|\hH\upm - H\upm\|^2$ in \fr{eq:davis-kahan}.
Denote 
\bes 
\tilH\upm = \sum_{s(l) = m} \hG\upl, \quad 
\tilH\upm - H\upm =  \sum_{s(l)=m} [\hG\upl - G\upl]. 
\ees
Observe that it follows from Theorem~4 of \cite{Lei_2023_JASA} and Theorem~\ref{thm:clust_between} that
\begin{align}  \nonumber 
& \PP  \lfi \|\tilH\upm - H\upm\|^2 \leq C\, L_m\,  \lkr \rhon^3 \, n^3\, \log(n + L_m) + \rhon^2\, n^2   \rkr \rfi \\
& \geq 1 - C \, n^{-(\tau - \tau_0)} - C (n + L_m)^{-(\tau -1)}.  \label{eq:noroozi}
\end{align}
Now, recall that, by Theorem~\ref{thm:clust_between}, with probability at least $1  - C  \, n^{-(\tau - \tau_0)}$
one has $\hs = s$, so that  $\tilH\upm = \hH\upm$, $\minM$. 
Hence, applying the union bound over $\minM$   and combining 
\fr{eq:davis-kahan}, \fr{eq:sigmin_Hm},  Lemma~\ref{lem:balanced_groups} in Section~\ref{sec:proof_suppl} and 
\fr{eq:noroozi}, we arrive at 
\be  \label{eq:errU1} 
\max_{\minM}\, \left\| \sin\Te \lkr \hU\upm,  U\upm \rkr \right\|  \leq 
C  \lkr \frac{\sqrt{M  \log n}}{\sqrt{\rhon  n   L}} + \frac{1}{n} \rkr
\ee 
with probability at least $1 - C \, n^{-(\tau - \tau_0)} - C\, M\, n^{-(\tau -1)}$, due to 
$\log(L_m + n) \propto \log(L+n) \propto \log n$.  
Now, in order the upper bound in \fr{eq:errU1} holds, one need the condition on $t$ in \fr{eq:new_t_cond} to be plausible,  
which is guaranteed by \fr{eq:L_lowbound} and Assumption~{\bf A4}.
The inequality \fr{eq:errU1}  implies that in \fr{eq:err_subspace_max}
\bes
\PP\, \lfi 
R_{U,max}  \leq 
\Ctau\,   \lkr \frac{\sqrt{M\, \log n} }{\sqrt{\rhon\, n \, L}}   + \frac{1}{n}  \rkr
\rfi   \geq 1 - \frac{C}{n^{\tau - \tau_0 - 1}},
\ees
and that the first inequality in \fr{eq:error_Save_max} holds. The validity of the second inequality 
follows from the fact that $R_{U,ave} \leq \sqrt{K}\, R_{U,max}$.
\\


\noindent
{\bf Proof of Theorem~\ref{th:error_2inf} \ }
Note that validity of the Theorem follows immediately from Corollary~\ref{cor:error_2inf}. 
In order to prove \fr{eq:error_2inf}, use Corollary~\ref{cor:error_2inf} with $L = L_m$ for each of the groups of layers, 
and recall that $L_m \geq C L/M$ by Lemma~\ref{lem:balanced_groups}.
The proof is completed by application of the union bound.


\subsection{Proofs of Corollaries~\ref{cor:error_2inf} and \ref{cor:clust_between} }
\label{sec:proof_coroll}


\noindent
{\bf Proof of Corollary~\ref{cor:error_2inf}  \ }
Let  assumptions of Theorem~\ref{th:error_Um_est} hold with $M=1$.
 the setting of the paper reduces to the COSIE model with $K_1 = K$. 
Denote  
\be \label{eq:X_hX}
X = \sumlL  ( P\upl)^2, \quad
\hX = \sumlL  \lkv (A\upl)^2 - \hD\upl \rkv
\ee
where $\hD\upl$  is defined in \fr{eq:hDl}. Then, the SVDs of $X$ and $\hX$ can be written, respectively, as 
\be  \label{eq:X_hX_SVD}
X = U \Psi U^T, \quad 
\hX = \hU \hPsi \hU^T + \hU_{\perp} \hPsi_{\perp} \hU^T_{\perp}
\ee
where $U, \hU \in \calO_{n,K}$, $\hU_{\perp}  \in \calO_{n,n-K}$ and 
$\hU^T \hU_{\perp} =0$.  Denote $\Xi\upl = A\upl - P\upl$.
Note that  $\diag(\hX)=0$, due to $(A\upl(i,j))^2 = |A\upl(i,j)|$.
For any square matrix $Y$, denote $\scrH (Y) = Y - \diag(Y)$, the so called hollowed matrix $Y$.
With this notation, observe that $\hX - X$ can be rewritten as 
\be \label{eq:S_first}
\Del =  \hX - X = \scrH (\hX) - X = \scrH(\hX - X) - \diag(X).
\ee
Hence, we can decompose the error as  $\Del =  S_0 + S_1 + S_2$,  where, 
\begin{align}
S_0 & = \sumlL \scrH \lkr P\upl \Xi\upl + \Xi\upl P\upl \rkr, \label{eq:S0} \\
S_1 & = \sumlL  \scrH  \lkr \Xi\upl\rkr^2, \label{eq:S1} \\
S_2 & = - \sumlL   - \diag(|X\upl|), \label{eq:S2} 
\end{align}
Since $\Del$ and $X$ are the same as in Theorem~\ref{th:error_Um_est}, 
with probability at least $1 - C\, n^{-(\tau - (\tau_0 \vee 1))}$ 
\be \label{eq:Del_opernorm}
\| \Del\| \leq C \lkv  (\rhon\, n)^{3/2} \, \sqrt{L\, \log n}  + n\, \rhon^2\, L \rkv.
\ee
Moreover, it is known that, for any square matrix $Y$  
\be \label{eq:scrH} 
\|\scrH(Y)\| \leq 2 \|Y\|.
\ee

In order to simplify the narrative, in what follows, we use $C$ as a generic constant
that depends only on the constants in our assumptions, $\Ctau$  as a generic constant
that depends only on the constants in our assumptions and $\tau$, and introduce a 
set $\Omtaun$,  on which all inequalities take place, and such that 
\be \label{eq:Omtaun}
\PP(\Omtaun) \geq   1 - C\, n^{-(\tau - (\tau_0 + 1))}.  
\ee
For $\om \in \Omtaun$, due to Assumptions~{\bf A2}, {\bf A3} and {\bf A5}, derive that 
\be  \label{eq:lammin_Psi}
\lam_{\min}(X) = \min_{i \in [K]} \lkr \Psi_{i,i} \rkr \geq C \rhon^2 n^2 L.
\ee
Therefore, it follows from \fr{eq:rhon_cond} and \fr{eq:Del_opernorm}   that
$\lam_{\min} \lkr \hPsi \rkr \geq C \rhon^2 n^2 L$ 
with probability at least $1 - C\, n^{-(\tau - (\tau_0\vee 1))}$ .
As a result, for $\om \in \Omtaun$ obtain 
\be \label{eq:hPsi}
\|\hPsi^{-1}\| \leq  C (\rhon^2 n^2 L)^{-1}.
\ee
In addition, observe that by \fr{eq:tilUX} and Lemma~\ref{lem:SighSigrel} with $M=1$ 
\be \label{eq:U2_inf}
\|U\|_{2,\infty} \leq C \, (K/n)^{1/2}.
\ee
Finally, combination of \fr{eq:Del_opernorm}, \fr{eq:lammin_Psi} and Davis-Kahan theorem yields that,
for $\om \in \Omtaun$, 
\be \label{eq:sinTeU_hU}
\|\sinTe (\hU,U)\|  \leq C \,  
\lkv \frac{\sqrt{\log n}}{\sqrt{\rhon n L}} + \frac{1}{n} \rkv.
\ee

Let $U^T\hU = U_H \Lam_H V_H^T$ be the SVD of $U^T\hU$. 
Denote  $W_U =  U_H  V_H^T \in \calO_K$.    
Instead of bounding above $R_{U,2,\infty}$, following \cite{Cape_AOS2019},
we bound $\|\hU - U W_U \|\tinf$. For this purpose,  we use Corollary 3.3 of \cite{Cape_AOS2019} which states that
\begin{align}  
\hU - U W_U & = (I - U U^T) \Del U W_U \hPsi^{-1} \nonumber\\
& + (I - U U^T) \Del (\hU - U W_U) \hPsi^{-1} \label{eq:cape_expansion} \\
& + (I - U U^T) X (\hU - U U^T \hU) \hPsi^{-1} \nonumber\\
& + U (U^T \hU - W_U). \nonumber
\end{align}
Due to \fr{eq:X_hX_SVD}, the third term in this expansion is equal to zero.
Moreover, it is easy to see that 
\bes
\sumlL (I - U U^T) P\upl \Xi\upl =0,   
\ees
so that one can replace $S_0$ in \fr{eq:S0} by 
\be \label{eq:tS0}
\tilS_0 =   \sumlL \lkv     \Xi\upl P\upl   - 2\, \diag(\Xi\upl P\upl) \rkv.  
\ee
After that, replace $\Del$ by $S = \tilS_0 + S_1 + S_2$, and obtain from \fr{eq:cape_expansion}
that 
\be  \label{eq:R}
R = \|\hU - U W_U\|\tinf \leq R_1 + R_2 + R_3  + R_4 + R_5
\ee
with
\begin{align*} 
& R_1 = \|U\|_{2,\infty}\, \|S U \| \, \|\hPsi^{-1}\|, \\
& R_2 =  \|S U\|\tinf \|\hPsi^{-1}\|, \\
& R_3 = \|U\|_{2,\infty}\, \|S \| \, \|\hU - U W_U\|\, \|\hPsi^{-1}\|, \\
& R_4 =   \|S \|\tinf \, \|\hU - U W_U\|\, \|\hPsi^{-1}\|,  \\
& R_5 = \| U \|\tinf \|(U^T \hU - W_U)\|. 
\end{align*}
It follows from  \fr{eq:W_U} of Lemma~\ref{lem:op_2infsuppl},  
and \cite{Cape_AOS2019} that
\begin{align*} 
& \|\hU - U W_U\| \leq \sqrt{2} \, \|\sinTe (\hU,U)\|,\\ 
& \|(U^T \hU - W_U)\| \leq \|\sinTe (\hU,U)\|^2.
\end{align*}
Note that 
\begin{align*}
\|\tilS_0 \| & \leq \left\|   \sumlL      \Xi\upl P\upl \right\| +  
2\, \left\| \diag\lkr \sumlL\, \Xi\upl P\upl  \rkv \right\|\\
&  \leq 3\, \left\|   \sumlL      \Xi\upl P\upl \right\|.
\end{align*}
Then, by the same argument that was used to obtain the upper bound for $\|S_0\|$,
obtain that, for $\om \in \Omtaun$, 
\bes 
\|\tilS_0 \| \leq \Ctau\, (\rhon^3\, n\, L\, \log L)^{1/2}.
\ees
Since $\|S_1\|$ and $\|S\|_2$ have the same upper bounds as before (up to a constant), 
derive  that
\be \label{eq:S_opernorm}
\| S \| \leq C \lkv  (\rhon\, n)^{3/2} \, \sqrt{L\, \log n}  + n\, \rhon^2\, L \rkv.
\ee
Denote 
\be \label{eq:epsrn}
\epsrn = (\rhon\, n\, L)^{-1/2}\, (\log n)^{1/2} + n^{-1} 
\ee
and observe that 
\begin{align} 
& \|S \| \,   \|\hPsi^{-1}\| \leq \Ctau\, \epsrn, \quad
\|\hU - U W_U\| \leq \Ctau\, \epsrn, \nonumber \\ 
& \|(U^T \hU - W_U)\| \leq  \Ctau\, \epsrn.  \label{eq:bound2}
\end{align} 
Then, we need to obtain upper bounds for $\|S U \|$, $\|S U \|\tinf$ and $\|S  \|\tinf$.
Due to $S = \tilS_0 + S_1 + S_2$, we develop those bounds for $\tilS_0$, $S_1$ and $S_2$,
separately.
\\


\underline{Bounds for the norms of $\tilS_0$.}\ 
In order to derive upper bounds for $\|\tilS_0 U \|\tinf$ and $\|\tilS_0   \|\tinf$,
denote $\xi^{(i,l)}:= \Xi\upl(:,i)$, $i \in [n]$, and observe that, due to symmetry,
\bes
\| \tilS_0  U  \|\tinf \leq  \| \tilS_0  U  \|\tinf  = \max_{i \in [n]}\, \left\| \sumlL P\upl \xi^{(i,l)} \right\|.
\ees
Here, for a fixed value of $i$, components of vectors $\xi^{(i,l)}$ are independent Bernoulli errors, 
and application of Theorem~3 of  \cite{Lei_2023_JASA} with $j=0,1$,  in a view of 
$\displaystyle \sumlL \| P\upl\|^2_F \leq C \rhon^2 n^2 L$ and 
$\displaystyle  \max_l \|P\upl\|\tinf \leq C \rhon \sqrt{n}$, 
yields
\begin{align*} 
\PP \lfi \| \sumlL P\upl \xi^{(i,l)} \| \geq t \rfi \leq 
& 2(n+1) \lkv   \exp \lkr - \frac{c t^2}{L n^2 \rhon^3} \rkr \right.\\
& \left. + \exp \lkr -\frac{Ct}{\rhon \sqrt{n}} \rkr \rkv.
\end{align*}
Set $t = \Ctau \, \sqrt{\rhon^3 n^2 L \log n}$. Then, taking a union bound over $i \in [n]$ 
and choosing an appropriate constant $\Ctau$, which depends on $\tau$, derive that,
for $\om \in \Omtaun$,
\be \label{eq:tS0tinf}
\| \tilS_0\, U\|\tinf \leq  \| \tilS_0\|\tinf \leq C \, \sqrt{\rhon^3 n^2 L \log n}.
\ee


\underline{Bounds for the norms of $S_1$.}\ 
Let $\xi^{(i,l)}:= \Xi\upl(:,i)$, as before. Consider matrices $\Xiii\upl \in \RR^{n \times n}$,
where the $i$-th row and the  $i$-th column are replaced by zeros. It is easy to see that for any $i$ and $l$,
vectors $\xiil$ and matrices $\Xiii\upl$ are independent. Observe also that 
\bes
S_1(i,:) = \sumlL\, \lkv (\xiil)^T\, \Xiii\upl +  \Xi\upl(i,i) (\xiil)^T  \rkv.
\ees
Since $A\upl(:,:)=0$, one has $\Xi\upl(i,i) = - P\upl(i,i)$, so that, for $j=0,1$, one has 
\bes 
S_1(i,:) U^j = \sumlL\, \lkv (\xiil)^T\, \Xiii\upl \, U^j -  P\upl(i,i) (\xiil)^T U^j \rkv.
\ees
Note that, for  $\om \in \Omtaun$ and $j=0,1$, one has 
\be \label{eq:extra_term}
\max_{i \in [n]} \left\|\sumlL\, P\upl(i,i) (\xiil)^T U^j \right\| \leq \Ctau\, \sqrt{\rhon^3\, n \, L\, \log n}.
\ee
Therefore, due to symmetry, obtain, for  $\om \in \Omtaun$ and $j=0,1$,
\begin{align*} 
\|S_1\, U^j\|\tinf & \leq \maxin\, \lnor \sumlL (\xiil)^T\, \Xiii\upl \, U^j \rnor \\
&  + 
\Ctau\, \sqrt{\rhon^3\, n \, L\, \log n}.
\end{align*} 
In order to obtain an upper bound for the first term in the last inequality,   
apply Theorem~3 of  \cite{Lei_2023_JASA} with $j=0,1$, conditional on $\Xiii\upl$, $\linL$. Obtain
\begin{align} \label{eq:probab_cor}
& \PP \lfi \lnor \sumlL (\xiil)^T\, \Xiii\upl \, U^j \rnor \geq t \Big| \Xiii\upl, \, \linL \rfi \\
& \leq 2 \, (n+1) \, \lfi  \exp\! \lkr - \frac{t^2}{8   \rhon\, \Del_1(j)} \rkr + \exp\! \lkr - \frac{t}{4   \Del_2(j)} \rkr \rfi,
\nonumber 
\end{align} 
where 
\bes
\Del_1 (j) = \sumlL \|\Xiii\upl\, U^j \|^2_F, \  
\Del_2 (j) = \max_{\linL}\, \|\Xiii\upl\, U^j \|\tinf.
\ees
The upper bounds for the above quantities are provided by  Lemma~\ref{lem:Del12_bounds},
formally stated and proved in the next section. Specifically, Lemma~\ref{lem:Del12_bounds}
implies that, with probability at least  $1 - C \, n^{-(\tau -  (\tau_0 \vee 1))}$, 
\begin{align*}
& \max (\Del_1 (0), \Del_1 (1)) \leq \Ctau\, \sqrt{n  \rhon \log n}, \\
& \Del_2 (0) \leq \Ctau \, \rhon  n^2  L, \quad 
\Del_2 (1) \leq \Ctau \, \rhon n  K   L.
\end{align*}
Plug  those quantities into \fr{eq:probab_cor} and set $t=t_0$ and $t = t_1$ for $j=0$ and $j=1$, respectively, where
\bes
t_0 = \Ctau\,  \sqrt{\rhon^2\, n^2 \, L}, \quad 
t_1 = \Ctau\,  \sqrt{\rhon^2\,  n\, K\,  \, L}. 
\ees
Obtain that, for  $\om \in \Omtaun$,
\be \label{eq:S1tinf}
\|S_1\|\tinf  \leq \Ctau \sqrt{\rhon^2  n^2  L}, \ 
\|S_1  U \|\tinf  \leq \Ctau   \sqrt{\rhon^2   n  K L}.
\ee


\underline{Bounds for the norms of $S_2$.}\   
Since $S_2$ is the diagonal matrix, obtain that 
$\|S_2\|\tinf = \|S_2\|_{1, \infty} = \|S_2\|$, and it follows from 
Proposition~6.5 of  \cite{Cape_AOS2019} that 
\be \label{eq:S2tinf}
\|S_2\|\tinf  \leq \Ctau\,  \rhon^2\, n \, L, \quad 
\|S_2  U \|\tinf  \leq \Ctau\,  \rhon^2\, \sqrt{n\, K}\, L.
\ee 


\underline{Overall upper bounds.}\ 
Now, from  \fr{eq:tS0tinf}, \fr{eq:S1tinf} and  \fr{eq:S2tinf},  for    $\om \in \Omtaun$, obtain that 
\begin{align*}
\|S \|\tinf  & \leq \Ctau\, \lkv \sqrt{\rhon^2\, n^2 \, L} + \rhon^2\, n \, L \rkv,\\
\|S\, U \|\tinf & \leq \Ctau\,  \lkv \sqrt{\rhon^3 n^2 L \log n} + \sqrt{\rhon^2\,  n\, K\,  \, L} +
 \rhon^2\, \sqrt{n\, K}\, L \rkv,
\end{align*}
Therefore, \fr{eq:hPsi} yields 
\begin{align} 
\|S \|\tinf \, \,   \|\hPsi^{-1}\| & \leq \Ctau\, \lkv \frac{\sqrt{\log n}}{\rhon\, n\, \sqrt{L}} + \frac{1}{n}\rkv, \label{eq:bou3}\\
\|S \, U \|\tinf \, \,   \|\hPsi^{-1}\| & \leq \Ctau\, \lkv \frac{\sqrt{K}\, \epsrn}{\sqrt{n}}   + 
\frac{\sqrt{\log n}}{\sqrt{n}\, \sqrt{\rhon\,n\, L}} \rkv, \label{eq:bou4}
\end{align}
where $\epsrn$ is defined in \fr{eq:epsrn}.

Hence, using \fr{eq:bound2}, \fr{eq:U2_inf},  \fr{eq:bou3} and \fr{eq:bou4}, and the 
the fact   that the second term in \fr{eq:bou4} is asymptotically smaller than the first,
we obtain upper bounds for the components $R_i$, $i \in [5]$, of $R$ in \fr{eq:R}:
\begin{align*}
R_1 & \leq \Ctau\, K^{1/2}\, n^{-1/2}\, \epsrn, \\
R_2 & \leq \Ctau\, K^{1/2}\, n^{-1/2}\, \epsrn, \\
R_3 & \leq \Ctau\, K^{1/2}\, n^{-1/2}\, \epsrn^2, \\
R_4 & \leq \Ctau\, [n^{-1} + (\rhon\, n)^{-1}\, L^{-1/2}\, (\log n)^{1/2}]\, \epsrn, \\
R_5 & \leq \Ctau\, K^{1/2}\, n^{-1/2}\, \epsrn^2. 
\end{align*}
Since $\epsrn = o(1)$, we arrive at \fr{eq:error_2inf_COSIE}.
\\


\noindent
{\bf Proof of Corollary~\ref{cor:clust_between}  \ }
Validity of Corollary~\ref{cor:clust_between}   follows directly from Theorem~\ref{thm:clust_between}.



\subsection{Proofs of supplementary statements }
\label{sec:proof_suppl}


\begin{lem} \label{lem:balanced_groups}  
Let Assumption~{\bf A1} hold.  Denote 
\be \label{eq:hnk_hLm}
L_m =  \sum_{l=1}^M I(s(l)=m), \quad   m \in [M]. 
\ee 
If $t \leq (\lowc_{\pi})^2\, L/(2 M^2 \log L) -1$, then,  
\be \label{eq:hLm}
\PP  \lfi \bigcap_{m=1}^M\, \lkr   
\frac{\lowc_{\pi}\, L}{2\, M}  \leq  L_m \leq \frac{3\, \highc_{\pi}\ L}{2\, M} \rkr
\rfi  \ge 1- 2 L^{-t}.
\ee 
\end{lem}


\noindent
{\bf Proof of Lemma~\ref{lem:balanced_groups}.\ }
Note that  $L_m  \sim \text{Binomial}(\pi_m , L)$  for a fixed $m$. By Hoeffding inequality, for any $x > 0$
\bes
\PP \left \{ \left |L_m/L  - \pi_m \right | \ge x \right \} \le 2 \exp\{-2L x^2\}
\ees
Then, applying Assumption~{\bf A1} and the union bound, obtain
\begin{align*}
& \PP \lfi  \bigcap_{m=1}^M\,  \lkr  \lowc_{\pi}\,L/M - L x \leq L_m  \leq   \highc_{\pi}\,\,L/M + L x \rkr  \rfi \\
& \ge 1- 2 M\, \exp\{-2L x^2\}
\end{align*}
Now, set $x=\sqrt{\log (L^t\, M)/(2 L)}$, which is equivalent to $M\, \exp\{-2L x^2\} = L^{-t}$.
If  let $L$ is large enough, so that $x < \lowc_{\pi}/(2\, M)$, then \fr{eq:hLm} holds. 
The latter    is guaranteed by $t+1 \leq (\lowc_{\pi})^2\, L/(2 M^2 \log L)$.
 \\


\begin{lem} \label{lem:SighSigrel}  
Let Assumptions~{\bf A1}--{\bf A4} hold. 
Let $n$ be large enough, so that 
\be \label{eq:n_lowbound}
n \geq 2\, (C_\tau)^2\, (\highc)^2 \, (\lowc)^{-4}\, (K + \tau \log n)
\ee
where $C_\tau$ is the constant in \fr{eq:hsig_oper_er}
Then, for any fixed constant $\tau$,    one has  
\be \label{eq:laminhSig }
\PP \lfi \min_{\minM}\ \lam_{\min} (\hSig\upm)  \geq \frac{\lowc}{2} \rfi 
\geq 1 - 2  \, M\, n^{-\tau}
\ee  
\end{lem}


\noindent
{\bf Proof of Lemma~\ref{lem:SighSigrel}.\ }
Apply Theorem 6.5 of \cite{wainwright_2019} which states that, under conditions of the Lemma, 
with probability at least $ 1 - 2\,  n^{-\tau}$, one has
\be \label{eq:hsig_oper_er}
\|\hSig\upm - \Sig\upm\| \leq  \frac{C_\tau\, \highc}{\lowc} \, \sqrt{\frac{K + \tau \log n}{n}}.
\ee  
Since $\lam_{\min} (\hSig\upm) \geq \lam_{\min} (\Sig\upm) - \|\hSig\upm - \Sig\upm\|$,
by Assumption~{\bf A3},  obtain that, with probability at least $ 1 - 2\,  n^{-\tau}$ 
\be  \label{eq:temp}
\lam_{\min} (\hSig\upm)  \geq \lowc \lkv 1 -   \frac{C_\tau\, \highc}{\lowc^2} \, \sqrt{\frac{K + \tau \log n}{n}}\rkv. 
\ee 
If $n$ satisfies \fr{eq:n_lowbound}, the expression in the square brackets in \fr{eq:temp} is bounded below by 1/2 and,
by taking the union bound in \fr{eq:temp}, we complete the proof.
\\



\begin{lem} \label{lem:true_Delta}  
Let Assumptions~{\bf A1}--{\bf A4} hold. 
Then, for any fixed constant $\tau$,    one has  
\be \label{eq:XiXj}
\PP \lfi \! \max_{i \neq j}  \| (\tilX^{(i)})^T   \tilX^{(j)}  \|   \leq C_1   \sqrt{n  \log n} \rfi 
\geq 1 - C_2     n^{-\tau},
\ee  
where $C_1$ and $C_2$  in \fr{eq:XiXj} depend on $\tau$   and constants in Assumptions~{\bf A1}--{\bf A4} only.
\end{lem}


\noindent
{\bf Proof of Lemma~\ref{lem:true_Delta}.\ }
Note that 
$\tilX^{(i)} = X\upi - \Pi X\upi =  (X\upi - 1_n \mu\upi) + (1_n \mu\upi - \Pi X\upi)$.
Introduce matrices 
\begin{align} \label{eq:xi_eta}
& \xi\upi = X\upi - 1_n \mu\upmi, \quad \eta\upi = 1_n \mu\upi - \Pi X\upi, \\
& \xi\upi,  \eta\upi \in \RR^{n \times K_i}, \ \     i  \in [M] \nonumber
\end{align}
Observe that, by \fr{eq:Xm_gen}, one has $\EE \xi\upi = \EE \eta\upi = 0$ and that
pairs $(\xi\upi, \eta\upi)$ and $(\xi\upj, \eta\upj)$ are independent for $i \neq j$. 
Note also that 
\bes 
\Del^{(i,j)}= \| (\tilX^{(i)})^T \, \tilX^{(j)}  \|\leq \Del_1^{(i,j)} + \Del_2^{(i,j)} + \Del_3^{(i,j)}+\Del_4^{(i,j)},
\ees
where 
$\Del_1^{(i,j)} = \|(\xi\upi)^T \xi\upj \|$,
$\Del_2^{(i,j)} = \|(\xi\upi)^T \eta\upj \|$,
$\Del_3^{(i,j)} = \|(\eta\upi)^T \xi\upj \|$ and $\Del_4^{(i,j)} = \|(\eta\upi)^T \eta\upj \|$.
Below, we construct upper bounds for $\Del_k$, $k =  1,2,3,4$.
\\

\noindent
\underline{An upper bound for $\Del_1^{(i,j)}$.}  \ 
Define matrix $\Psi = (\xi\upi)^T \xi\upj \! \in \! \RR^{K_i \times K_j}$ and note that
\bes
\Psi = \sum_{t=1}^n \Psi_t, \quad   \Psi_t  = (\xi\upi(t,:))^T\, \xi\upj(t,:), \quad \EE  \Psi_t = 0, 
\ees
where rank one matrices $\Psi_t$ are i.i.d. for different values of $t \in [n]$.
We apply matrix Bernstein inequality from \cite{Tropp_2011}
\be \label{eq:matr_Bernstein}
\PP \lfi \|\Psi\| \geq  z\rfi \leq (K_i + K_j)   \exp \lfi - \frac{z^2}{ 2 ( n  \del_1 +   z  \del_2/3)}\rfi,
\ee 
with 
$\del_1 = \max \lkv \|\EE(\Psi_1 \Psi_1^T)\|, \|\EE(\Psi_1^T \Psi_1)\|\rkv$ and 
$\del_2 = \displaystyle \max_t \|\Psi_t\|$. It is easy to see that
$\del_2 = \|\xi\upi(t,:)\|  \|\xi\upj(t,:)\| \leq 1$   by Assumption~{\bf A2}.
In order to upper-bound $\del_1$, note that
\begin{align*}
\EE(\Psi_1 \Psi_1^T) & = \EE \| \xi\upj(t,:)\|^2 \, \EE \lkv (\xi\upi(1,:))^T\, \xi\upi(1,:)  \rkv\\
& \leq  \EE \lkv (\xi\upi(1,:))^T\, \xi\upi(1,:)  \rkv
\end{align*}
and, by  \fr{eq:Xm_gen},  $ \EE \lkv (\xi\upi(1,:))^T\, \xi\upi(1,:)  \rkv = \Sig\upi$.
Therefore, $\| \EE(\Psi_1 \Psi_1^T) \| \leq \highc$.
Similarly,
\begin{align*} 
\EE(\Psi_1^T \Psi_1) & = \EE \lkv \xi\upi(1,:) \,  \EE \lkr (\xi\upj(1,:))^T \, \xi\upj(1,:)  \rkr (\xi\upi(1,:))^T \rkv \\
& =    \EE \lkv \xi\upi(1,:) \,   \Sig\upi (\xi\upi(1,:))^T \rkv \ 
\end{align*}
and, hence
\bes 
\| \EE(\Psi_1^T \Psi_1 )  \| \leq \highc\,    \EE \|\xi\upi (1,:)\| ^2 \leq  \highc
\ees
Therefore, combination of  \fr{eq:matr_Bernstein}
and the union bound over $i,j \in [M]$ yield, for any $\tau >0$,
\be \label{eq:Del1_bound}
\PP \lfi \max_{i \neq j}\ \Del_1^{(i,j)} \leq C \, \sqrt{n\, \log (M^2\, n)} \rfi \geq  1 -   n^{-\tau}.
\ee


Now, consider $\Del_2^{(i,j)} = \|(\xi\upi)^T \eta\upj \|$. Observe that, due to $n^{-1} 1_n^T X\upi = \barX\upi$,
one has
\begin{align*} 
(\xi\upi)^T \eta\upj & = (X\upi - 1_n \mu\upmi)^T  (1_n \mu\upj - \Pi X\upj) \\
& = n (\barX\upi - \mu\upi)^T (\barX\upj - \mu\upj),
\end{align*}
and a similar calculation applies to  $(\eta\upi)^T \xi\upj$ and $(\eta\upi)^T \eta\upj$.
Therefore,  for $i,j \in [M]$
\be \label{eq:Del234}
\Del_k^{(i,j)} \leq n \, \|\barX\upi - \mu\upi\| \,  \|\barX\upj - \mu\upj\|, \quad k=2,3,4.
\ee
It is known (see, e.g., \cite{wainwright_2019})  that if $\zeta$ is a zero mean sub-gaussian $k$-dimensional 
random vector with the sub-gaussian parameter $\sig$, then 
\be \label{eq:subgaus_vect}
\PP \lfi \|\zeta\| \leq 4 \sig \sqrt{k} + 2 \sig \sqrt{\log(1/\eps)} \rfi \geq 1 - \eps.
\ee
Note that vectors $(\Sig\upi)^{-1/2} \,  (X(j,:) - \mu\upi)$ are anisotropic $K_i$ dimensional random vectors 
of length at most $1 /\sqrt{\lowc}$. Hence, vectors $(\Sig\upi)^{-1/2} \,  (\barX\upi - \mu\upi)$
are anisotropic sub-gaussian with   $\sig \leq 2/\sqrt{\lowc\, n}$, so that  
\begin{align} 
\hspace*{-4mm} \PP \lfi \|\barX\upi - \mu\upi\| \right. & \left. \leq \frac{4 \sqrt{\highc} [ 2 \, \sqrt{K} + \sqrt{\log (M^2\, n^{\tau})}] }{\sqrt{n\, \lowc}} \rfi
\nonumber\\
& \geq 1 -    M^{-2}\,   n^{-\tau}. \label{eq:barXupi}
\end{align}
Combining   \fr{eq:Del234}, \fr{eq:barXupi} and the union bound over $i,j \in [M]$, $i \neq j$, obtain, for $k = 2,3,4$:
\be \label{eq:Delk_bound}
\PP \lfi \max_{i \neq j}\ \Del_k^{(i,j)} \leq C \,   \sqrt{K +  \log (M^2\, n)} \rfi \geq  1 -   n^{-\tau}, 
\ee
which, together with \fr{eq:Del1_bound} and   $\log(M^2 \, n) \propto \log n$,  complete  the proof.

\medskip


\begin{lem} \label{lem:U_hU}  
Let Assumptions~{\bf A1}--{\bf A5} and \fr{eq:rhon_cond}   hold. 
Then, if $n$ is large enough, for any fixed constant $\tau$,  with probability at least $1-  2\, n^{-(\tau - \tau_0)}$,  one has  
\be \label{eq:U_hU}
\max_{\linL}\, \left\| \sinTe (\hU_A\upl, \tilU_P\upl) \right\|_F \leq C\, K^{1/2}\, (n \rhon)^{-1/2}.
\ee 
\end{lem} 
  

\noindent
{\bf Proof of Lemma~\ref{lem:U_hU}.\ }
It follows from \cite{levina_vershynin2018} that, under condition \fr{eq:rhon_cond},
one has
\bes
\EE \|A\upl - P\upl\| \leq C \lkr \sqrt{n \rhon}+ \sqrt{\log n} \rkr.
\ees
Also, by formula (2.4) of \cite{Benaych_Georges_2020}, derive that, for any $t>0$,
and some absolute constant $\tilC$ one has  
\bes
\PP \lfi \|A\upl - P\upl\|  - \EE \|A\upl - P\upl\| \geq t \rfi \leq 2\, \exp(- \tilC\, t^2).
\ees
Setting $t = \sqrt{\tilC^{-1}\, \log(n^\tau\, L)}$, combining the two inequalities above and applying the union bound,
obtain
\begin{align} \label{eq:probab_APupl}
\PP \lfi \|A\upl - P\upl\|  \right. \leq \left. C \lkr \sqrt{n \rhon} + \sqrt{\log(n^\tau\, L)} \rkr \rfi \\
\geq 1 - 2\, n^{-\tau},  \nonumber
\end{align}
where the constant $C$ in \fr{eq:probab_APupl} depends on $\tau$ and the constants in Assumptions~{\bf A1}--{\bf A5}.
Taking into account that\\ $\|\tilA\upl - \tilP\upl\| \leq \|A\upl - P\upl\|$ and that, under assumptions of the Lemma,
$\log(n^\tau\, L)  \leq C \log n \leq C n \rhon$, derive that 
\be \label{eq:probab_tilAPupl}
\PP \lfi \|\tilA\upl - \tilP\upl\|  \leq C   \sqrt{n \rhon}   \rfi \geq 1 - 2\, n^{-\tau}.
\ee
By Davis-Kahan Theorem (see, e.g., \cite{10.1093/biomet/asv008}), if $s(l) = m$, obtain
\be  \label{eq:Davis-Kahan}
 \left\| \sinTe (\hU_A\upl, \tilU_P\upl) \right\|_F \leq 
\frac{ 2\, \sqrt{K_m} \|\tilA\upl - \tilP\upl\|}{\sig_{K_m} (\tilP\upl)}.
\ee
Here, by \fr{eq:tilbp1} and \fr{eq:X_P_svds}, 
$\tilP\upl = \tilX\upm   B\upl (\tilX\upm)^T = 
\rhon\, \tilU\upm \tilD\upm  (\tilO\upm)^T B_0\upl (\tilD\upm  (\tilO\upm)^T)^T (\tilU\upm)^T$,
so that, by   \fr{eq:hSig_def},  
\bes 
\tilP\upl = (n-1)\, \rhon\, \tilU\upm \sqrt{\hSig\upm}\, B_0\upl\, (\sqrt{\hSig\upm})^T \, (\tilU\upm)^T.
\ees
Therefore, Assumption~{\bf A5} and Lemma~\ref{lem:SighSigrel} imply that
\bes
\sig_{K_m} (\tilP\upl) \geq C n \rhon 
\ees
Combining the latter with \fr{eq:probab_tilAPupl} and \fr{eq:Davis-Kahan}, using the union bound over 
$\linL$,  and taking into account that $K_m \leq K$, obtain \fr{eq:U_hU}.

\medskip


\begin{lem} \label{lem:op_2infsuppl}  
Let  $\tH = U_H \Lam_H V_H^T$ be the SVD of $\tH = U^T\hU$. 
Denote  $W_U =  U_H  V_H^T \in \calO_K$.    Then,
\be \label{eq:W_U}
\|U^T\hU - W_U\| \leq \| \sin\Te (U,\hU)\|^2. 
\ee
\end{lem} 


\noindent
{\bf Proof of Lemma~\ref{lem:op_2infsuppl}.\ }
Observe that $\Lam_H = \cos(\Te)$, where $\Te$ is the diagonal matrix of the principal angles between the subspaces.
Therefore, due to inequality $\sin(\te/2) \leq \sin \te/\sqrt{2}$ for $0 \leq \te \leq \pi/2$, one has
\begin{align*}
\|W_U - \tH\| & = \|U_H (\cos \Te - I_K) V_H^T \| = 2 \|\sin^2 (\Te/2)\| \\
& \leq \| \sin \Te\|^2 = \| \sin\Te (U,\hU)\|^2.
\end{align*}

\medskip


\begin{lem} \label{lem:Del12_bounds}  
Let  assumptions of Theorem~\ref{th:error_Um_est} hold with $M=1$.
Then, for any $\tau >0$, and absolute constant  $\Ctau$ that depends on $\tau$ and constants 
in Assumptions~{\bf A1}--{\bf A5} only,  with probability at least 
$1 - C \, n^{-(\tau -   (\tau_0 + 1))}$, one has for $j=0,1$
\begin{align}  
& \max_{\linL}\, \max_{j=0,1}\, \|\Xi\upl\, U^j \|\tinf \leq \Ctau \, \sqrt{n\, \rhon\, \log n},   \label{eq:Ub1} \\
& \sumlL \|\Xiii\upl  \|^2_F  \leq \Ctau \, \rhon\, n^2 \, L, \label{eq:Ub2}\\
& \sumlL \|\Xiii\upl\, U \|^2_F  \leq \Ctau \, \rhon\, n\, K\,  \, L. \label{eq:Ub3}
 \end{align}
\end{lem} 


\noindent
{\bf Proof of Lemma~\ref{lem:Del12_bounds}.\ }
Note that the first inequality follows from Theorem 5.2 of \cite{lei2015} and  the fact that 
\bes
\|\Xi\upl\, U^j \|\tinf   \leq \|\Xi\upl\| \|U^j\| \leq  \|\Xi\upl\|.
\ees
In order to prove \fr{eq:Ub2}, note that 
\be \label{eq:expect_Xiupl}
\EE \lkr \sumlL \lnor \Xi \upl \rnor^2_F \rkr \leq L \, n^2 \,\rhon
\ee
Furthermore, introduce random variables $\etaijl = \lkv \Xi\upl(i,j)\rkv^2    - \EE\lkv \Xi\upl(i,j)\rkv^2$.
Observe that   $\etaijl$ are independent zero mean random variables,  for $1 \leq i \leq j \leq n$ and $\linL$,
with $|\etaijl| \leq 1$ and $\Var (\etaijl) \leq \rhon$. Then, application of the Bernstein inequality yields
\begin{align*} 
\PP \lfi \sumlL\, \sum_{1 \leq i \leq j \leq n} \, \etaijl \geq t_1 \rfi & \leq
2\, \exp \lkr - \frac{C\, t_1^2}{L \, n^2 \,\rhon} \rkr \\
& + 2\, \exp \lkr -C \, t_1\rkr.
\end{align*}
Setting $t_1 = \Ctau \, \sqrt{\rhon\, n^2\, L\, \log n}$ and taking into account \fr{eq:expect_Xiupl},
obtain, with probability at least $1 - C \, n^{-\tau}$, that
\bes 
\sumlL\, \|\Xiii\upl  \|^2_F  \leq L \, n^2 \,\rhon + \Ctau \, \sqrt{\rhon\, n^2\, L\, \log n}.
\ees
Then,  \fr{eq:Ub2} follows from the fact that, in the last inequality, the first term dominates the second,
and from the fact that $\|\Xiii\upl  \|^2_F \leq \|\Xi\upl  \|^2_F$ .

In order to assert the validity of  \fr{eq:Ub3}, observe that $\rank\lkr \Xiii\upl\, U\rkr \leq K$,
and, therefore, by Theorem 5.2 of \cite{lei2015}, derive
\bes
\|\Xiii\upl\, U  \|^2_F \leq C\, K\, \|\Xi\upl\|^2 \leq C\, K\, n\, \rhon.
\ees



\section*{Acknowledgment}

The   author of the paper gratefully acknowledges partial support by National Science Foundation 
(NSF) grants DMS-2014928 and   DMS-2310881.

\bibliographystyle{IEEEtranS.bst}
\bibliography{MultilayerNew.bib}

\ifCLASSOPTIONcaptionsoff
  \newpage
\fi

%

\begin{IEEEbiography}[{\includegraphics[width=1in,height=1.25in,clip,keepaspectratio]{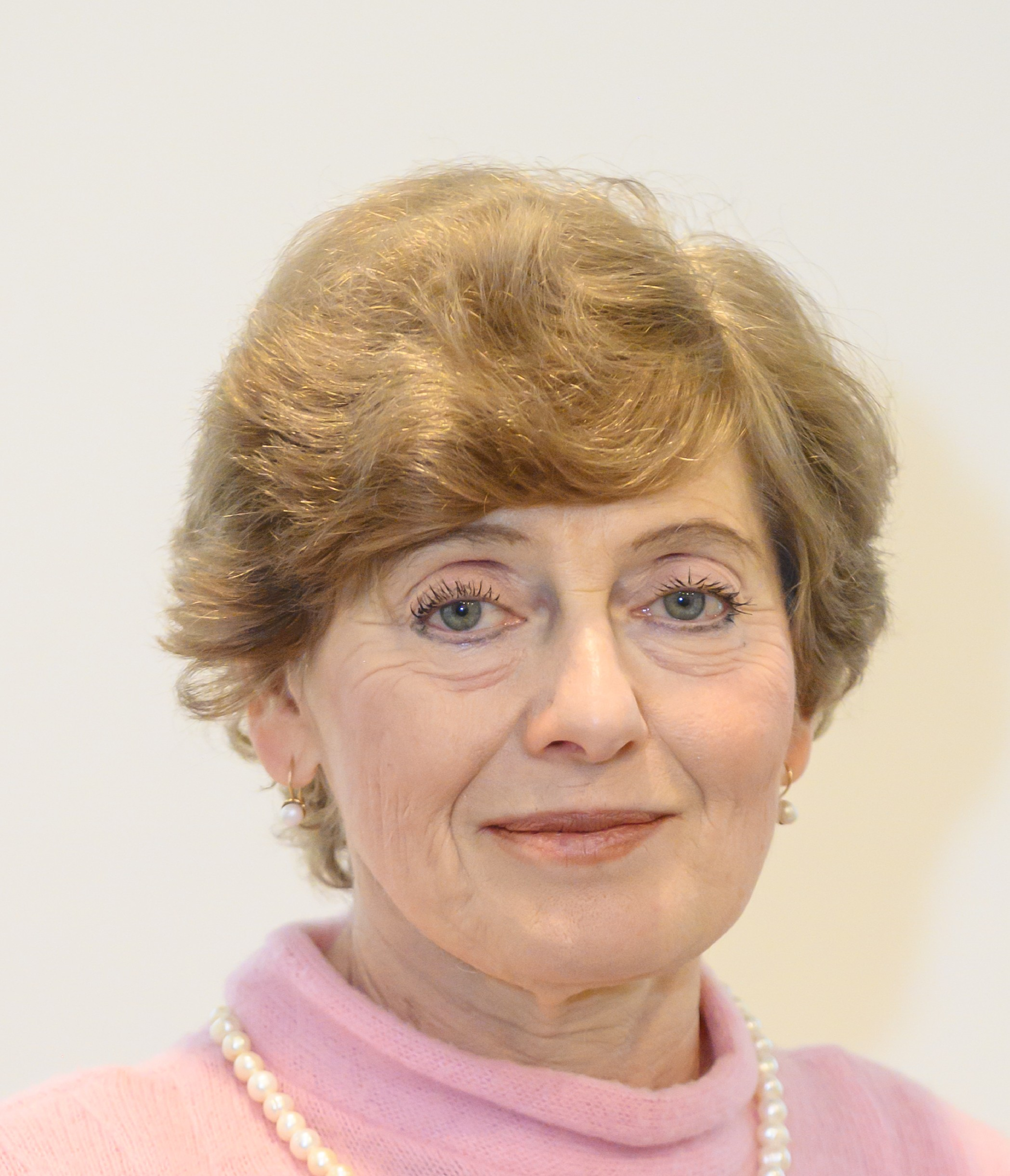}}]{Marianna Pensky}
 is a Professor with the Department of Mathematics, University of Central Florida. 
She is a Fellow of the Institute of Mathematical Statistics and a Fellow of the American Statisical Association. 
She is also  an executive editor of the Journal of the Statistical Planning and Inference. 
\end{IEEEbiography}



\end{document}